\documentclass[final,5p,times,twocolumn]{elsarticle}

\usepackage{amssymb}

\usepackage[displaymath]{lineno}

\usepackage{XiGrpCommon}
\graphicspath{{fig/}}
\usepackage{multirow}
\usepackage{tabularx}
\usepackage{longtable}
\usepackage{rotating}
\usepackage{amsmath}

\renewcommand{\RevisedText}[1]{#1}

\journal{Journal of Non-Newtonian Fluid Mechanics}

\begin{document}

\begin{frontmatter}



\title{Inertia-driven and elastoinertial viscoelastic turbulent channel flow simulated with a hybrid pseudo-spectral/finite-difference numerical scheme}


\author[]{Lu Zhu}
\author[]{Li Xi\corref{cor1}}
\cortext[cor1]{corresponding author,
	E-mail: xili@mcmaster.ca;
	web: www.xiresearch.org}
\address[]{Department of Chemical Engineering, McMaster Universtiy, Hamilton, Ontario L8S 4L7, Canada}
\date{\today}

\begin{abstract}
Numerical simulation of viscoelastic flows is challenging because of the hyperbolic nature of viscoelastic constitutive equations.
Despite their superior accuracy and efficiency, pseudo-spectral methods require the introduction of artificial diffusion (AD) for numerical stability in hyperbolic problems, which alters the physical nature of the system.
This study presents a hybrid numerical procedure that integrates an upwind total variation diminishing (TVD) finite-difference scheme, which is known for its stability in hyperbolic problems, for the polymer stress convection term into an overall pseudo-spectral numerical framework.
Numerically stable solutions are obtained for Weissenberg number well beyond $\RevisedText{\bigO}(100)$ without the need for either global or local AD.
Side-by-side comparison with an existing pseudo-spectral code reveals the impact of AD, which is shown to differ drastically between flow regimes.
Elastoinertial turbulence (EIT) becomes unphysically suppressed when AD, at any level necessary for stabilizing the pseudo-spectral method, is used. This is attributed to the importance of sharp stress shocks in its self-sustaining cycles.
Nevertheless, in regimes dominated by the classical inertial mechanism for turbulence generation, there is still an acceptable range of AD that can be safely used to predict the statistics, dynamics, and structures of drag-reduced turbulence.
Detailed numerical resolution analysis of the new hybrid method, especially for capturing the EIT states, is also presented.
\end{abstract}


\end{frontmatter}



\section{Introduction}\label{Sec_intro_c6}
Research on the friction drag reduction (DR) in turbulent flows is of immense interest to the fluid mechanics community for its theoretical and practical implications.
A well-known technique to induce DR is through the addition of polymer additives, which, under certain conditions, can result in up to $80\%$ of friction drag reduction in pipeline systems~\citep{Virk_AIChEJ1975,Myska_Stern_CollPolySci1998}.
Such extraordinary DR performance has attracted great attention over the decades~\citep{Xi_POF2019}. 

Significant progress has been made in uncovering the nature of polymer-induced DR using both experimental and numerical tools~\citep{Virk_AIChEJ1975,Sureshkumar_Beris_POF1997,Xi_POF2019}.
It has been widely accepted that polymer stress can reduce drag by lowering turbulent intensity~\citep{Sureshkumar_Beris_POF1997,Xi_Graham_JFM2010,Housiadas_Beris_POF2003,Li_Kawaguchi_POF2005,Kim_Adrian_JFM2007}.
In this scenario, same as Newtonian flow, turbulence is still driven by flow instabilities associated with strong inertial effects. This classical type of turbulence will be referred to as inertia-driven turbulence (IDT) hereinafter.
Polymer stress can subdue turbulent fluctuations by suppressing coherent flow structures, which is the main DR mechanism in the IDT regime~\citep{DeAngelis_Piva_CompFl2002,Sibilla_Baron_POF2002,Kim_Adrian_JFM2007,Zhu_Xi_POF2019,Zhu_Xi_JNNFM2019}.
It became later known that when polymer elasticity, measured by the Weissenberg number $\mathrm{Wi}$, is sufficiently strong, it can turn into to a driving force for flow instability~\citep{Min_Choi_JFM2003a,Min_Choi_JFM2003b,Dallas_Vassilicos_PRE2010}.
Evidence for this so-called elastoinertial turbulence (EIT) emerged more recently in experiments~\citep{Samanta_Hof_PNAS2013,Choueiri_Hof_PRL2018,Chandra_Shankar_JFM2018,Chandra_Shankar_JFM2020}.
At high levels of DR when polymer stress is strong enough to interrupt the self-sustaining cycles of IDT, EIT steps in to keep turbulence sustained~\citep{Zhu_Xi_arXiv2020}.

Direct numerical simulation (DNS) has been pivotal to our fundamental understanding of polymer DR for over two decades.
DNS of viscoelastic turbulence requires coupling the Navier-Stokes (N-S) equation with a constitutive equation describing the polymer stress field under flow.
Since the seminal work by \citet{Sureshkumar_Beris_POF1997}, the FENE-P (finitely-extensible nonlinear elastic dumbbell model with the Peterlin closure approximation) model~\citep{Bird_Curtis_1987} has been the most widely used constitutive equation in the literature, which is also more suited for dilute polymer solutions, although Oldroyd-B and Giesekus models have also been used~\citep{Dimitropoulos_Beris_JNNFM1998, Min_Choi_JFM2003a, Yu_Kawaguchi_JNNFM2004}.

(Pseudo-)spectral methods (SMs) are widely used in DNS of flow turbulence in canonical flow geometries for their high efficiency and accuracy (exponential convergence), in which numerical differentials are evaluated using Fourier and Chebyshev transforms while nonlinear operators are still calculated in the physical domain~\citep{gottlieb1977numerical,Kim_Moin_JFM1987,Canuto_Hussaini_1988,rogallo1984numerical}.
However, in the case of viscoelastic turbulence, the hyperbolic nature (due to the lack of a diffusion term) of polymer constitutive equations make numerical stability practically unattainable at high $\mathrm{Wi}$.
Adding an artificial diffusion (AD) term is a common practice for numerical stability of SMs in such problems. \citet{Sureshkumar_Beris_JNNFM1995} initiated the practice of using AD in the DNS of viscoelastic flow. It was argued that if the AD magnitude decreases in proportion to $\delta_x^2/\delta_t$ ($\delta_x$ and $\delta_t$ are the mesh size and time step, respectively), the numerical solution converges to that of the unaltered equation system at the $\delta_x\sim\delta_t\to0$ limit~\citep{Sureshkumar_Beris_POF1997,Dimitropoulos_Beris_JNNFM1998}.
This practice of altering the physical equation for numerical expediency has been subject to scrutiny over the years.
The extra AD term is known to smear the polymer stress profile and reduce the stress gradient at shocks -- sharp discontinuity-like stress changes~\citep{Min_Choi_JNNFM2001,Vaithianathan_Collins_JNNFM2006}.
It was believed that this effect can cause the underprediction of the level of DR and overprediction of the $\mathrm{Wi}$ for the onset of DR $\mathrm{Wi}_\text{onset}$, which was indeed seen when AD was imposed on finite-difference (FD) DNS solvers~\citep{Yu_Kawaguchi_JNNFM2004,Vaithianathan_Collins_JNNFM2006}.
Nevertheless, DNS using SMs with AD has reproduced key experimental observations and has thus been widely used by researchers with the general belief that the flow physics is not significantly altered as long as AD is kept small in magnitude~\citep{Housiadas_Beris_POF2003,Ptasinski_Nieuwstadt_JFM2003,Li_Khomami_JNNFM2006,Xi_Graham_JFM2010,Thais_Mompean_IJHFF2013,Lopez_Hof_JFM2019}.

Questions about AD usage became reignited in recent years as the focus in the area shifts towards EIT, which is believed to be associated with maximum drag reduction (MDR) -- the asymptotic upper limit of DR by polymers~\citep{Samanta_Hof_PNAS2013,Choueiri_Hof_PRL2018,Zhu_Xi_arXiv2020}.
Flow instabilities relying on elasticity as a driving force often involve stress shocks and are thus more susceptible to numerical artifacts of AD~\citep{Xi_Graham_JFM2009,Thomases_JNNFM2011,Gupta_Vincenzi_JFM2019}.
Such effects are well studied for low-$\mathrm{Re}$ elastic instabilities but are often considered less significant in high-$\mathrm{Re}$ flow.
However, a recent DNS study using a FD solver found that adding AD can suppress EIT (which is a high-$\mathrm{Re}$ and high-$\mathrm{Wi}$ phenomenon) solutions unless the AD magnitude is at least 1-2 orders of magnitude smaller than the level required to stabilize SMs~\citep{Sid_Terrapon_PRFluids2018}. Even for smaller AD, EIT solution continues to depend on this artificial numerical parameter.

Finite difference methods (FDMs) provide an alternative for spatial discretization. With proper selection of the discretization scheme for the convection term in the constitutive equation, the use of AD can be minimized and, sometimes, avoided altogether.
The earliest attempt to simulate viscoelastic turbulence with a FDM was reported by \citet{Min_Choi_JNNFM2001,Min_Choi_JFM2003a} who adopted a third-order compact upwind difference (CUD3) scheme to treat the convection term. AD is still required but only at grid points where the polymer conformation tensor looses its positive definiteness (one of its fundamental physical attributes), which account for a very small fraction of the domain.
Usage of local AD (LAD) is not possible in SMs, where numerical discretization is intrinsically global and only global AD (GAD) can be applied.
\citet{lee2017simulations} used a third-order weighted essentially non-oscillatory (WENO) scheme~\citep{shu1998essentially,shu2009high} for the convection term, which again required LAD to stabilize local numerical oscillations.
WENO schemes provide high orders of accuracy at the expense of computational efficiency.
A faster second-order upwind total variation diminishing (TVD) scheme was used by \citet{Yu_Kawaguchi_JNNFM2004}. (TVD schemes are a classical and widely adopted group of discretization schemes for hyperbolic problems~\citep{harten1983high,sweby1984high,leveque1996high}.)
Its excellent numerical stability allowed them to completely eliminate AD for $\mathrm{Wi}$ up to $45$ reported in the study (corresponding to $53\%$ of DR).
\citet{Vaithianathan_Collins_JNNFM2006} adapted the \citet{kurganov2000new} scheme to viscoelastic DNS and achieved numerical stability without AD in homogeneous shear flow. Their method was shown to analytically preserve the positive definiteness of the polymer conformation tensor with reasonably small time steps.
On the other hand, for FENE-P, the polymer conformation tensor must also satisfy the finite extensibility (upper boundedness) constraint -- i.e., total polymer extension cannot exceed its contour length, which can be analytically guaranteed using an implicit time-stepping formulation for the trace of the polymer conformation tensor~\citep{Vaithianathan_Collins_JCompPhys2003,Dubief_Lele_FTC2005}.

Improved numerical stability in those FDM approaches comes at the expense of either numerical accuracy or efficiency. High-order FDM schemes have been proposed, which can reduce the accuracy gap with SMs at least in solving the N-S equation~\citep{laizet2009high,Dallas_Vassilicos_PRE2010}.
However, most researchers opt to use second-order central difference schemes for the rest of the equation system other than the polymer convection term, to avoid further increasing the computational burden in an already demanding numerical problem.
Since the convection term is the primary source of numerical instability, it is natural to expect that FD discretization does not need to be applied for the whole equation system to reap its numerical advantage.
A similar idea was used in \citet{Vaithianathan_Collins_JNNFM2006} in a 3D periodic domain, which suggested that numerical stability is apparently insensitive to the discretization scheme of other terms.

In this study, we present a pseudo-spectral/finite-difference hybrid method (HM) for DNS of viscoelastic channel flow, which is spatially periodic in two dimensions but bounded in the wall-normal direction.
The convection term is discretized with a second-order conservative TVD FD scheme
\RevisedText{while the SM is used for all other terms, which maximally preserves its benefits including higher accuracy and efficiency.
The algorithm is found to be numerically stable without the need of AD, either local or global, for $\mathrm{Wi}$ well beyond $\RevisedText{\bigO}(100)$, which is at least one order of magnitude higher than that tested in the previous TVD study~\citep{Yu_Kawaguchi_JNNFM2004}.
Furthermore, success of TVD in an overall SM algorithm framework, which we will demonstrate, is practically important.
It}
means an existing SM code can be adapted, with relative ease, to this new HM and break away from its reliance on AD.
Likewise, an existing Newtonian DNS code using a SM, which is most common in canonical flows, can be expanded for viscoelastic DNS following this paradigm.
\RevisedText{The overall computational cost of the new HM algorithm is even lower than that of a pure SM approach, as avoiding the AD term simplifies the time-stepping algorithm of the constitutive equation.}

\RevisedText{Another motivation of us is to perform a thorough investigation into the effects of AD on DNS solutions in both IDT and EIT regimes.}
Previous studies on this issue all added a diffusion term to a pure FD algorithm framework, which, strictly speaking, measured the effects of AD on FDMs~\citep{Min_Choi_JNNFM2001,Yu_Kawaguchi_JNNFM2004,Dubief_Lele_FTC2005,Vaithianathan_Collins_JNNFM2006,Sid_Terrapon_PRFluids2018}.
\RevisedText{Comparison between the new HM (no AD) and the traditional SM+GAD approaches (both codes now available in our group) offers the most direct insight into the AD effects on the SM.
This investigation is particularly relevant after the discovery of EIT and, in particular, its numerical suppression by AD.
Since SM+GAD has been the most widely used numerical approach for over two decades, it is important to understand whether such artifacts are limited to the EIT regime only and to what extent results from SM+GAD can be trusted in the IDT regime.}

In the following, we will first describe the HM numerical procedure, compare it with the SM approach, and provide all simulation parameters in \cref{Sec_method_c6}.
In \cref{Sec_STG}, correctness of our implementation is validated by comparing with an established SM code based on transient trajectories from a steak transient growth (STG) simulation.
Performance of these two algorithms and the effects of AD in the statistical steady state of turbulence, in both IDT and EIT regimes, are compared and discussed in \cref{Sec_ResSS}.
Finally in \cref{Sec_Mesh}, we will discuss the mesh resolution sensitivity of the new HM scheme (in both IDT and EIT regimes).
The paper is concluded in \cref{Sec_conclude}.
\ref{appendix:tau} provides details on satisfying the divergence-free constraint and no-slip boundary condition in the flow field.
In \ref{Sec_schmComp}, the rationale of choosing the TVD scheme for the convection term is discussed and the performance (speed and shock-capturing capability) of various schemes is compared in a simple benchmark problem.

\section{Methodology}\label{Sec_method_c6}

\subsection{Computational domain and governing equations}\label{Sec_GovEqu}

\begin{figure}
	\centering				
	\includegraphics[width=.6\linewidth, trim=0mm 0mm 0mm 0mm, clip]{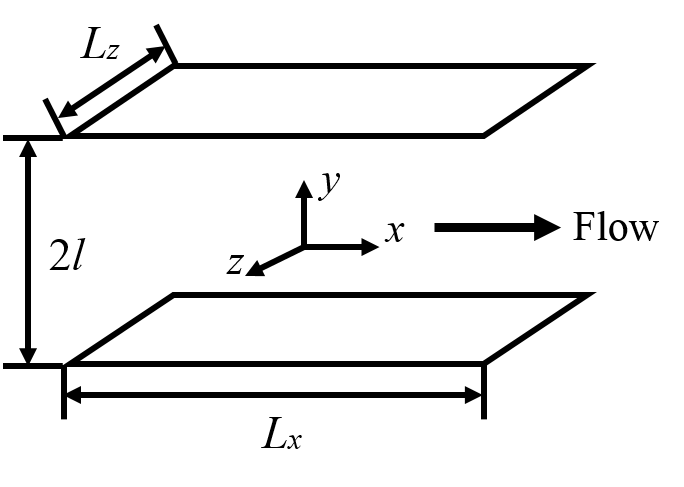}			
	\caption{Schematic of the flow geometry.}\label{fig:geomery}
\end{figure}

Our HM is implemented for viscoelastic plane Poiseuille flow. The geometry of the computational domain is illustrated in \cref{fig:geomery}. The incompressible fluid is driven by a constant mean pressure gradient in the $x$-direction (streamwise). Two parallel walls are located in the $y$-direction (wall-normal) with a separation of $2l$.
Periodic boundary conditions are applied to $x$- and $z$-directions (spanwise) with the periods of $L_x$ and $L_z$, respectively. The no-slip boundary condition is applied to the walls. The half-channel height $l$ and Newtonian laminar centerline velocity $U$ are used to nondimensionalize all flow length and velocity quantities, respectively.
Pressure $p$ and time $t$ are scaled by $\rho U^2$ ($\rho$ is the total density of the solution) and $l/U$, respectively. 

The viscoelastic DNS algorithm solves an equation system that couples the momentum balance (\cref{equ_momentum_c6}) and continuity equations (\cref{equ_continuity_c6}) with the FENE-P constitutive equations (\cref{equ_Fene1_c6,equ_Fene2_c6}), as respectively given by
\begin{gather}
	\frac{\partial\boldsymbol{v}}{\partial t}+\boldsymbol{v}\cdot \boldsymbol{\nabla}\boldsymbol{v}=-\boldsymbol{\nabla}p+\frac{\beta}{\mathrm{Re}}\nabla^2\boldsymbol{v}+\frac{2\left(1-\beta\right)}{\mathrm{Re}\mathrm{Wi}}\left(\boldsymbol{\nabla}\cdot\boldsymbol{\tau }_p\right),
	\label{equ_momentum_c6}
	\\
	\boldsymbol{\nabla}\cdot\boldsymbol{v}=0
	\label{equ_continuity_c6}
\end{gather}
and
\begin{gather}
\begin{split}
	\frac{\partial\boldsymbol{\alpha }}{\partial t}
		+\boldsymbol{v}\cdot\boldsymbol{\nabla}\boldsymbol{\alpha }
		&-\boldsymbol{\alpha }\cdot\boldsymbol{\nabla} \boldsymbol{v}-\left(\boldsymbol{\alpha }\cdot\boldsymbol{\nabla} \boldsymbol{v}\right)^\mathrm{T}
		\\
		&= \frac{2}{\mathrm{Wi}}(-\frac{\boldsymbol{\alpha }}{1-\frac{\mathrm{tr}\left(\boldsymbol{\alpha }\right)}{b}} +\frac{b\boldsymbol{\delta}}{b+2}),
\end{split}
	\label{equ_Fene1_c6}
	\\
	\boldsymbol{\tau}_p=\frac{b+5}{b}\left(\frac{\boldsymbol{\alpha}}{1-\frac{\mathrm{tr}\left(\boldsymbol{\alpha }\right)}{b}}-\left(1-\frac{2}{b+2}\right)\boldsymbol{\delta }\right).
	\label{equ_Fene2_c6}
\end{gather}
Here, $\boldsymbol\delta$ denotes the Kronecker delta tensor.
The Reynolds number is defined as $\mathrm{Re}\equiv \rho Ul/\eta$ (where $\eta$ is the total zero-shear rate viscosity of the fluid). 
\RevisedText{%
Note that the velocity scale $U$ is obtained from a Newtonian laminar flow of the same pressure drop (or same mean wall shear stress $\tau_w$) and does not directly measure turbulent velocity.
As a result,}
$\mathrm{Re}$ is directly related to the friction Reynolds number $\mathrm{Re}_\tau\equiv \rho u_\tau l/\eta$ ($u_\tau\equiv \sqrt{\tau_w/\rho}$ is the friction velocity\RevisedText{)}
through $\mathrm{Re}=\mathrm{Re}_\tau^2/2$.
The Weissenberg number, defined as $\mathrm{Wi}\equiv 2\lambda U/l$, is the product of the polymer relaxation time $\lambda$ and the characteristic shear rate (using the wall shear rate of Newtonian flow) $2U/l$;
$\beta \equiv \eta_s/\eta$ is the ratio of the solvent viscosity $\eta_s$ to the total solution viscosity $\eta$ at the zero-shear limit (for dilute solutions, $1-\beta$ is proportional to polymer concentration).
The effect of polymers on the flow is accounted for by the last term on the right-hand side (RHS) of the momentum balance equation (\cref{equ_momentum_c6}). Here, $\boldsymbol{\tau}_p$ is the polymer stress tensor and is modeled by the FENE-P constitutive equations~\citep{Bird_Curtis_1987} which describe a polymer chain as a finitely extensible nonlinear elastic (FENE) dumbbell. 
(``P'' stands for the Peterlin approximation that allows the mathematical closure of the model.)
\Cref{equ_Fene1_c6} solves for the polymer conformation tensor $\boldsymbol{\alpha}\equiv\langle\boldsymbol{QQ}\rangle$ (where $\boldsymbol{Q}$ is the non-dimensional end-to-end vector of dumbbells), which is used to calculate the polymer stress tensor $\boldsymbol{\tau}_p$ through \cref{equ_Fene2_c6}.
The model enforces the finite extensibility of polymers by ensuring that $\mathrm{tr}(\boldsymbol{\alpha})\leq b$ is always satisfied: note that as $\mathrm{tr}(\boldsymbol{\alpha})\to b$, $\boldsymbol\tau_p$ diverges (\cref{equ_Fene2_c6}) and the relaxation term in FENE-P (RHS of \cref{equ_Fene1_c6}) also approaches infinity.
The physical basis of using FENE-P for modeling the flow of dilute polymer solutions was discussed in \citet{Xi_POF2019}.

\subsection{Numerical procedure of the hybrid method (HM)}
\label{Sec_NumAppro}
The new HM algorithm is implemented by adapting an existing pseudo-spectral DNS code that has been extensively used and validated in a large number of previous studies over the course of 10 years.
The earliest version of the SM code was developed by coupling a custom FENE-P solver with the open-source Newtonian DNS code \texttt{Channelflow} by \citet{Gibson_ChFlowCode} (also see \citet{Gibson_Cvitanovic_JFM2008}), which was first used in \citet{Xi_Graham_JFM2010} (see algorithmic details in \citet{Xi_PhD2009}).
That pseudo-spectral FENE-P solver was later parallelized and coupled with a newer parallel version of \texttt{Channelflow~2.0} by \citet{Gibson_ChFlowCode2p0}, which was first used in \citet{Zhu_Xi_JNNFM2018}.
The current HM algorithm is implemented within the framework of the existing SM code with minimal change to the program architecture.
Only the convection term in \cref{equ_Fene1_c6} needs to be discretized using a TVD scheme whereas all other terms retain their pseudo-spectral treatment.
As a result of removing the AD term (which was otherwise added in \cref{equ_Fene1_c6} for the SM), its time-integration algorithm is also simplified.
We expect that existing SM codes from other groups can be similarly adapted.


\subsubsection{Time integration of the Navier-Stokes equation} 
\label{Sec_TimeIntNS}
The numerical procedure for solving the N-S equation (\cref{equ_momentum_c6,equ_continuity_c6}) largely preserves the original algorithm in \texttt{Channelflow} except for the extra polymer stress term.
Without loss of generality, velocity and pressure are first decomposed into base-flow and perturbation components
\begin{gather}
	\boldsymbol{v}=\mathscr{U}\boldsymbol{e}_x+\boldsymbol{v}^\dagger
	\label{eq:decomp:v}\\
	p=\Pi x+p^\dagger
	\label{eq:decomp:p}
\end{gather}
where ``\dag'' indicates the perturbation component and $\boldsymbol{e}_x$ is the unit vector in the $x$-direction. The non-dimensional Newtonian laminar velocity profile $\mathscr{U}(y)=1-y^2$ (the walls are at $y=\pm 1$) and imposed mean pressure gradient $\Pi=-2/\mathrm{Re}$ (per the constant pressure gradient constraint) are chosen as the base-flow velocity and pressure gradient, respectively.
After decomposition, \cref{equ_momentum_c6} is rewritten as, 
\begin{equation}
	\frac{\partial \boldsymbol{v}^{\dagger}}{\partial t}=
		-\boldsymbol{N}-\boldsymbol{\nabla}p^{\dagger}
		+L\boldsymbol{v}^\dagger+\boldsymbol{C}+\boldsymbol{S}
\label{equ_NS_decomposed}
\end{equation}
where
\begin{gather}
	\boldsymbol{N}\equiv \boldsymbol{v}\cdot \boldsymbol{\nabla}\boldsymbol{v},
	\label{eq:N:conv}\\
	L\boldsymbol{v}^\dagger\equiv \frac{\beta}{\mathrm{Re}}\nabla^2 \boldsymbol{v}^{\dagger},\\
	\boldsymbol{C}\equiv
		\left(\frac{\beta}{\mathrm{Re}}\frac{d^2\mathscr{U}}{d y^2}-\Pi\right)
		\boldsymbol{e}_x,\\
	\boldsymbol{S}\equiv \frac{2(1-\beta)}{\mathrm{ReWi}} 
		\boldsymbol{\nabla}\cdot \boldsymbol{\tau}_p
	\label{eq:S}
\end{gather}
are the inertial (nonlinear), viscous (linear), base-flow (constant), and polymer terms, respectively ($L$ is the linear operator).
In practice, it is known that time integration using \cref{eq:N:conv} for the inertial term is numerically unstable~\citep{Zang_ANM1991}.
In our algorithm, the convection form of \cref{eq:N:conv} alternates with the divergence form of
\begin{gather}
	\boldsymbol{N}= \boldsymbol\nabla\cdot(\boldsymbol v\boldsymbol v)
	\label{eq:N:div}
\end{gather}
between consecutive time steps.

Applying fast Fourier transform (FFT), in $x$- and $z$-directions, to all terms, we obtain
\begin{equation}
	\frac{\partial\tilde{\boldsymbol{v}}^{\dagger}}{\partial t}=
		-\widetilde{\boldsymbol{N}}-\widetilde{\boldsymbol{\nabla}}\tilde p^{\dagger}
		+\widetilde L\tilde{\boldsymbol{v}}^\dagger
		+\widetilde{\boldsymbol{C}}+\widetilde{\boldsymbol{S}}
	\label{eq:ns:decomp:fpf}
\end{equation}
where $\tilde\cdot$ indicates variables in the Fourier space.
Differential operators in this space are defined as
\begin{gather}
	\widetilde{\mbf \nabla}\equiv 2\pi i \frac{k_x}{L_x}\mbf e_x
		+ \frac{\partial}{\partial y}\mbf e _y + 2\pi i \frac{k_z}{L_z}\mbf e_z
	\label{eq:diffop:grad}
	\\
	\widetilde{\nabla}^2\equiv \frac{\partial ^2}{\partial y ^2}
		- 4\pi^{2}(\frac{k_x^2}{L_x^2}+\frac{k_z^2}{L_z^2})
	\label{eq:diffop:lap}
    \\
	\widetilde{L}\equiv\frac{\beta}{\mathrm{Re}}\widetilde{\nabla}^2
	\label{equ_NS_FCF}
\end{gather}
where $k_x$ and $k_z$ are wavenumbers and $i$ is the imaginary unit.

\begin{table}
	\begin{center}
		\caption{Numerical coefficients for the third-order Adams-Bashforth/backward-differentiation temporal discretization~\citep{Peyret_2002}.}
		\begin{tabular}{c|ccc|ccc}
			\hline	
			$\zeta$	&	$a_0$	&	$a_1$	&	$a_2$		&	$b_0$	&	$b_1$	&	$b_2$\\
			\hline
			$11/6$		&	$-3$	&	$3/2$	&	$-1/3$		&	$3$		&	$-3$	&	$1$	\\
			\hline
		\end{tabular}
		\label{tab:BDAB3_coeff}
	\end{center}	
\end{table}

A semi-implicit third-order Adams-Bashforth/backward-differentiation (AB/BD3) scheme is used for time integration, where the linear terms $\widetilde L\tilde{\mbf v^\dagger}$ and $\widetilde{\mbf\nabla}\tilde p^\dagger$ are treated implicitly to allow less restrictive numerical stability conditions~\citep{Peyret_2002}.
After temporal discretization, \cref{eq:ns:decomp:fpf} is rearranged into
\begin{equation}
\begin{split}
	\frac{\zeta}{\delta_t}\tilde{\boldsymbol{v}}^{\dagger,n+1}&
		-\widetilde L\tilde{\boldsymbol{v}}^{\dagger,n+1}
		+\widetilde{\boldsymbol{\nabla}}\tilde{p}^{\dagger,n+1}\\
		&=
		-\sum^{2}_{j=0}\left(
			\frac{a_j}{\delta_t}\tilde{\boldsymbol{v}}^{\dagger,n-j}
			+b_j\left(\widetilde{\boldsymbol{N}}^{n-j}-\widetilde{\boldsymbol{S}}^{n-j}\right)
		\right)+\widetilde{\boldsymbol{C}}\\
		&\equiv\widetilde{\mbf R}^{n}
\end{split}
\label{equ_NS_discrete}
\end{equation}
where $\zeta$, $a_j$ and $b_j$ are numerical coefficients given in \cref{tab:BDAB3_coeff}, $n$ and $n+1$ are the indices for the current and next time steps, and $\delta_t$ is time step size.
For the third order scheme, solutions at the current step ($n$) and two earlier steps ($n-2$, $n-1$) must be stored. 
$\widetilde{\mbf R}^{n}$ is a shorthand symbol for the RHS which can be calculated from information available by step $n$.

After expanding the linear operator in \cref{equ_NS_discrete}, we can write the whole equation set (including boundary conditions) to be solved for each $(k_x,k_z)$ wavenumber pair as
\begin{gather}
	\begin{split}
	\frac{\beta}{\mathrm{Re}}\frac{d^2}{dy^2}\tilde{\mbf v}^{\mathrm{\dag}, n+1}
		-\left(
			4\pi^2\frac{\beta}{\mathrm{Re}}\left(\frac{k_x^2}{L_x^2}+\frac{k_z^2}{L_z^2}\right)
			+\frac{\zeta}{\delta_t}
		\right)\tilde{\mbf v}^{\mathrm{\dag}, n+1}\quad&
		\\
		- \widetilde{\mbf\nabla}\tilde{p}^{\mathrm{\dag},n+1}
		= - \widetilde{\mbf R}^{n}&
	\end{split}
	\label{eq:tau:vp}\\
	\widetilde{\mbf\nabla}\cdot\tilde{\mbf v}^{\dagger,n+1}=0
	\label{eq:tau:divv}\\
	\left.\tilde{\mbf v}^{\dagger,n+1}\right\vert_{y=\pm 1}=0
	\label{eq:tau:bc}
\end{gather}
where \cref{eq:tau:divv,eq:tau:bc} come from applying \cref{eq:decomp:v} to \cref{equ_continuity_c6} and the no-slip boundary condition, respectively.
Note that the partial differential operator $\partial$ is replaced by $d$ because, after discretization in $x$, $z$, and $t$, $y$ is the only remaining continuous independent variable.
Given $k_x$ and $k_z$, the equations solve for $\tilde{\mbf v}^{\dagger,n+1}(y)$ and $\tilde p^{\dagger,n+1}(y)$, which is nontrivial since there are no explicit differential equation and boundary condition(s) for $\tilde p^{\dagger,n+1}$.
The influence-matrix method by \citet{Kleiser_Schumann_Proc3GAMM1980} is used here, which satisfies the continuity constraint (\cref{eq:tau:divv}) and no-slip boundary conditions (\cref{eq:tau:bc}) with analytical accuracy in time.
Spatial discretization in $y$ uses Chebyshev expansion on a Gauss-Lobbato grid~\citep{Canuto_Hussaini_1988}.
Detailed procedures for solving \cref{eq:tau:vp,eq:tau:divv,eq:tau:bc} are provided in \ref{appendix:tau}.

\subsubsection{Time integration of the FENE-P equation}\label{sec:fenep}
Similarly, \cref{equ_Fene1_c6} is re-written into
\begin{gather}
	\frac{\partial\mbf\alpha}{\partial t} = -\mbf N_\alpha+\mbf C_\alpha
		-\frac{2}{\mathrm{Wi}}\left(\frac{\mbf\alpha}{1-\frac{\mathrm{tr}(\mbf\alpha)}{b}}\right)
	\label{eq:fenep:hm}
\end{gather}
where
\begin{gather}
	\boldsymbol{N}_\alpha\equiv
		\boldsymbol{v}\cdot\boldsymbol{\nabla}\boldsymbol{\alpha }-\left(
			\boldsymbol{\alpha}\cdot\boldsymbol{\nabla}\boldsymbol{v}
			+\left(\boldsymbol{\alpha}\cdot\boldsymbol{\nabla}\boldsymbol{v}\right)^\mathrm{T}
		\right)
	\label{eq:Nalpha}
	\\
	\boldsymbol{C}_\alpha\equiv\frac{2}{\mathrm{Wi}}\left(\frac{b\boldsymbol{\delta}}{b+2}\right).
	\label{eq:Calpha}
\end{gather}
Within $\mbf N_\alpha$, the convection term $\mbf v\cdot\mbf\nabla\mbf\alpha$ is calculated with the TVD scheme described shortly below in \cref{Sec_PolyConv}.
Calculation of the polymer stretching terms $\boldsymbol{\alpha}\cdot\boldsymbol{\nabla}\boldsymbol{v}+\left(\boldsymbol{\alpha}\cdot\boldsymbol{\nabla}\boldsymbol{v}\right)^\mathrm{T}$ requires the velocity gradient tensor $\mbf\nabla\mbf v$, which is obtained by applying the gradient operator \cref{eq:diffop:grad} to velocity $\tilde{\mbf v}$ in the Fourier-Chebyshev-Fourier space, followed by inverse Fourier and Chebyshev transforms.

Temporal discretization with the AB/BD3 scheme leads to
\begin{equation}
\begin{split}
	&\frac{\zeta}{\delta_t}\boldsymbol{\alpha}^{n+1}
	\\&\quad
		= -\sum^{2}_{j=0}\left(
			\frac{a_j}{\delta_t}\boldsymbol{\alpha}^{n-j}+b_j\boldsymbol{N}^{n-j}_\alpha
		\right)
		+\boldsymbol{C}_\alpha
		-\frac{2}{\mathrm{Wi}}\left(
			\frac{\boldsymbol{\alpha}^{n+1}}{1-\frac{\mathrm{tr}(\boldsymbol{\alpha}^{n+1})}{b}}
		\right)
	\\&\quad
		= \mbf R^{n}_\alpha
		-\frac{2}{\mathrm{Wi}}\left(
			\frac{\boldsymbol{\alpha}^{n+1}}{1-\frac{\mathrm{tr}(\boldsymbol{\alpha}^{n+1})}{b}}
		\right)
\end{split}
\label{equ_Fene_discrete}
\end{equation}
where
\begin{gather}
		\mbf R^{n}_\alpha\equiv-\sum^{2}_{j=0}\left(
			\frac{a_j}{\delta_t}\boldsymbol{\alpha}^{n-j}+b_j\boldsymbol{N}^{n-j}_\alpha
		\right)
		+\boldsymbol{C}_\alpha
\end{gather}
is again a shorthand variable representing quantities known by step $n$.
The relaxation (last) term is treated implicitly to enforce the finite extensibility (upper-boundedness of $\mathrm{tr}(\mbf\alpha^{n+1})$ constraint, which has also been used in earlier studies~\citep{Vaithianathan_Collins_JCompPhys2003,Dubief_Lele_FTC2005}.
Although finite extensibility is mathematically upheld by \cref{equ_Fene1_c6}, an explicit time-stepping algorithm may cause its numerical violation, which will lead to catastrophic breakdown of the simulation.

The procedure solves for $\mathrm{tr}(\mbf\alpha^{n+1})$ first before individual components of $\mbf\alpha^{n+1}$. 
Adding up the diagonal components of \cref{equ_Fene_discrete} gives
\begin{gather}
	\frac{\zeta}{\delta_t}\mathrm{tr}(\mbf\alpha^{n+1})
		= \mathrm{tr}(\mbf R^{n}_\alpha)
		-\frac{2}{\mathrm{Wi}}\left(
			\frac{\mathrm{tr}(\mbf\alpha^{n+1})}{1-\frac{\mathrm{tr}(\mbf\alpha^{n+1})}{b}}
		\right).
	\label{eq:fenep:tr}
\end{gather}
After a change of variable
\begin{gather}
	\omega\equiv1-\frac{\mathrm{tr}(\mbf\alpha^{n+1})}{b}
\end{gather}
\cref{eq:fenep:tr} can be rearranged into a quadratic equation
\begin{gather}
	A\omega^2+B\omega+C=0
\end{gather}
with
\begin{gather}
	A\equiv\frac{\zeta}{\delta_t}\\
	B\equiv\frac{\mathrm{tr}(\mbf R^n_\alpha)}{b}+\frac{2}{\mathrm{Wi}}-\frac{\zeta}{\delta_t}\\
	C\equiv-\frac{2}{\mathrm{Wi}}.
\end{gather}
Because $A>0$ and $C<0$ (thus $\sqrt{B^2-4AC}>|B|$), its roots
\begin{gather}
	\omega_\pm=\frac{-B\pm\sqrt{B^2-4AC}}{2A}
\end{gather}
must have opposite signs. Taking the positive root $\omega_+>0$ ensures that $\mathrm{tr}(\mbf\alpha^{n+1})<b$.
With $\mathrm{tr}(\mbf\alpha^{n+1})$ known, \cref{eq:fenep:tr} becomes a linear equation in $\mbf\alpha^{n+1}$ whose components can be readily solved for.

Earlier FD approaches often require LAD to maintain numerical stability~\citep{Min_Choi_JNNFM2001,Dubief_Lele_FTC2005,lee2017simulations}.
Typically, a $\kappa\delta^2\nabla^2\mbf\alpha$ ($\kappa$ is a numerical parameter; $\delta$ is the local spatial grid spacing) term is added to the RHS of \cref{equ_Fene1_c6} only at grid points where $\mbf\alpha$ loses its positive definiteness, which account for a small fraction of all grid points in the domain. 
Those FDM+LAD approaches, where the unphysical smearing of stress shocks is rather small, represent a significant improvement over the SM+GAD approach.
According to our practical experience from all simulation conditions tested so far, our TVD-based HM can maintain numerical stability without resorting to any AD -- either local or global.
Nevertheless, LAD, if required, can be easily integrated into our algorithm by adding the $\kappa\delta^2\nabla^2\mbf\alpha$ term to the RHS of \cref{equ_Fene1_c6} and absorbing it into $\mbf N_\alpha$ (by adding $-\kappa\delta^2\nabla^2\mbf\alpha$ to the RHS of \cref{eq:Nalpha}).
In $x$ and $z$ directions where the Fourier grids are uniform, a standard second-order central difference scheme can be used: e.g.,
\begin{gather}
	\left.\kappa\delta_x^2\frac{\partial^2\mbf\alpha}{\partial x^2}\right\vert_q=
		\kappa\left(\mbf\alpha_{q+1}-2\mbf\alpha_q+\mbf\alpha_{q-1}\right)
\end{gather}
where $q$ is the grid index.
In the $y$ direction, the Chebyshev-Gauss-Lobbato (CGL) grids are non-uniform and the corresponding FD expression becomes
\begin{gather}
\begin{split}
	\left.\kappa\delta_{y,q-1}\delta_{y,q}\frac{\partial^2\mbf\alpha}{\partial y^2}\right\vert_q
		=&
		\kappa\delta_{y,q-1}\delta_{y,q}\left(
			\frac{2\mbf\alpha_{q+1}}{\delta_{y,q}(\delta_{y,q-1}+\delta_{y,q})}
		\right.\\&\left.
			-\frac{2\mbf\alpha_{q}}{\delta_{y,q-1}\delta_{y,q}}
			+\frac{2\mbf\alpha_{q-1}}{\delta_{y,q-1}(\delta_{y,q-1}+\delta_{y,q})}
		\right)
\end{split}
\end{gather}
where
\begin{gather}
	\delta_{y,q}\equiv y_{q+1}-y_q
	\label{eq:deltayq}
	\\
	\delta_{y,q-1}\equiv y_{q}-y_{q-1}
\end{gather}
are the immediately neighboring grid spacings.
In all results presented in this study using the new HM, LAD is not used ($\kappa=0$).

\subsubsection{Spatial discretization of the convection term}
\label{Sec_PolyConv} 



Calculation of the $\mbf v\cdot\mbf\nabla\mbf\alpha$ term in \cref{eq:Nalpha} is critical to the numerical stability during the time integration of \cref{equ_Fene1_c6}.
There are a plethora of FD schemes designed for hyperbolic problems.
After balancing the considerations of numerical accuracy and efficiency, we choose a conservative second-order upwind TVD scheme~\citep{Zhang_Jiang_JCompPhys2015}.
Detailed comparison between various schemes based on a simple benchmark problem is discussed in \ref{Sec_schmComp}.

For any component of the conformation tensor $\alpha_{ij}$, the convection term is rewritten in terms of flux derivatives as
\begin{equation}
	\boldsymbol{v}\cdot\boldsymbol{\nabla}\alpha_{ij}
		=\boldsymbol{\nabla}\cdot\left(\boldsymbol{v}\alpha_{ij}\right)
		=\sum_{w=x,y,z}\frac{\partial}{\partial w}\left(v_w \alpha_{ij}\right)
\label{equ_Fene_conv}
\end{equation}
with the first equality coming from the divergence free condition \cref{equ_continuity_c6}.
In our HM, the same numerical grid is shared between spectral and FD discretization. The TVD method must be adapted to the gridding scheme in each dimension.
We will start with the calculation of contributions from $x$ and $z$ partial derivatives, before discussing how the procedure is adapted to the inhomogeneous $y$ direction.

\begin{figure}
	\centerline{\includegraphics[width=\linewidth, trim=0mm 0mm 0mm 0mm, clip]{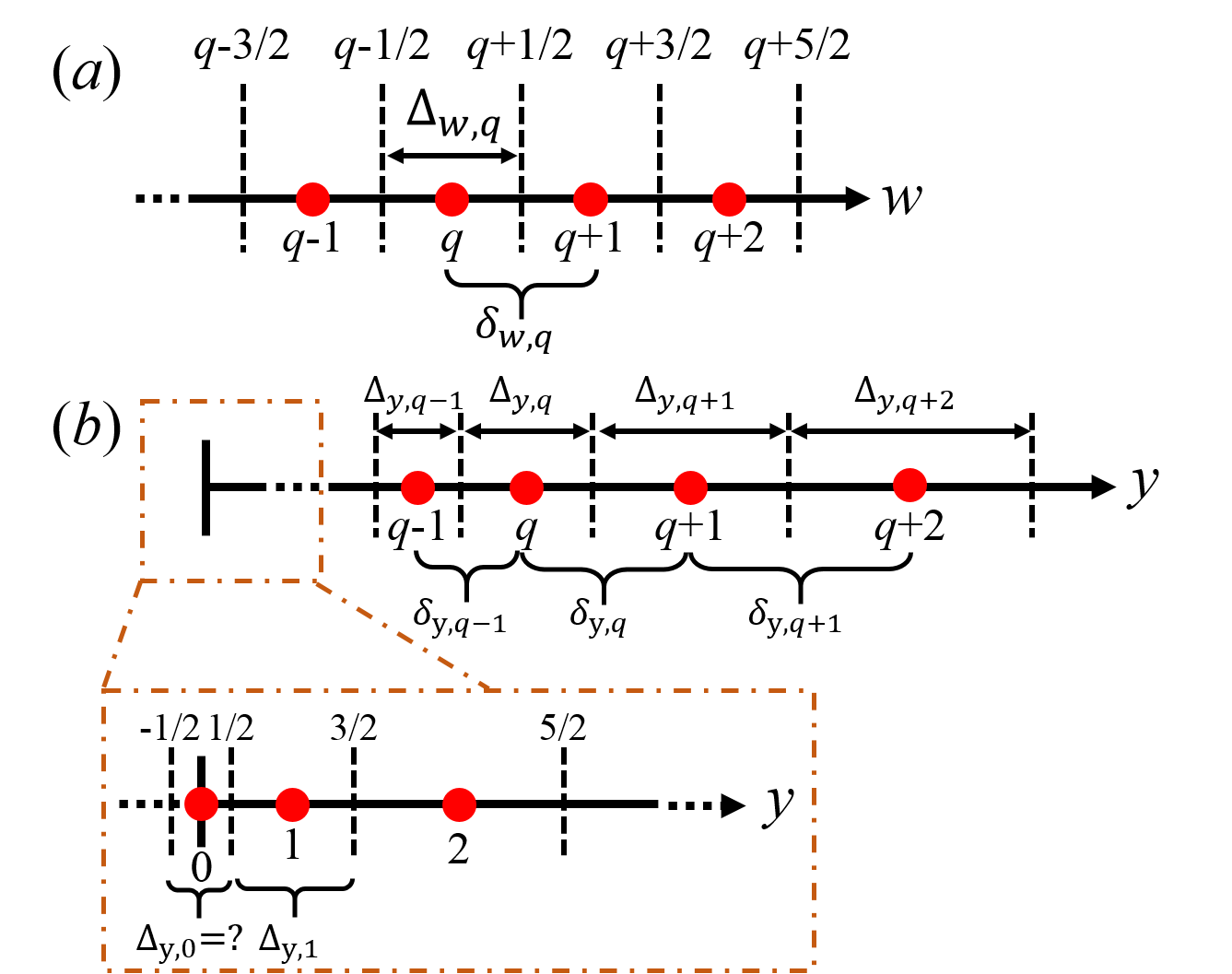}}
	\caption{Schematics of the numerical grid systems for the TVD scheme: (a) homogeneous directions with uniform grids ($w=x$ or $z$); (b) wall-bounded direction with a non-uniform grid ($y$).
	Round dots represent grid points and dashed lines show virtual cell edges.
	Integers index grid points; half-integers index cell edges.
	For $\Delta_{y,0}=0$ used in this study, cell edges at $q=\pm1/2$ collapse onto the grid point $q=0$ on the wall.}
	\label{fig:schem_grid}
\end{figure}

\paragraph{Discretization in homogeneous directions ($x$ and $z$)}
Let
\begin{gather}
	F\equiv v_w \alpha_{ij}
\end{gather}
be the convective flux of $\alpha_{ij}$ in the $w$ direction. Its partial derivative with respect to (w.r.t.) $w$ at the $q$-th grid point
\begin{equation}
	\left.\frac{\partial}{\partial w}(v_w\alpha_{ij})\right|_q
		=\left.\frac{\partial F}{\partial w}\right|_q
		=\frac{F_{q+1/2}-F_{q-1/2}}{\Delta_{w,q}}
	\label{equ_ben_conv2}
\end{equation}
is approximated by the FD expression between flux values at the edges (indexed by $q-1/2$ and $q+1/2$) of a surrounding virtual cell.
The cell is so defined that the corresponding grid point sits at its center (\cref{fig:schem_grid}(a)).
The $x$- and $z$-directions use uniform grids with periodic boundary conditions (as required by FFT). The cell size $\Delta_{w,q}$ is thus constant and equals the grid spacing $\delta_w$ used in $w$
\begin{gather}
	\Delta_{w,q}=\Delta_w=\delta_w.
\end{gather}

Before proceeding with the calculation procedure for edge fluxes $F_{q\pm 1/2}$, we first introduce the Lax-Friedrichs flux splitting (LFFS) approach~\citep{shu1998essentially}, which splits the numerical flux $F_q$ into a so-named positive and negative flux pair $F_q^\pm$:
\begin{equation}
	F_q=F_q^+ + F_q^-
\label{equ_ben_conv3}
\end{equation}
with
\begin{equation}
\begin{split}
	F_q^\pm&=\frac{1}{2}\left(F_q \pm |v_w|_\text{max} \alpha_{ij,q}\right)\\
		&=\frac{1}{2}\left(v_{w,q} \pm |v_w|_\text{max}\right)\alpha_{ij,q}
\end{split}
\label{equ_ben_conv4}
\end{equation}
where $|v_w|_\text{max}$ is the maximum $w$-component velocity magnitude $|v_w|$ along the one-dimensional $w$ grid line: e.g., for $w=x$, it is the maximum of $|v_x|$ among all $x$ grid points (i.e., for all $q$) at the given $y$ and $z$ coordinates.
By this definition, $|v_q|_\text{max}\geq v_q$: thus, $F_q^+\geq 0$ and $F_q^-\leq 0$ are fluxes pointing toward positive and negative $w$ directions, respectively.

The cell edge fluxes, needed in \cref{equ_ben_conv2}, are similarly decomposed
\begin{equation}
	F_{q\pm 1/2}=F_{q\pm 1/2}^+ + F_{q\pm 1/2}^-
\label{equ_ben_conv5}
\end{equation}
and their positive/negative flux components $F_{q\pm 1/2}^\pm$ are calculated from $F_q^\pm$ through numerical differentiation.
We use a TVD scheme that approximates the positive/negative flux pair at the $q+1/2$ edge (the procedure for  $q-1/2$ edge fluxes are similar) by
\begin{equation}
\begin{dcases}
	F^+_{q+1/2}=F^+_q+\frac{1}{2}\phi\left(r^+_{q+1/2}\right)\left(F^+_q-F^+_{q-1}\right)\\
	F^-_{q+1/2}=F^-_{q+1}+\frac{1}{2}\phi\left(r^-_{q+1/2}\right)\left(F^-_{q+1}-F^-_{q+2}\right)
\end{dcases}
\label{equ_ben_tvd}
\end{equation}
where a left-bias local stencil $[q-1, q, q+1]$ is used for $F^+_{q+1/2}$ and a right-bias local stencil $[q, q+1,q+2]$ is used for $F^-_{q+1/2}$~\citep{sweby1984high,Zhang_Jiang_JCompPhys2015}.
The flux limiter function $\phi$ can take several forms -- examples can be found in \citet{sweby1984high,waterson2007design}, and \citet{Zhang_Jiang_JCompPhys2015}.
We adopt the MINMOD limiter~\citep{roe1981approximate,Zhang_Jiang_JCompPhys2015}, which is widely used in TVD schemes and known to provide numerical stability in the DNS of viscoelastic turbulence~\citep{Yu_Kawaguchi_JNNFM2004}:
\begin{equation}
	\phi(r)\equiv\max\left[0,\min(1,r)\right]
\label{equ_ben_tvd2}
\end{equation}
where $r$ is the successive gradient ratio and calculated at the $q+1/2$ cell edge with
\begin{equation}
\begin{dcases}
	r^+_{q+1/2}=\frac{F^+_{q+1}-F^+_{q}}{F^+_{q}-F^+_{q-1}}\\
	r^-_{q+1/2}=\frac{F^-_{q}-F^-_{q+1}}{F^-_{q+1}-F^-_{q+2}}
\end{dcases}.
\label{equ_ben_tvd3}
\end{equation}
The definition of the flux limiter function ensures that $\phi(r)$ is in the range of $[0,1]$.
When spurious oscillations occur, typically around shocks, $F^\pm$ fluctuate between consecutive grid points. As a result, the $r$ ratio will be negative and $\phi(r)=0$. The edge flux expressions in \cref{equ_ben_tvd} then reduce from second-order to first-order accuracy, which will suppress numerical oscillations.
\Cref{equ_ben_conv5,equ_ben_tvd,equ_ben_tvd2,equ_ben_tvd3} enable the calculation of the net edge flux at $q+1/2$, $F_{q+1/2}$, from grid fluxes at $[q-1,q,q+1,q+2]$. $F_{q-1/2}$ is likewise calculated from the grid fluxes at $[q-2,q-1,q,q+1]$.
The original flux derivative \cref{equ_ben_conv2} is then calculated from these edge flux values.

\paragraph{Discretization in the inhomogeneous $y$ direction}
The $y$-direction requires special treatment not only as a result of the no-slip walls, but also for its non-uniform grid which is required by the Chebyshev transform used in the spectral part of the HM.
The CGL grid points are given by 
\begin{gather}
	y_{q}=\cos\left(\frac{q\pi}{N_y-1}\right)\quad(q=0,1,\ldots,N_y-1)
\end{gather}
where $N_y$ is the total number of grid points in $y$.
The first ($q=0$) point sits on the top wall ($y=1$) and the last ($q=N_y-1$) point sits on the bottom wall ($y=-1$). The grid is symmetric w.r.t. the central plane ($y=0$) and the grid points are denser near the walls than the center.

The expressions for positive/negative edge flux pairs (\cref{equ_ben_tvd,equ_ben_tvd3}) must be generalized for non-uniform grid systems as
\begin{gather}
\begin{dcases}
	F^+_{q+1/2}=F^+_q+\frac{\Delta _{y,q}}{2}\phi(r^+_{q+1/2})
		\frac{F^+_q-F^+_{q-1}}{\delta_{y,q-1}}\\
	F^-_{q+1/2}=F^-_{q+1}+\frac{\Delta _{y,q+1}}{2}\phi(r^-_{q+1/2})
		\frac{F^-_{q+1}-F^-_{q+2}}{\delta_{y,q+1}}
\end{dcases}
\label{eq:fpmplus:y}
\\
\begin{dcases}
	r^+_{q+1/2}=\frac{\left(F^+_{q+1}-F^+_{q}\right)/\delta_{y,q}}
		{\left(F^+_{q}-F^+_{q-1}\right)/\delta_{y,q-1}}\\
	r^-_{q+1/2}=\frac{\left(F^-_{q}-F^-_{q+1}\right)/\delta_{y,q}}
		{\left(F^-_{q+1}-F^-_{q+2}\right)/\delta_{y,q+1}}
\end{dcases}
\label{eq:rpmplus:y}
\end{gather}
where $\Delta_{y,q}$ denotes the size of the virtual cell containing grid point $q$ and $\delta_{y,q}$ is the grid spacing between points $q$ and $q+1$ (see \cref{fig:schem_grid}(b) and \cref{eq:deltayq}).

\begin{figure}
	\centerline{\includegraphics[width=0.8\linewidth, trim=0mm 0mm 0mm 0mm, clip]{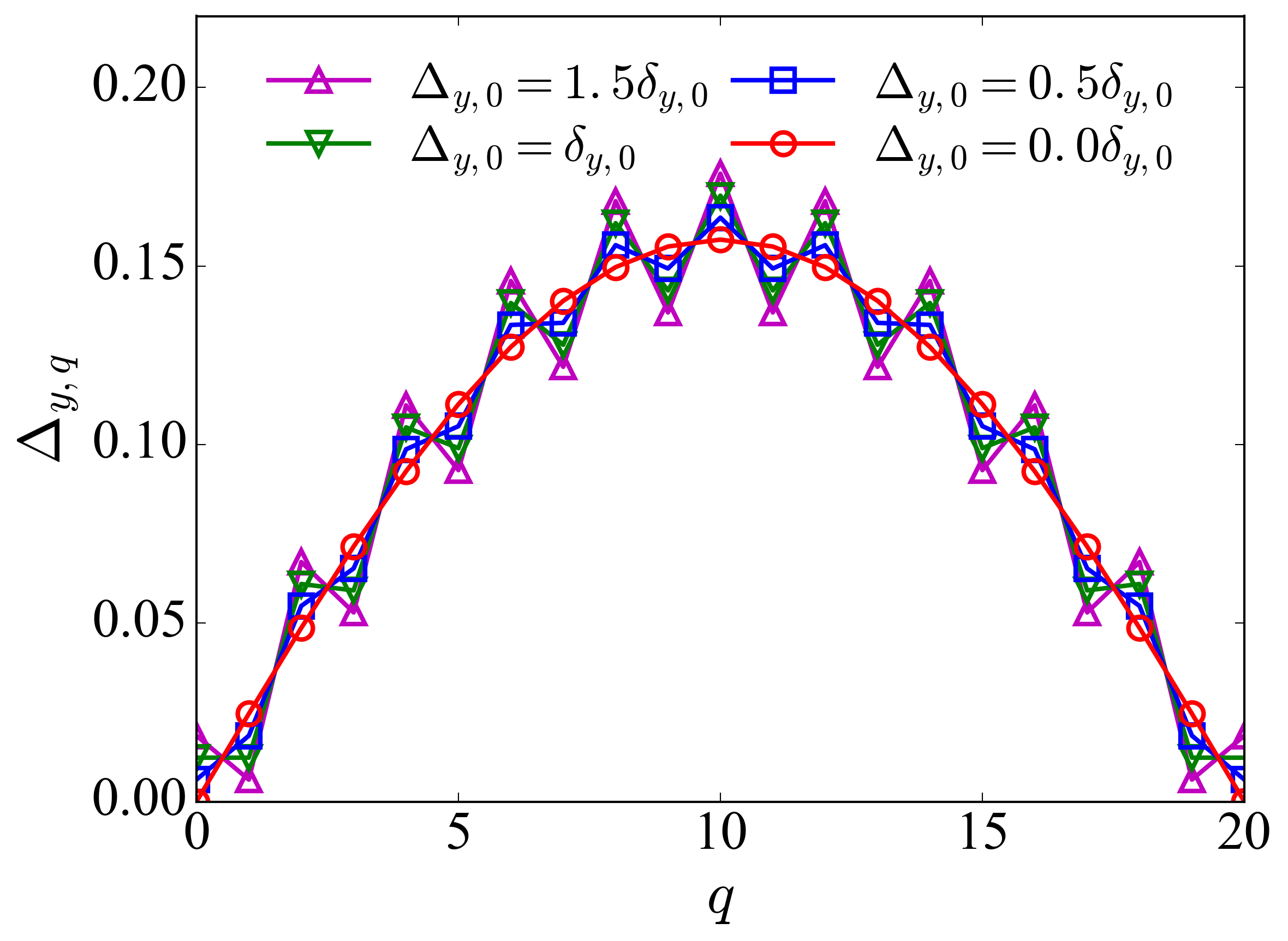}}
	\caption{Virtual cell size distribution under different $\Delta_{y,0}$ choices using the $N_y=21$ case as an example. Non-zero $\Delta_{y,0}$ choices all result in zigzag distribution.}
	\label{fig:grid_dist}
\end{figure}

To ensure that grid points are located at the center of the corresponding virtual cell, cell sizes are related to grid spacings through
\begin{gather}
\begin{array}{ccccc}
	\frac{1}{2}\left(\Delta_{y,0}\right.	&+	&\left.\Delta_{y,1}\right)
		&=	&\delta_{y,0}	\\
	\frac{1}{2}\left(\Delta_{y,1}\right.	&+	&\left.\Delta_{y,2}\right)
		&=	&\delta_{y,1}	\\
	\vdots	&	&\vdots		&	&\vdots		\\
	\frac{1}{2}\left(\Delta_{y,q}\right.	&+	&\left.\Delta_{y,q+1}\right)
		&=	&\delta_{y,q}	\\
	\vdots	&	&\vdots		&	&\vdots		\\
	\frac{1}{2}\left(\Delta_{y,N_y-2}\right.&+	&\left.\Delta_{y,N_y-1}\right)
		&=	&\delta_{y,N_y-2}.
\end{array}
\label{equ_Fene_grid}
\end{gather}
Since the first ($q=0$) and last ($q=N_y-1$) grid points sit on the walls, half of their corresponding cells extend outside the domain, as sketched in \cref{fig:schem_grid}(b).
With $N_y-1$ grid spacings $\delta_{y,q}$ known and $N_y$ cell sizes $\Delta_{y,q}$ to solve for, \cref{equ_Fene_grid} is underdetermined with one extra degree of freedom.
We choose to set $\Delta_{y,0}=0$, with which \cref{equ_Fene_grid} can be solved to obtain $\Delta_{y,q}$ for $q=1,2,\ldots,N_y-1$. Note that, by symmetry, the solution will give $\Delta_{y,N_y-1}=0$ at the opposite wall.
Having zero-sized wall cells greatly simplifies the boundary treatment,
as wall grid points and their corresponding cell edges merge. Take the $q=0$ point for instance: since $y_{-1/2}=y_0=y_{1/2}$ mark the same point, we have $F_{-1/2}=F_0=F_{1/2}=0$, where the boundary flux value of zero is due to the no-slip boundary condition.
At the wall, $\mbf v=0$ and thus $\mbf v\cdot\mbf\nabla\alpha_{ij}=\mbf\nabla\cdot(\mbf v\alpha_{ij})=0$ and $\partial(v_x\alpha_{ij})/\partial x=\partial(v_z\alpha_{ij})/\partial z=0$, which leads to the boundary flux derivative value $\partial F/\partial y|_0\equiv\partial(v_y\alpha_{ij})/\partial y\vert_{y=1}=0$.
For the next grid point $q=1$, calculating its flux derivative $\partial F/\partial y|_1$ requires its edge fluxes $F_{1/2}$ and $F_{3/2}$.
$F_{3/2}$ is still obtained with the standard numerical differentiation approach (\cref{eq:fpmplus:y,eq:rpmplus:y}), while its left edge flux $F_{q-1/2}=F_{1/2}=0$ is already known -- numerical differentiation, which would otherwise rely on local stencils extending to $q-2=-1$, is no longer needed.
As such, the choice of $\Delta_{y,0}=0$ allows the algorithm to be independent of any information beyond the wall boundary, which circumvents the complexity of adding ghost points.
As an added benefit, it can be shown that $\Delta_{y,0}=0$ leads to smoothly increasing cell size from the wall to the channel center whereas any non-zero choice would cause undesirable zigzag patterns (as shown in \cref{fig:grid_dist}).

Another difference in the $y$-direction discretization comes from its velocity inhomogeneity, which affects the choice of $|v_w|_\text{max}$ in LFFS (\cref{equ_ben_conv4}).
Unlike the $x$- and $z$-directions, where different grid points are statistically equivalent and choosing a global maximum $|v_w|$ among all $x$ or $z$ grid points is
sensible, in the $y$-direction, velocity magnitudes depend strongly on wall distance and a global LFFS approach is no longer appropriate.
For example, $|v_x|$ is typically highest near the channel center and decays to zero at the walls. Choosing the global maximum along the $y$-axis will cause the second term in \cref{equ_ben_conv4} to be much higher than the first term in near-wall regions, which causes numerical inaccuracy~\citep{delis2000evaluation}.
A local LFFS procedure is instead applied in the $y$-direction~\citep{chaudhuri2011numerical,liu1998convex}, which sets $|v_w|_\text{max}$ (for $w=x, y,$ or $z$) to the maximal $|v_w|$ value within the numerical stencils used for each cell edge.
That is, all grid-point fluxes used in calculating $F_\text{q+1/2}^\pm$ (\cref{eq:fpmplus:y,eq:rpmplus:y}) use the maximum $|v_w|$ among grid points $[q-1,q,q+1,q+2]$, while those used in calculating $F_\text{q-1/2}^\pm$ use the maximum among  $[q-2,q-1,q,q+1]$.

\subsubsection{Overall workflow}
\begin{figure}
	\centerline{\includegraphics[width=.8\linewidth, trim=0mm 0mm 0mm 0mm, clip]{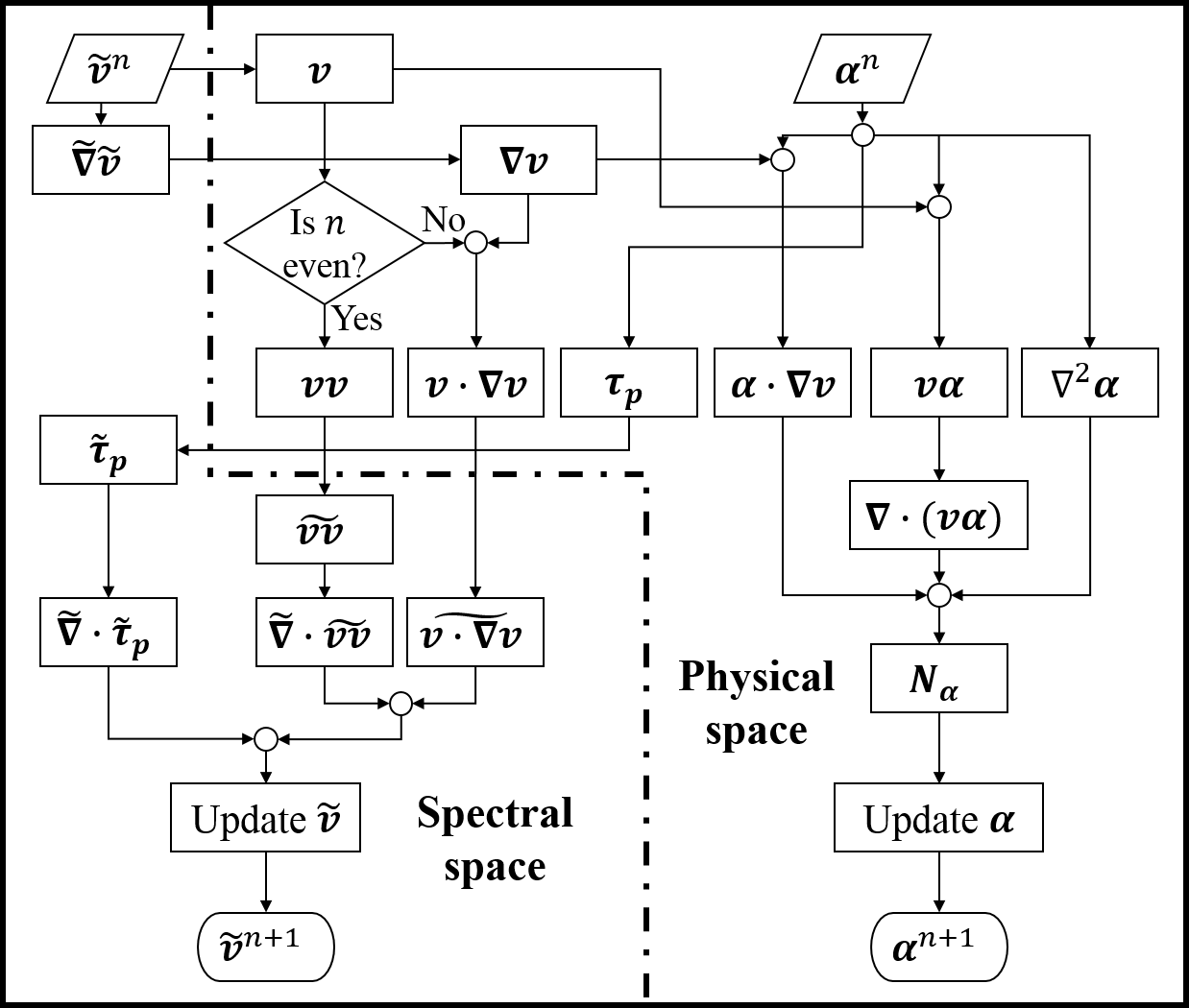}}
	\caption{Flow chart of the hybrid pseudo-spectral/finite-difference algorithm for updating the velocity $\mbf v$ and polymer conformation $\mbf\alpha$ fields to the $(n+1)$-th step. The AB/BD3 time-stepping scheme requires information at $(n-2)$-, $(n-1)$-, and $n$-th steps. For simplicity, only the $n$-th step is shown; required quantities at $(n-2)$- and $(n-1)$-th steps are stored from previous steps.
	Note: the $\nabla^2\mbf\alpha$ term is only needed in the LAD term, which is not used in simulations reported in this study.}
	\label{fig:scheme_flowchart}
\end{figure}

The overall workflow for implementing the HM is shown in \cref{fig:scheme_flowchart}.
The N-S equation is solved mainly in the spectral space (i.e., after Fourier transforms in $x$ and $z$ and Chebyshev tranform in $y$) with the exception of the inertia term $\mbf v\cdot\mbf\nabla\mbf v$, while FENE-P is solved in the physical space.
The same Fourier-Chebyshev-Fourier grid system is used for all variables using either spectral or FD discretization.
Upon completing time step $n$, the velocity gradient $\widetilde{\boldsymbol\nabla}\tilde{\mbf v}$ is calculated from the velocity field $\tilde{\mbf v}^{n}$ in the spectral space.
Inverse Fourier and Chebyshev transforms are then used to project both $\tilde{\boldsymbol{v}}$ and $\widetilde{\boldsymbol\nabla}\tilde{\mbf v}$ to the physical space, where they are needed for the nonlinear terms in FENE-P (\cref{eq:Nalpha}) --
$\mbf\nabla\mbf v$ is used for calculating $\mbf\alpha\cdot\mbf\nabla\mbf v$ at each grid point while $\mbf v$ goes to the convection term $\mbf v\cdot\mbf\nabla\mbf\alpha=\mbf\nabla\cdot(\mbf v\mbf\alpha)$ following the procedure in \cref{Sec_PolyConv}.
The LAD term $\kappa\delta^2\nabla^2\mbf\alpha$, if used, is calculated as described in \cref{sec:fenep}.
By now, all components in $\mbf N_\alpha$ are known and, with $\mbf\alpha^{n-2}$, $\mbf\alpha^{n-1}$, $\mbf N_\alpha^{n-2}$, and $\mbf N_\alpha^{n-1}$ stored from previous time steps, FENE-P can be advanced in time to obtain $\mbf\alpha^{n+1}$ following the procedure in \cref{sec:fenep}.
Polymer stress $\mbf\tau_p$ is calculated from $\mbf\alpha^{n}$ using \cref{equ_Fene2_c6}.
Its divergence is calculated after projecting $\mbf\tau_p$ to the spectral space ($\widetilde{\boldsymbol\nabla}\cdot\tilde{\boldsymbol\tau}_p$), from which $\widetilde{\mbf S}$ is available.
Calculation of the inertia term $\widetilde{\mbf N}$ alternates between two pathways as $n$ switches its parity.
In the convection form (\cref{eq:N:conv}), $\boldsymbol{v}\cdot \boldsymbol{\nabla v}$ is calculated in the physical space and then projected to the spectral space.
In the divergence form (\cref{eq:N:div}), $\mbf v\mbf v$ is calculated in the physical space and projected to the spectral space where its divergence $\widetilde{\mbf\nabla}\cdot\widetilde{\mbf v\mbf v}$ is calculated.
At this point, the N-S equation can be advanced to the $(n+1)$-th step following the procedure in \cref{Sec_TimeIntNS} ($\tilde{\mbf v}^{n-2}$, $\tilde{\mbf v}^{n-1}$, $\widetilde{\mbf N}^{n-2}$, $\widetilde{\mbf N}^{n-1}$, $\widetilde{\mbf S}^{n-2}$, and $\widetilde{\mbf S}^{n-1}$ again need to be stored from previous time steps).

\subsection{Numerical algorithm of the pseudo-spectral method (SM)}\label{sec:SM}
Our old pure SM algorithm was documented in detail in \citet{Xi_PhD2009} and extensively used and validated in a large number of previous studies~\citep{Xi_Graham_JFM2010,Xi_Graham_PRL2010,Xi_Graham_JFM2012,Xi_Graham_PRL2012,Wang_Graham_AIChEJ2014,Wang_Graham_JNNFM2017,Xi_Bai_PRE2016,Zhu_Xi_JNNFM2018,Zhu_Xi_JNNFM2019,Zhu_Xi_JPhysCS2018}. It is only recapitulated here for comparison with the new HM algorithm.
The procedure for integrating the N-S equation is identical to that of the HM, which was detailed in \cref{Sec_TimeIntNS}.
For FENE-P, a GAD term, necessary for numerical stability in SM,
\begin{gather}
	\frac{1}{\mathrm{Pe}}\nabla^2\mbf\alpha=\frac{1}{\mathrm{ScRe}}\nabla^2\mbf\alpha
	\label{eq:fenep:gad}
\end{gather}
is added to the RHS of \cref{equ_Fene1_c6},
where
\begin{gather}
	\mathrm{Pe}\equiv\frac{UL}{D}
\end{gather}
and
\begin{gather}
	\mathrm{Sc}\equiv\frac{\eta}{\rho D}
\end{gather}
are the Peclet and Schmidt numbers, respectively. Here, $D$ is the numerical diffusivity and $1/\mathrm{Pe}$ can be viewed as its nondimensional counterpart.
An abbreviated version of \cref{equ_Fene1_c6} with the added GAD term is written as
\begin{gather}
	\frac{\partial\mbf\alpha}{\partial t} = -\mbf N'_\alpha+\mbf C_\alpha+L_\alpha\mbf\alpha
	\label{eq:fenep:sm}
\end{gather}
where
\begin{gather}
	\mbf N'_\alpha\equiv
		\boldsymbol{v}\cdot\boldsymbol{\nabla}\boldsymbol{\alpha }-\left(
			\boldsymbol{\alpha}\cdot\boldsymbol{\nabla}\boldsymbol{v}
			+\left(\boldsymbol{\alpha}\cdot\boldsymbol{\nabla}\boldsymbol{v}\right)^\mathrm{T}
		\right)
		+\frac{2}{\mathrm{Wi}}\left(\frac{\mbf\alpha}{1-\frac{\mathrm{tr}(\mbf\alpha)}{b}}\right)
	\label{eq:NalphaSM}
	\\
	L_\alpha\equiv\frac{1}{\mathrm{ScRe}}\nabla^2
\end{gather}
and $\mbf C_\alpha$ is the same as \cref{eq:Calpha}.
Applying AB/BD3 discretization in time and projecting the variables to the Fourier space gives
\begin{gather}
\begin{split}
	\frac{\zeta}{\delta_t}\tilde{\mbf\alpha}^{n+1}-\widetilde L_\alpha\tilde{\mbf \alpha}^{n+1}
		=& -\sum^{2}_{j=0}\left(
			\frac{a_j}{\delta_t}\tilde{\mbf\alpha}^{n-j}
				+b_j\widetilde{\mbf N}^{\prime,n-j}_\alpha
		\right)
		+\widetilde{\mbf C}_\alpha
	\\
		\equiv&\widetilde{\mbf R}^{\prime, n}_\alpha.
\end{split}
\label{eq:fenep:sm:fft}
\end{gather}
Different from the HM algorithm (\cref{sec:fenep}), here, the polymer relaxation term is treated explicitly and thus absorbed into $\mbf N'_\alpha$ (last term in \cref{eq:NalphaSM}), while the added GAD term is treated implicitly.
Expanding the Laplacian operator in
\begin{gather}
	\widetilde L_\alpha\equiv\frac{1}{\mathrm{ScRe}}\widetilde\nabla^2
\end{gather}
with \cref{eq:diffop:lap},
\cref{eq:fenep:sm:fft} becomes
\begin{gather}
	\frac{1}{\mathrm{ScRe}}\frac{d^2}{dy^2}\tilde{\mbf\alpha}^{n+1}
		-\left(
			\frac{4\pi^2}{\mathrm{ScRe}}\left(\frac{k_x^2}{L_x^2}+\frac{k_z^2}{L_z^2}\right)
			+\frac{\zeta}{\delta_t}
		\right)\tilde{\mbf\alpha}^{n+1}
		= - \widetilde{\mbf R}^{\prime,n}_\alpha
	\label{eq:fenep:sm:helmholtz}
\end{gather}
which is an inhomogeneous Helmholtz equation, solved by the Chebyshev-tau method~\citep{Canuto_Hussaini_1988}, for each $\tilde{\mbf\alpha}^{n+1}$ component and each $(k_x,k_z)$ wavenumber pair.
The GAD term changes the equation from a hyperbolic form to a parabolic form, which now requires boundary conditions at the walls.
Fortunately, the equation is numerically stable without GAD at the no-slip walls. Boundary values of $\tilde{\mbf\alpha}^{n+1}$ can be obtained by marching \cref{eq:fenep:sm:fft} without the GAD term $\widetilde L_\alpha\tilde{\mbf \alpha}^{n+1}$ at the walls ($y=\pm1$). The results are then used as boundary conditions to solve \cref{eq:fenep:sm:helmholtz} for the rest of the flow domain.

Time integration of both N-S and FENE-P equations is performed in the spectral space. At the beginning of each time step, inverse Fourier and Chebyshev transforms are applied to obtain $\mbf v$, $\mbf\nabla\mbf v$, $\mbf\alpha$, and $\mbf\nabla\mbf\alpha$ in the physical space, which are needed for the calculation of $\mbf N$, $\mbf S$, and $\mbf N'_\alpha$ (\cref{eq:N:conv,eq:N:div,eq:S,eq:NalphaSM}).
In particular, the convection term $\mbf v\cdot\mbf\nabla\mbf\alpha$ is calculated directly at each grid point.
The calculated nonlinear terms are projected back to the spectral space for time integration.

\subsection{Parallelization}
\begin{figure}
	\centering
	\includegraphics[width=.96\linewidth, trim=25 15 5 25, clip]{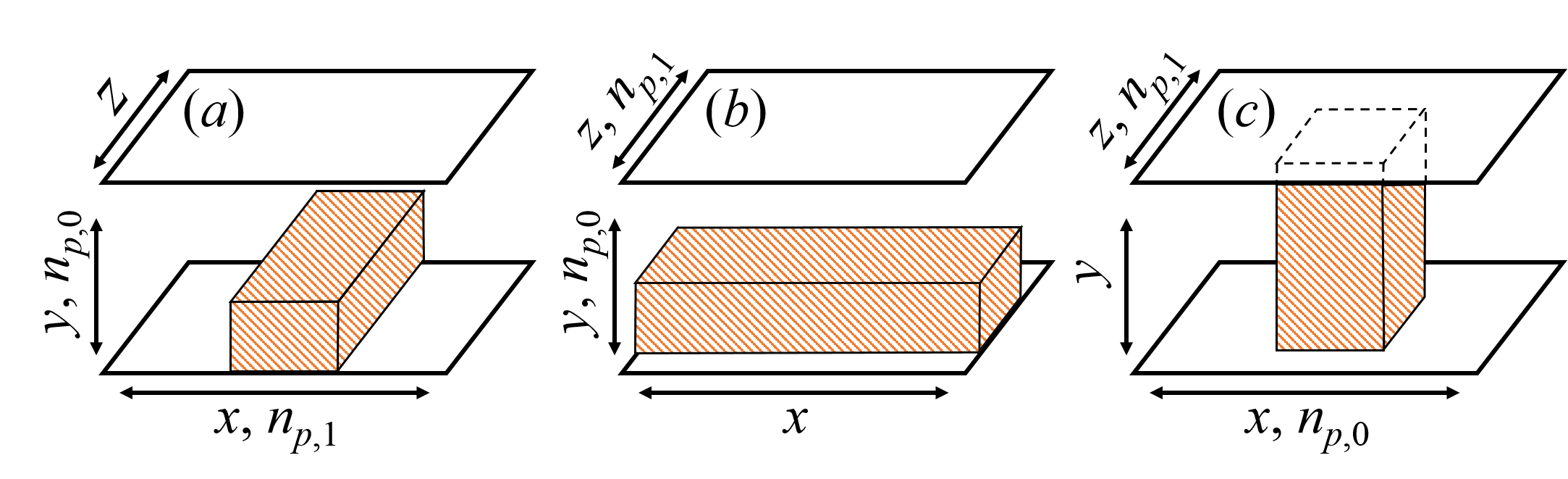}
	\caption{Illustration of different modes of 2D pencil domain decomposition:
	(a) $xy$-decomposition,
	(b) $yz$-decomposition, and
	(c) $xz$-decomposition.
	In each mode, a 2D plane is divided into $n_p=n_{p,0}\times n_{p,1}$ equisized pieces.}
	\label{fig:parallel_sch}
\end{figure}

The code is written in \texttt{C++} and parallelized with MPI (message passing interface). Spectral (Fourier or Chebyshev) transforms are by nature 1D global operations that require data of all grid-points along the direction of transform to be kept on the same processor.
Therefore, a rotating 2D ``pencil'' domain decomposition strategy is used.
For example, transforming a 3D field to the spectral space starts with an $xy$-decomposition (\cref{fig:parallel_sch}(a)): the domain is partitioned into $n_p=n_{p,0}\times n_{p,1}$ ($n_p$ is the number of processors) equisized pieces in the $xy$-plane. Each subdomain covers the entire $z$ axis and thus 1D $z$-direction FFT can be applied for each $(x,y)$ grid point.
This is followed by $yz$-decomposition (\cref{fig:parallel_sch}(b)) for the $x$-direction FFT and finally $xz$-decomposition (\cref{fig:parallel_sch}(c)) for the Chebyshev transform in $y$.
Inverse transforms are done in the reverse order ($xz$-, $yz$-, and then $xy$-decomposition).

FD discretization, used for $\mbf v\cdot\mbf\nabla\mbf\alpha$ in HM, is instead local. It only requires information within a stencil around each grid point. This makes 3D decomposition a viable option, since only data at a limited number of grid points beyond the subdomain boundary need to be copied from other processors.
We, however, still use the same rotating 2D decomposition approach (as that of the spectral part of the algorithm) mostly out of convenience.
Compared with the 3D decomposition approach, rotation between different directions of 2D decomposition adds a small overhead of redistributing data between processors after each direction.
On the other hand, this approach stores all data points along the direction of discretization on the same processor -- there is no overlap between information required by different processors. It thus avoids the duplication in both data storage and calculation.

Time integration in the spectral space (for N-S in HM and for both N-S and FENE-P in SM), which comes down to solving $y$-dependent Helmholtz equations, is performed using $xz$-decomposition.
Time integration in the physical space (for FENE-P in HM), as well as pointwise arithmetic operations (calculation of all nonlinear terms at each grid point in the physical space), can use any mode of decomposition and $xy$-decomposition is used in our code. 

\begin{figure}
	\centering
	\includegraphics[width=\linewidth, trim=0mm 0mm 0mm 0mm, clip]{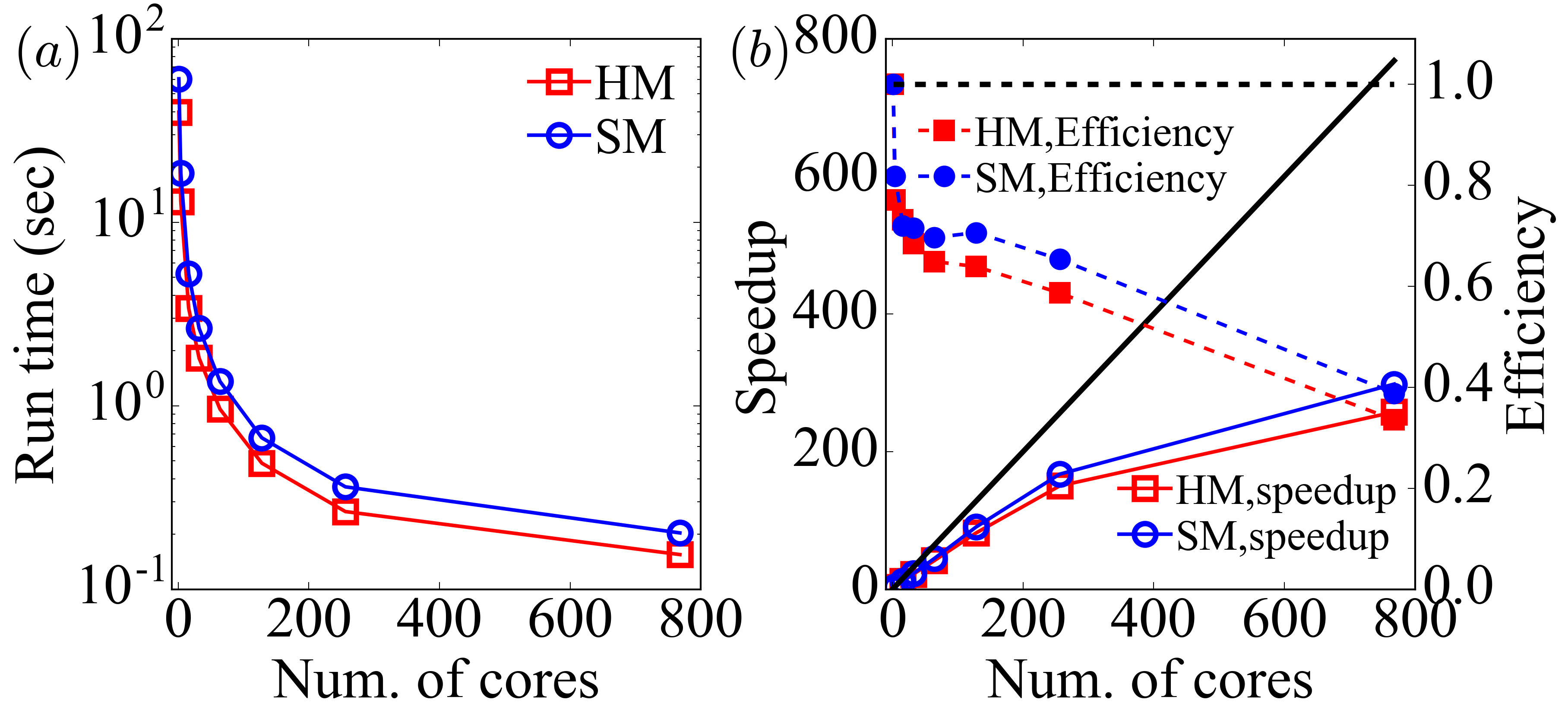}
	\caption{Parallelization performance metrics as functions of the number of processor cores:
		(a) average run time for one time step (in real time);
		(b) speedup ratio and parallelization efficiency.
		Solid straight line shows ideal speedup.
		The benchmark simulation uses an $N_x\times N_y\times N_z = 144\times 391\times 142$ grid.
		Both HM and SM cases include the AD term for equal comparison.}
	\label{fig:parallel_eff}
\end{figure}

Performance of the parallelized code is benchmarked on the \texttt{Graham} (\texttt{graham.sharcnet.ca}) system of SHARCNET (Shared Hierarchical Academic Research Computing Network -- a local consortium of high-performance computing and part of the Compute Canada network).
Each node used in the benchmark runs has 32 Intel\textsuperscript\textregistered{} E5-2683 v4 (Broadwell) \SI{2.10}{GHz} CPU cores and \SI{125}{GB} total memory. 
In \cref{fig:parallel_eff}(a), we compare the computational cost of the HM and SM algorithms.
HM is computationally less demanding than SM -- for serial simulation on a single processor, the average time for one time step is \si{40}{s} and \si{60}{s} for HM and SM, respectively.
(Note that \cref{fig:parallel_eff} reports upper-bound estimates for HM, as not only the LAD term is switched on in benchmark runs but AD is calculated for each grid point regardless of the sign of $\det(\mbf\alpha)$.)
As shown in \ref{Sec_schmComp}, the cost of TVD discretization of the convection term is comparable to that of SM.
Meanwhile, comparing \cref{sec:fenep} and \cref{sec:SM}, the time integration procedure of HM is simplified by avoiding the implicit treatment of the diffusion term (even when LAD is included).
There is thus a net reduction in computational cost compared with SM.

The speedup enabled by parallelization is quantified by the
\begin{gather}
	\text{speedup ratio}\equiv
		\frac{\text{run time of serial simulation}}{\text{run time of parallel simulation}}
\end{gather}
and
\begin{gather}
	\text{parallel efficiency}\equiv
		\frac{\text{speedup ratio}}{N_\text{core}}
\end{gather}
(the number of cores $N_\text{core}$ equals the ideal speedup ratio when there is no loss of efficiency in parallelization) as shown in \cref{fig:parallel_eff}(b).
After an initial decline, the parallel efficiency stabilizes at a roughly constant level ($\approx 70\%$ for SM and $\approx 65\%$ for HM) for $N_\text{core}\approx$ \numrange{16}{128}.
The efficiency deteriorates at $N_\text{core}\gtrsim 256$ and at the highest $N_\text{core}=768$, it drops to around $40\%$ for both codes.
This may be attributed to the relatively small size of the benchmark system ($N_x\times N_y\times N_z = 144\times 391\times 142$). For example, the $xz$ 2D decomposition of the benchmark system between $N_\text{core}=768$ cores assigns only $<27$ $(x,y)$ (or $(k_x,k_y)$) pairs to each core.
The overhead of inter-core communication thus becomes significant in comparison with the cost of computation itself.
We expect that the efficiency can stay at the $60\sim 70\%$ level for a wider $N_\text{core}$ range if larger simulation systems are tested (which is not possible for the limitation of computational resources).
Finally, we note that the parallel efficiency of HM is slightly lower than that of SM. This is attributed to the extra communication cost during the FD calculation of the $\mbf v\cdot\mbf\nabla\mbf\alpha$ term, which uses the rotating 2D pencil decompositions and requires inter-core data transfer between decomposition steps.
Also considering its lower computational cost, the communication to computation ratio is higher in HM, which explains its lower parallel efficiency.

\subsection{Parameters and numerical settings}
\begin{table*}
		\caption{Summary of geometry and numerical parameters used in 3D DNS runs for the IDT regime: $\mathrm{Wi}=23$ in the MFU and $\mathrm{Wi}\leq 96$ in the large box.}
		\begin{minipage}{\linewidth}
		\centering
		\begin{tabular}{c|c|c|c|c|c|c}
			\hline	
			&\multirow{2}{*}{$L_x^+\times L_y^+\times L_z^+$}
				&\multirow{2}{*}{$N_x\times N_y\times N_z$}
				&\multirow{2}{*}{$\delta_x^+\times(\delta_{y,\min}^+\sim\delta_{y,\max}^+)\times\delta_z^+$}
				&\multirow{2}{*}{$\delta_t$}	&\multicolumn{2}{c}{$\mathrm{Sc}$}\\
			\cline{6-7}
			&&&&&	SM	&HM	\\
			\hline
			STG	&$360\times 169.71\times 250$	&$80\times 145\times 95$
				&$4.5\times(0.02\sim 1.85)\times 2.63$
				&$0.01$	&\num{e5}	&\multirow{7}{*}{$\infty$}	\\
			\cline{1-6}
			MFU	&$360\times 169.71\times 250$	&$128\times 131\times 126$
				&$2.81\times(0.025\sim 2.05)\times 1.98$
				&$0.005$	&$0.5$	&\\
			\cline{1-6}
			Extended	&$4000\times 169.71\times 800$	&$440\times 97\times 147$
				&$9.09\times(0.046\sim 2.81)\times 5.44$
				&$0.02$	&$0.5$\footnote{Also tested: $0.02$, $0.08$, and $0.2$.}	&\\
			\cline{1-6}
			\multirow{4}{*}{Resolution Test}	&\multirow{4}{*}{$360\times 169.71\times 250$}
				&$40\times 73\times 46$	&$9.0\times(0.081\sim 3.70)\times 5.43$
					&$0.02$	&\multirow{4}{*}{N/A}\\
			&	&$64\times 109\times 62$	&$5.63\times(0.036\sim 2.47)\times 4.03$
					&$0.01$	&\\
			&	&$80\times 145\times 94$	&$4.5\times(0.020\sim 1.85)\times 2.66$
					&$0.01$	&\\
			&	&$128\times 217\times 126$	&$2.81\times(0.0090\sim 1.23)\times 1.98$
					&$0.005$	&\\
			\hline	
		\end{tabular}
		\end{minipage}
		\label{tab:resolution_3d}
\end{table*} 

\begin{table*}
		\caption{Summary of geometry and numerical parameters used in 2D DNS runs for the EIT regime: $\mathrm{Wi}=64$ and $800$.}
		\centering
		\begin{tabular}{c|c|c|c|c|c|c}
			\hline	
			&\multirow{2}{*}{$L_x^+\times L_y^+$}	&\multirow{2}{*}{$N_x\times N_y$}
				&\multirow{2}{*}{$\delta_x^+\times(\delta_{y,\min}^+\sim \delta_{y,\max}^+)$}
				&\multirow{2}{*}{$\delta_t$}	&\multicolumn{2}{c}{$\mathrm{Sc}$}\\
			\cline{6-7}
			&&&&&SM	&HM	\\
			\hline
			Standard	&$720\times 169.71$	&$1280\times 369$
				&$0.563\times(0.0031\sim 0.72)$
				&$0.001$	&0.5	&\multirow{8}{*}{$\infty$}\\
			\cline{1-6}
			\multirow{6}{*}{Resolution Test} &\multirow{7}{*}{$720\times 169.71$}
				&$288\times 97$		&$2.50\times(0.045\sim 2.78)$	&$0.005$
				&\multirow{7}{*}{N/A}
				&\\
			&	&$288\times 369$	&$2.50\times(0.0031\sim 0.72)$	&$0.005$	&&\\
			&	&$512\times 185$	&$1.41\times(0.012\sim 1.45)$	&$0.0025$	&&\\
			&	&$512\times 369$	&$1.41\times(0.0031\sim 0.72)$	&$0.0025$	&&\\
			&	&$512\times 731$	&$1.41\times(0.00079\sim 0.37)$	&$0.0025$	&&\\
			&	&$1280\times 369$	&$0.563\times(0.0031\sim 0.72)$	&$0.001$	&&\\
			&	&$2560\times 369$	&$0.281\times(0.0031\sim 0.72)$	&$0.0005$	&&\\
			\hline	
		\end{tabular}
		\label{tab:resolution_2d}
\end{table*} 

All results reported in this study are obtained at $\mathrm{Re}=3600$ ($\mathrm{Re}_\tau=84.85$), $\beta=0.97$, and $b=5000$, with different $\mathrm{Wi}$.
Numerical parameters for 3D and 2D DNS runs are summarized in \cref{tab:resolution_3d} and \cref{tab:resolution_2d}, respectively.
Hereinafter, ``+'' indicates quantities scaled by the friction velocity $u_\tau$ and viscous length scale $l/u_\tau$ (or ``wall unit'') -- turbulent inner scales~\citep{Pope_2000}.
$N_w$ is the number of grid points in the $w$ direction; $\delta^+_x$ and $\delta^+_z$ are uniform mesh sizes used in $x$ and $z$, whereas the CGL grid in $y$ is nonuniform with the smallest meshes ($\delta^+_{y,\min}$) at the walls and largest ones ($\delta^+_{y,\max}$) at the channel center.
Time step is adjusted in accordance with spatial resolution based on the Courant-Friedrichs-Lewy (CFL) stability condition.

Three types of simulation are performed.
\begin{itemize*}
	\item Streak transient growth (STG) simulation that follows the transient development of turbulence from an imposed disturbance (\cref{Sec_STG}).
	\item Statistically converged flow in the IDT regime in 3D domains, including a minimal flow unit (MFU) and an extended flow cell.
	\item Statistically converged flow in the EIT regime in an $xy$-2D flow domain.
\end{itemize*}
MFU is used when the focus is on the temporal evolution and intermittency of flow structures, which are conveniently reflected in time series of spatial average quantities when the domain is sufficiently small~\citep{Jimenez_Moin_JFM1991,Gibson_Cvitanovic_JFM2008,Xi_Graham_JFM2010}.
It is also preferred for numerical resolution tests because of its lower computational cost.
Extended-domain simulation is usually used for more accurate flow statistics. It is also used in our study because it supports IDT for a wider $\mathrm{Wi}$ range~\citep{Xi_Graham_JFM2010,Wang_Graham_AIChEJ2014}.
Since IDT is only sustained in 3D domains but EIT also exists in 2D domains~\citep{Sid_Terrapon_PRFluids2018,Shekar_Graham_PRL2019,Zhu_Xi_arXiv2020}, 2D DNS is used to obtain pure EIT states without the interference from IDT.

The purpose of STG is to validate the correctness of the new HM code by comparing with the established SM code.
Negligibly low GAD, with $\mathrm{Sc}=\num{e5}$ ($1/\mathrm{Pe}=\num{2.78e-9}$), is thus used in STG with SM (compared with $\mathrm{Sc}=\infty$ of HM). We are not using strictly zero GAD in SM because that would require non-trivial changes of the SM code.
In statistically converged turbulence, effects of AD are studied in both IDT and EIT regimes by comparing the statistical results from SM and HM.
For SM, a GAD term is used, while for HM, neither GAD nor LAD is used and thus $\mathrm{Sc}=\infty$.
The standard $\mathrm{Sc}=0.5$ used in most SM runs of this study is consistent with our earlier work, which corresponds to $1/\mathrm{Pe}=1/(\mathrm{ScRe})=\num{5.56e-4}$. 
This is lower than $1/\mathrm{Pe}=\RevisedText{\bigO}(\num{e-2})$ often seen in the literature~\citep{Sureshkumar_Beris_POF1997,Ptasinski_Nieuwstadt_JFM2003,Housiadas_Beris_POF2005,Li_Khomami_JNNFM2006,Li_Khomami_PRE2015}.

Dependence on numerical resolution is studied in both IDT and EIT regimes for the new HM (\cref{Sec_Mesh}).
The standard resolutions used in our extended-domain DNS for IDT and 2D DNS for EIT reflect the results of our resolution analysis.
For standard MFU, resolution much higher than the recommended level for IDT (also compare with $\delta_x^+=8.57$, $\delta_z^+=5.0\sim 5.5$, and $N_y=73$ used in \citet{Xi_Graham_JFM2010} for the same $\mathrm{Re}_\tau$) is used to evaluate the effects of AD on small scales.

\section{Results and Discussion}\label{Sec_results}
\subsection{Comparison of HM and SM codes in STG simulation}\label{Sec_STG}
\begin{table}
	\begin{center}
		\caption{Parameters of the initial condition for STG simulations.}
		\def~{\hphantom{0}}	
		\begin{tabular}{c|c|c|c|c}
			\hline	
			$\beta_g$	&	$A_s$	&	$\beta_s$	&	$A_p$	&	$\alpha_p$ \\[5pt]
			\hline	
			9.0	&	0.0707	&	2		&	0.0707	&	1\\	
			\hline	
		\end{tabular}
		\label{tab:init_STG}
	\end{center}	
\end{table} 

To validate the correctness of the new HM code, direct comparison with the established SM code is made through simulation from the same initial condition (IC).
The streak transient growth (STG) simulation is frequently used to study the transient development of turbulence from a well-defined initial disturbance.
The velocity IC is a superposition of a base flow and a perturbation velocity field.
The base flow is given by 
\begin{gather}
	U_b(y,z)=U_m(y)+U_s(z)g(y'), V_b=W_b=0
\end{gather}
where
\begin{gather}
	U_s(z)=A_s\cos\left(\frac{2\pi\beta_s}{L_z}\left(z-z_\beta\right)\right)\\
	g(y')=\frac{y'\exp\left(-\beta_g y'^2\right)}
		{\max\left(y'\exp\left(-\beta_g y'^2\right)\right)}
\end{gather}
($y'\equiv y+1$ is distance to the bottom wall).
$U_b$, $V_b$, and $W_b$ are the $x$-, $y$-, and $z$-components of the base-flow velocity, respectively.
$U_m(y)$ is the mean velocity profile of Newtonian turbulent channel flow at the same $\mathrm{Re}$.
$U_s(z)g(y')$ defines streamwise velocity streaks with their amplitude, spacing, and phase of the wave adjusted by $A_s$, $\beta_s$, and $z_\beta$, respectively.
Due to the periodicity in $z$, $z_\beta=0.5 L_z$ is used without loss of generality.
Function $g(y')$ provides $y$-dependence and $\beta_g$ is adjusted to set the peak of $g(y')$ (normalized to 1) at $y^+=20$.
The perturbation velocity
\begin{gather}
	v_{p,x}=v_{p,y}=0, v_{p,z}=A_p\sin(\frac{2\pi\alpha_p x}{L_x})g(y')
\end{gather}
further introduces streamwise variation to the disturbance, which is necessary to trigger turbulence at least in Newtonian flow.
Parameters $A_p$ and $\alpha_p$ adjust the amplitude and wavelength, respectively.
All STG parameters used in this study are provided in \cref{tab:init_STG}. More details on our STG simulation are found in \citet{Zhu_Xi_JFM2019}.
For viscoelastic STG simulation, the IC sets all components of $\mbf\alpha$ to 1.
This arbitray choice is inconsequential as far as our purpose of comparing two codes is concerned. (A more reasonable IC would be setting $\mbf\alpha=(b/(b+5))\mbf\delta$, which is the equilibrium solution to FENE-P \cref{equ_Fene1_c6}.)


\begin{figure}
	\centering
	\includegraphics[width=.75\linewidth, trim=0mm 0mm 0mm 0mm, clip]{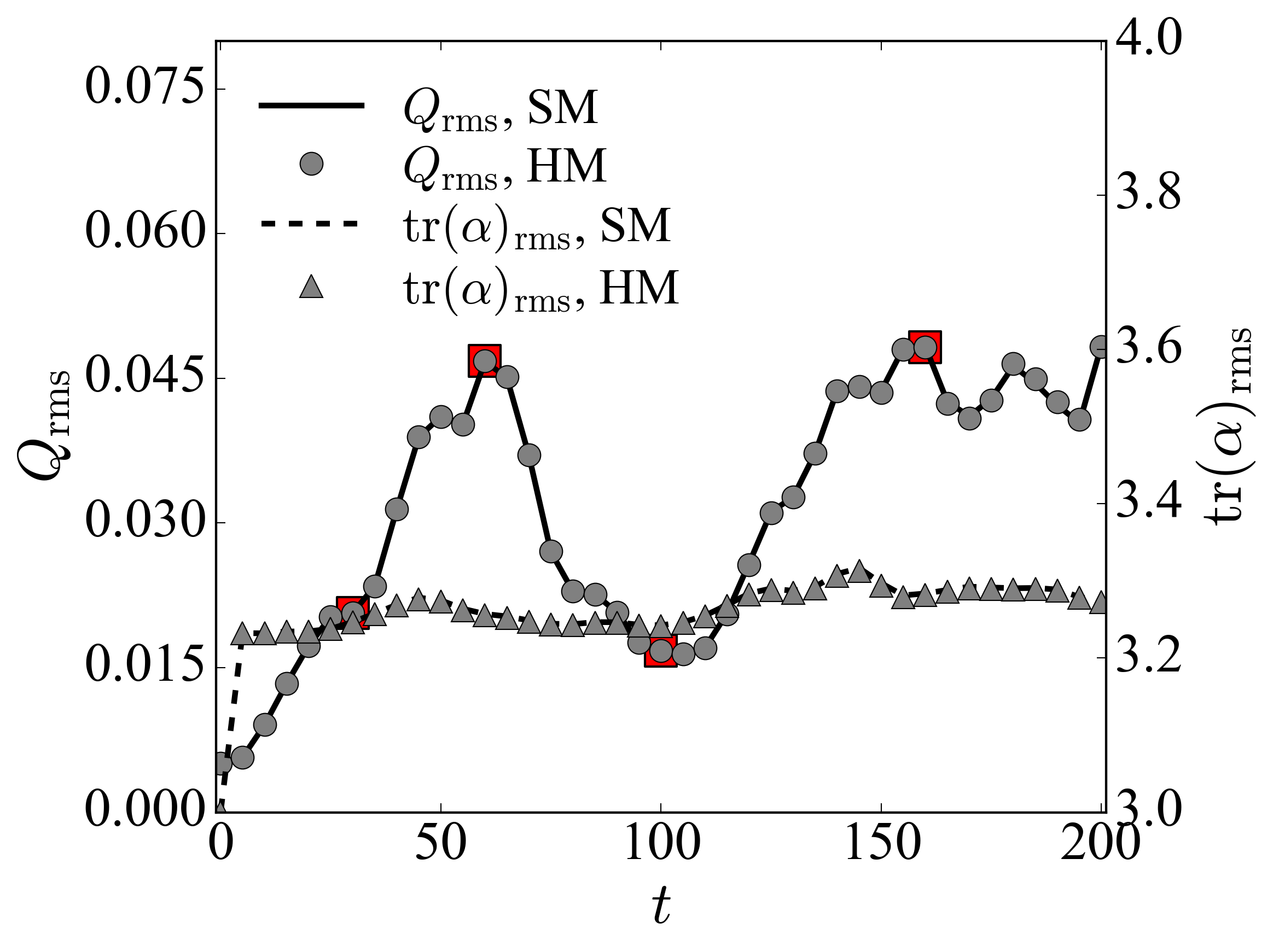}
	\caption{Time series of the r.m.s. of $Q$ and $\mathrm{tr}(\boldsymbol{\alpha})$ in STG simulations ($\mathrm{Wi}=1$). Red squares mark data points corresponding to the snapshots shown in \cref{fig:Q_STG}. }
	\label{fig:tim_STG}
\end{figure}

\begin{figure}
	\centering
	\includegraphics[width=\linewidth, trim=0mm 0mm 0mm 0mm, clip]{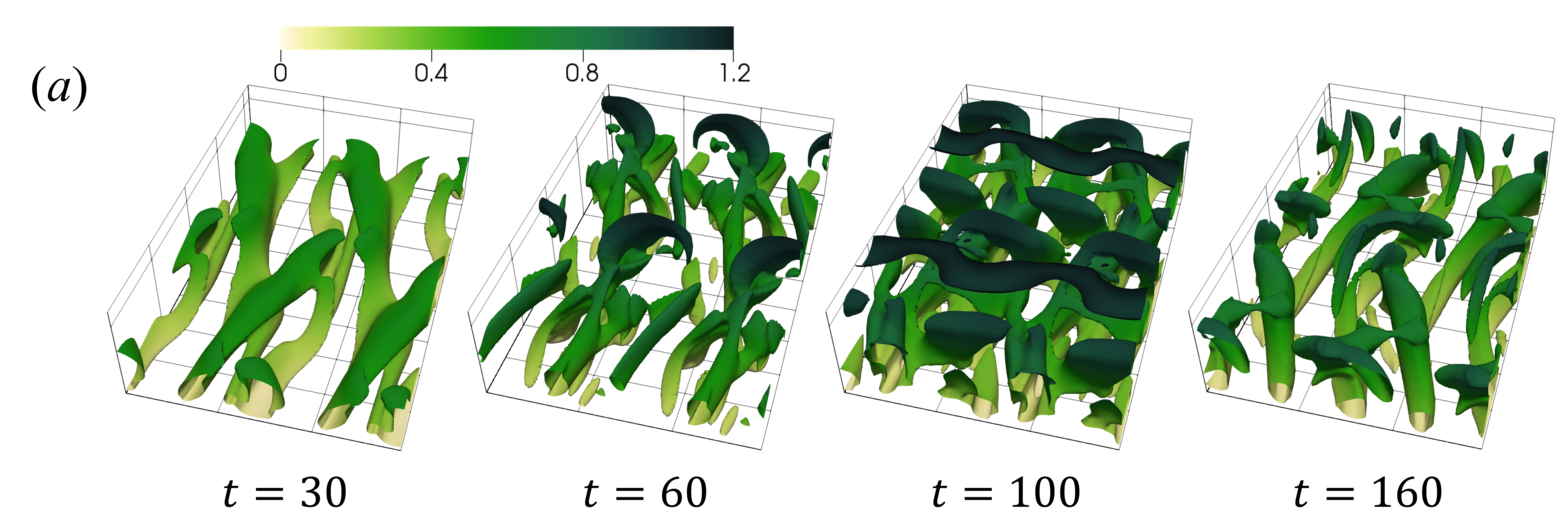}
	\includegraphics[width=\linewidth, trim=0mm 0mm 0mm 0mm, clip]{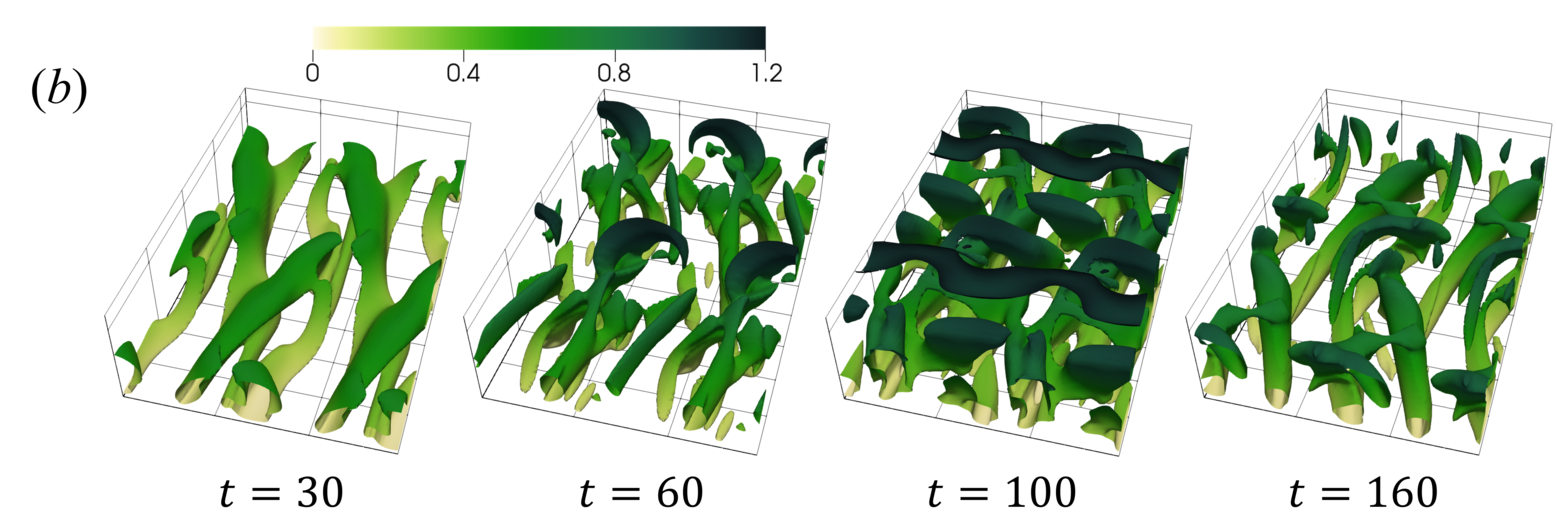}	
	\caption{Vortex configuration identified by $Q=Q_\mathrm{rms}$ in STG simulations ($\mathrm{Wi}=1$): (a) HM; (b) SM.
	Color varies from light to dark with the distance from the bottom wall.
	Time stamps match \cref{fig:tim_STG}.}
	\label{fig:Q_STG}
\end{figure}

\begin{figure}
	\centering
	\includegraphics[width=\linewidth, trim=0mm 0mm 0mm 0mm, clip]{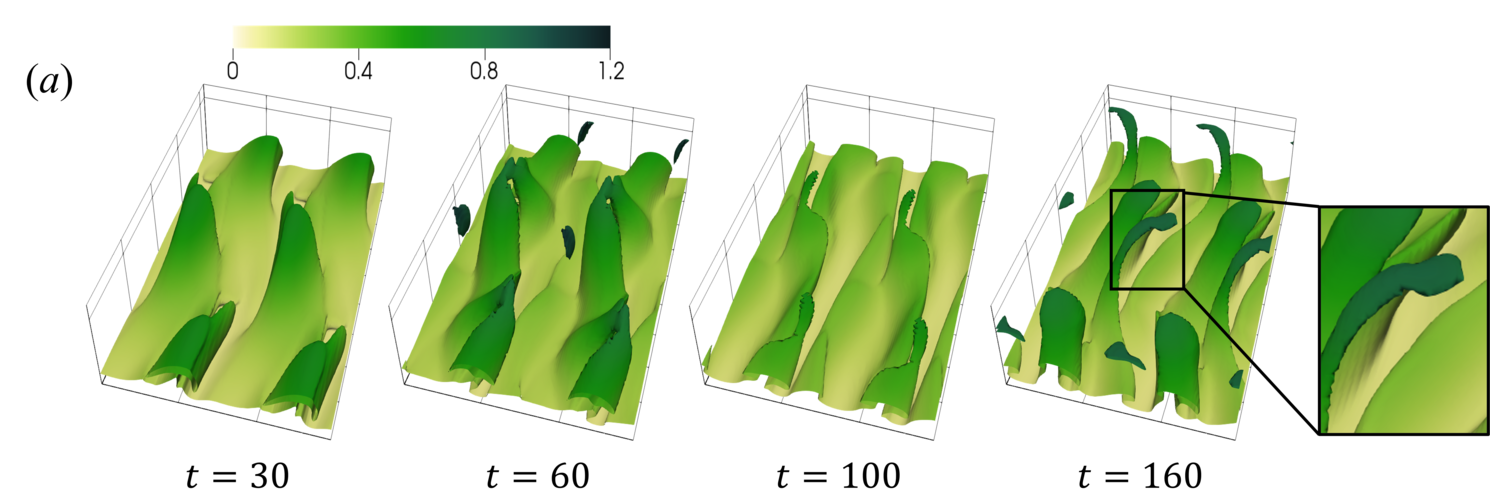}
	\includegraphics[width=\linewidth, trim=0mm 0mm 0mm 0mm, clip]{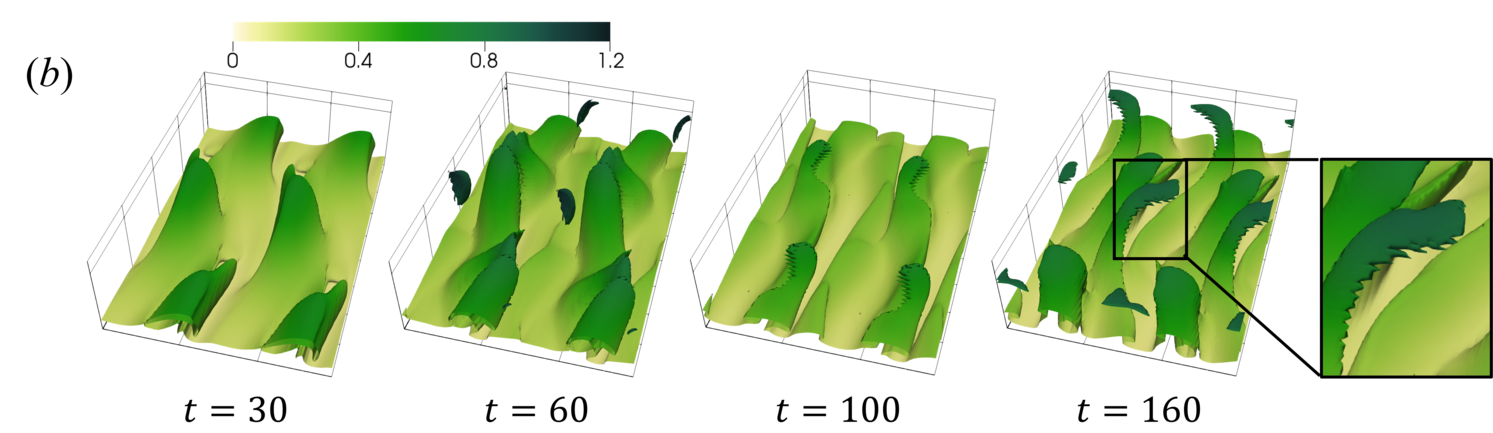}
	\caption{Iso-surfaces of $\mathrm{tr}(\boldsymbol{\alpha})=\mathrm{tr}(\boldsymbol{\alpha})_\mathrm{rms}$ in STG simulations ($\mathrm{Wi}=1$): (a) HM; (b) SM.
	Color varies from light to dark with the distance from the bottom wall.
	Time stamps match \cref{fig:tim_STG}.}
	\label{fig:trA_STG}
\end{figure}

For direct comparison with HM simulation (which uses no AD), an extremely high $\mathrm{Sc}=\num{e5}$ is used in SM.
With this negligibly low GAD, comparison can only be made at very low $\mathrm{Wi}$ ($\mathrm{Wi}=1$ is used here) where numerically stable solution is still possible.
In \cref{fig:tim_STG}, time series of the instantaneous root-mean-square (r.m.s.) value of $Q$, as in the $Q$-criterion for vortex identification~\citep{Hunt_Wray_CTR1988}, and that of the trace of the polymer conformation tensor $\mathrm{tr}(\boldsymbol{\alpha})$ in STG simulations using HM and SM are compared. 
According to the $Q$-criterion, regions with
\begin{equation} 
	Q\equiv\frac{1}{2}\left(\|\mbf\Omega\|^2-\|\mbf S\|^2\right)>0
\label{equ_Q}
\end{equation}
are considered to be dominated by vortex flow, where
\begin{gather}
	\mbf\Omega\equiv\frac{1}{2}\left(\mbf{\nabla}\mbf{v}-\mbf{\nabla}\mbf{v}^T\right)\\
	\mbf\Gamma\equiv\frac{1}{2}\left(\mbf{\nabla}\mbf{v}+\mbf{\nabla}\mbf{v}^T\right)
\end{gather}
are the vorticity and rate-of-strain tensors, respectively, and $\Vert\cdot\Vert$ denotes the Frobenius tensor norm.
Isosurfaces $Q=Q_\mathrm{rms}$ and $\mathrm{tr}(\boldsymbol{\alpha})=\mathrm{tr}(\boldsymbol{\alpha})_\mathrm{rms}$ of selected moments are shown in \cref{fig:Q_STG,fig:trA_STG}.

Starting from the IC, vortex strength, as measured by $Q_\mathrm{rms}$, initially grows and reaches its first peak at $t=60$ (\cref{fig:tim_STG}).
During this period, quasi-streamwise vortices sweep in the spanwise direction ($t=30$ in \cref{fig:Q_STG}). Their heads bend sideways to form spanwise arms, which lift up as the vortices appear hook-shaped ($t=60$).
After the peak, $Q_\mathrm{rms}$ decreases until reaching a minimum near $t=100$ when vortices dilate and interconnect as they weaken.
$Q_\mathrm{rms}$ bounces again and reaches its second peak at $t=160$, during which quasi-streamwise legs of the vortices strengthen but their spanwise arms continue to decay.
Comparing the two numerical algorithms, excellent agreement is found in both time series and detailed flow images for the entire duration of this complex sequence of dynamical events.

For $\mathrm{tr}(\boldsymbol{\alpha})_\text{rms}$, there is much less variation in the time series after the initial rapid rise that describes the stretching of polymers by the flow from their initial conformations (\cref{fig:tim_STG}). Both HM and SM again render indistinguishable temporal profiles. 
Instantaneous images also appear highly similar between the two methods at $t=30$ and $60$ (\cref{fig:trA_STG}).
Some subtle differences become discernible at later times.
At $t=100$, the HM case shows elongated structures of high $\mathrm{tr}(\boldsymbol{\alpha})$ between vortices protruding out in the shape of index fingers, which become stretched to thin curved threads at $t=160$.
The same structures are captured by SM as well but they appear more inflated and less stretched in shape. In addition, numerical oscillation starts to occur, as reflected by the jagged edges of those structures.
At low $\mathrm{Wi}$, these differences do not affect the flow field in any appreciable way, resulting in nearly identical $Q$ fields between HM and SM (\cref{fig:Q_STG}).
The overall agreement between polymer conformation fields captured by the two methods proves that the new HM algorithm was correctly implemented.
Discrepancies in detailed features reflect their different numerical performance, especially in terms of numerical stability in regions with strong polymer stress variations, which will be further evaluated below in statistically-converged flow.

\subsection{DNS of IDT and EIT: effects of artificial diffusion}\label{Sec_ResSS}
We now turn to DNS solutions where the dynamics has converged (in the statistical sense) with time. Our focus here is on the comparison between HM and SM algorithms, through which we are especially interested in the effects of GAD in SM on the results.
Considering the recent discovery and understanding of EIT as a fundamentally different stage of turbulence~\citep{Samanta_Hof_PNAS2013,Choueiri_Hof_PRL2018,Xi_POF2019}, numerical solutions in IDT and EIT regimes will be differentiated in this discussion.

\begin{figure}
	\centering
	\includegraphics[width=\linewidth, trim=0 0 0 0, clip]{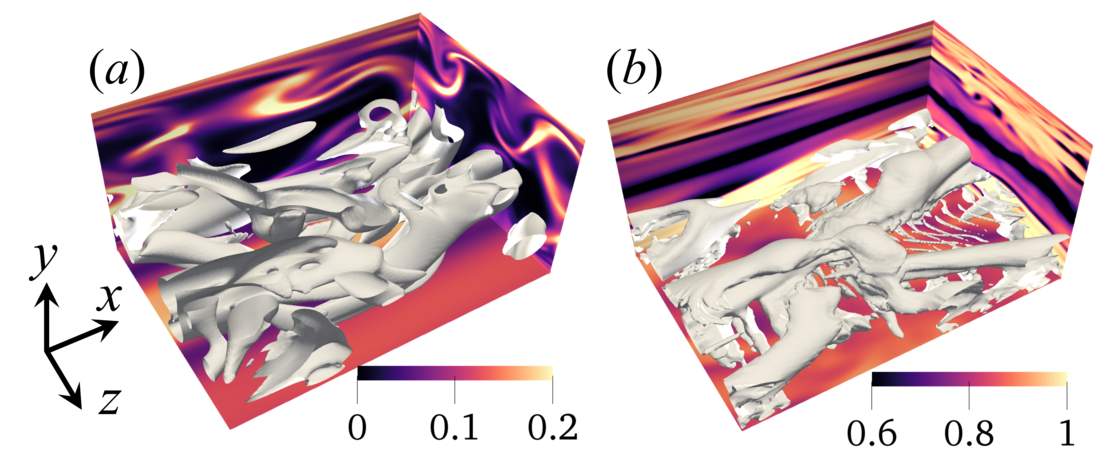}
	\caption{\RevisedText{Representative snapshots of 3D solutions using HM (no AD): (a) $\mathrm{Wi}=23$ at IDT; (b) $\mathrm{Wi}=800$ at EIT.
	Color scale maps to $\mathrm{tr}(\mbf\alpha)/b$; iso-surfaces are for $Q=0.005$ (bottom half of the channel only).}}
	\label{fig:eit:3Dimages}
\end{figure}

IDT is better known to researchers, which shares similar flow structures and an analogous self-sustaining mechanism with Newtonian turbulence~\citep{Xi_POF2019}.
\RevisedText{\Cref{fig:eit:3Dimages}(a) shows a typical snapshot of IDT that represents a moment of strong turbulent activities -- the so-called active turbulence~\citep{Xi_Graham_PRL2010,Xi_Graham_JFM2012,Xi_POF2019}.
The flow field is strongly three-dimensional with quasi-streamwise vortices populating the near wall region.
These counter-rotating vortices sweep up slower-moving fluids upwards (away from the wall) and generate streamwise wavy low-speed velocity streaks
(shown as high-$\mathrm{tr}(\mbf\alpha)$ strands stretching from the wall towards the channel center in \cref{fig:eit:3Dimages}(a)) in between.
Such vortex-streak coherent structures are quasi-repetitive in both space and time, which allows}
the key dynamics to be captured in small basic units known as MFUs~\citep{Jimenez_Moin_JFM1991,Xi_Graham_JFM2010}\RevisedText{: e.g., solutions shown in \cref{fig:eit:3Dimages} are from a MFU.}
\RevisedText{Meanwhile,}
longer-range correlations between such units in realistic flows can only be captured in more extended flow domains~\citep{Li_Khomami_JNNFM2006}.
Regardless of the domain size, IDT can only sustain in 3D flow~\citep{Waleffe_POF1997}.
Sufficient polymer stress weakens vortex structures of IDT, which leads to the onset of DR~\citep{Li_Graham_JFM2006}.
More recently, it was shown that at higher $\mathrm{Wi}$, polymer stress can suppress the vortex lift up process and interrupt its dependent vortex regeneration pathway~\citep{Zhu_Xi_JNNFM2018,Zhu_Xi_POF2019}, causing distinct differences between low- and high-extent DR (LDR and HDR) regimes~\citep{Warholic_Hanratty_EXPFL1999}.

\begin{figure}
	\centering
	\includegraphics[width=\linewidth, trim=20 24 6 12, clip]{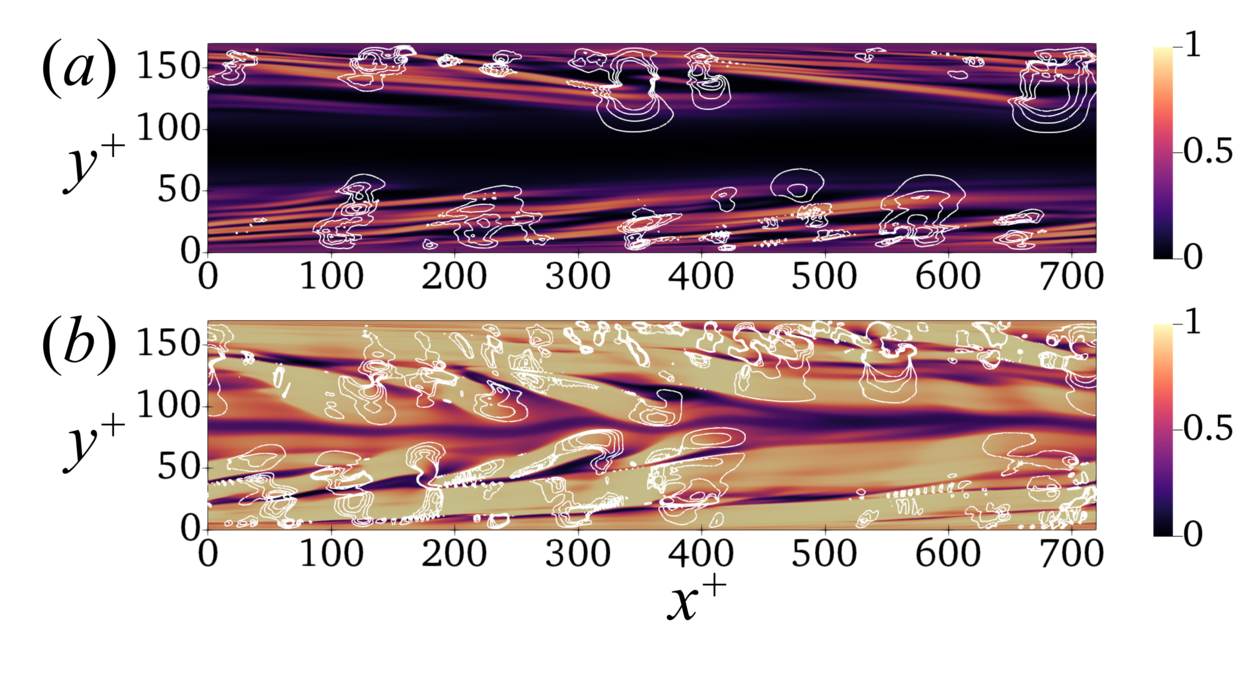}
	\caption{Representative snapshots of 2D EIT solutions using HM (no AD): (a) $\mathrm{Wi}=64$; (b) $\mathrm{Wi}=800$.
	Color scale maps to $\mathrm{tr}(\mbf\alpha)/b$; contour lines are for $Q=0.005, 0.01, 0.015$ and $0.02$.}
	\label{fig:eit:images}
\end{figure}

EIT is a developing concept with new understanding \RevisedText{continuously emerging.}
Its underlying self-sustaining mechanism, in which polymer stress feeds into (instead of merely suppressing) flow instability, differs fundamentally from that of IDT.
EIT is typically observed at higher $\mathrm{Wi}$ and can be triggered and sustained independent of IDT~\citep{Samanta_Hof_PNAS2013,Choueiri_Hof_PRL2018,Xi_POF2019}.
It is characterized by distinct spanwise vortices and titled sheets of highly-stretched polymers, which is thus easily distinguishable  from IDT.
It became recently known that EIT is self-sustaining in an $xy$-2D plane~\RevisedText{\citep{Sid_Terrapon_PRFluids2018,Shekar_Graham_PRL2019,Zhu_Xi_arXiv2020}},
making it possible to isolate EIT from IDT structures which may intermittently occur in 3D flow even at high $\mathrm{Wi}$.
Representative images from 2D DNS of such EIT states, using our new HM code, are shown in \cref{fig:eit:images}.
Spanwise vortices appear as circular $Q$ patterns originating from the walls and come in two size groups: large rolls with diameter up to $\sim 50$ wall units and thin threads with diameter lower than $\sim 10$ wall units.
Polymer structures are dominated by layered sheets of high $\mathrm{tr}(\mbf\alpha)$ titled at an acute angle from the $x$ direction.
At higher $\mathrm{Wi}$ (\RevisedText{$\mathrm{Wi}=800$ in} \cref{fig:eit:images}(b)), the overall $\mathrm{tr}(\mbf\alpha)$ magnitude is higher and vortices also spread closer to the channel center.

\RevisedText{%
DNS solutions of 3D flow have also be obtained for $\mathrm{Wi}$ up to $800$ and, same as the 2D case, solutions using the new HM algorithm remain numerically stable without AD (local or global) for all $\mathrm{Wi}$ attempted.
As shown in \cref{fig:eit:3Dimages}(b) for $\mathrm{Wi}=800$, 3D solutions in the EIT regime have a strong presence of spanwise vortices, including both rolls and threads, that resemble 2D EIT structures.
Meanwhile, quasi-streamwise vortices are also observed and}
some level of three-dimensionality \RevisedText{clearly} persists\RevisedText{. Similarity of these quasi-streamwise vortices with IDT structures once led to the speculation that the 3D solution is simply}
a hodgepodge of \RevisedText{the} pure-form 2D EIT and intermittent eruptions of IDT structures \RevisedText{ -- the latter were presumed to diminish with increasing $\mathrm{Wi}$}~\citep{Sid_Terrapon_PRFluids2018}\RevisedText{.
On the contrary, our latest study found that three dimensionality persists with increasing $\mathrm{Wi}$~\citep{Zhu_Xi_arXiv2020}. The nature of quasi-streamwise vortices at high $\mathrm{Wi}$ differs from those in IDT -- they are indeed an integral part of a new type of dynamics.
Spanwise 2D-like structures are nonetheless still critical in 3D flow in the EIT regime, without which it is well known that the flow would laminarize at $\mathrm{Wi}\gtrsim 30$ for the current simulation domain and parameters~\citep{Xi_Graham_JFM2010}.}

The focus of this study is on the numerical performance of the new HM algorithm.
\RevisedText{Therefore, we will avoid the complexity of 3D turbulent dynamics in the EIT regime, which is as yet not fully understood, and limit our discussion on EIT solutions to 2D DNS.
These solutions contain all key features of EIT, especially its}
stress shocks -- sharp edges of polymer sheets\RevisedText{, that make it numerically challenging.
It is thus adequate as a test case for numerical methods.}
Likewise,
\RevisedText{although DNS has been performed using HM for $\mathrm{Wi}$ up to $800$, detailed investigation of the effects of AD and numerical resolution on 2D EIT is only} performed
\RevisedText{at $\mathrm{Wi}=64$, since the fundamental understanding of the $\mathrm{Wi}$-dependence of those solutions is also limited~\citep{Zhu_Xi_arXiv2020}.}

\begin{figure}
	\centering
	\includegraphics[width=0.75\linewidth, trim=0 0 0 0, clip]{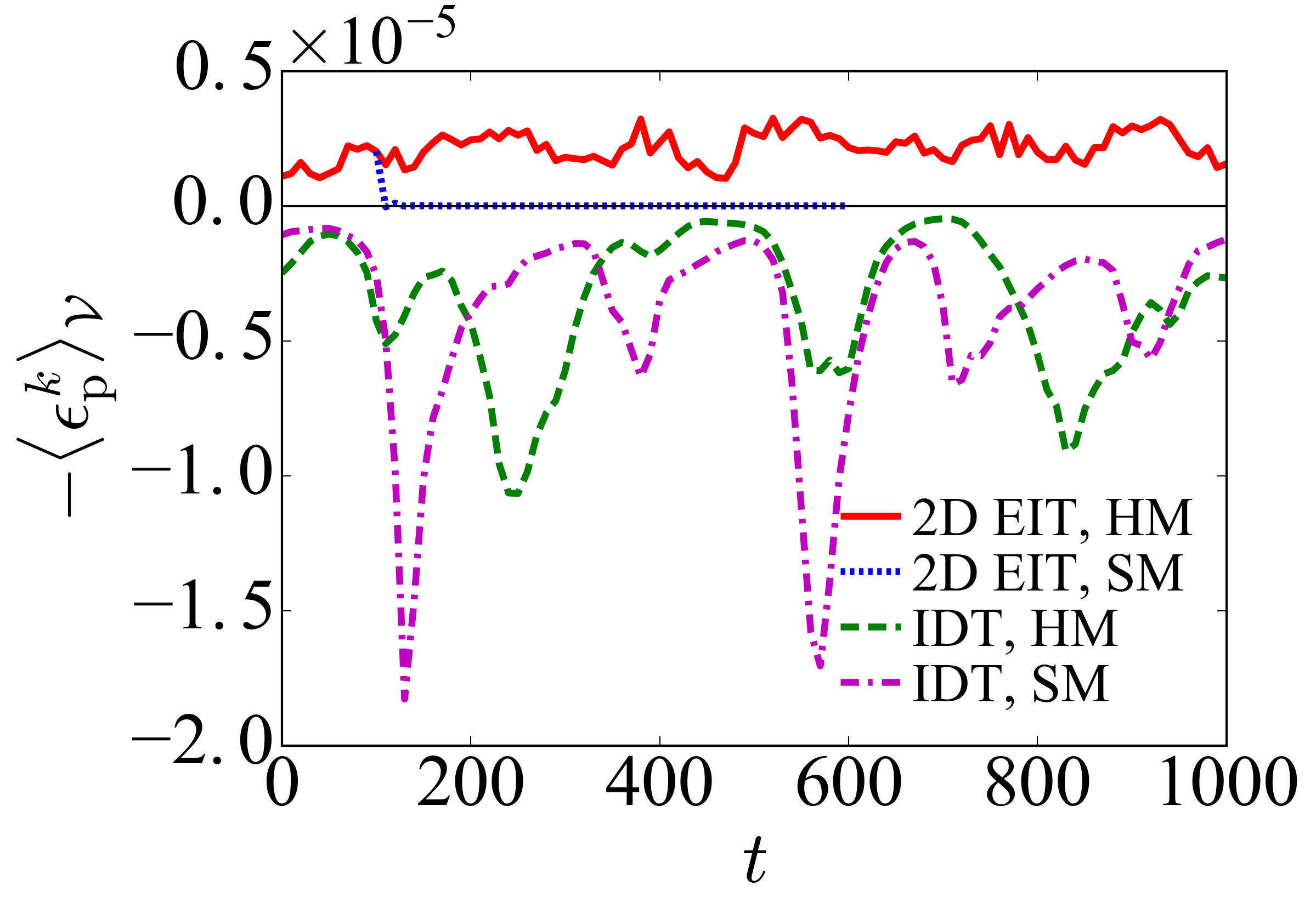}
	\caption{Time series of the elastic conversion rate of IDT (3D MFU solution at $\mathrm{Wi}=23$) and EIT (2D solution at $\mathrm{Wi}=64$) using HM (no AD) and SM ($\mathrm{Sc}=0.5$).}
	\label{fig:epsilonp}
\end{figure}

Whether polymer stress suppresses or enhances turbulence and instability is most straightforwardly seen from the balance of turbulent kinetic energy (TKE) $k$:
\begin{gather}
	\frac{\partial\langle k\rangle_\mathcal{V}}{\partial t}
	= \langle\mathcal{P}^k\rangle_\mathcal{V}
		-\langle\epsilon_\text{v}^k\rangle_\mathcal{V}
		-\langle\epsilon_\text{p}^k\rangle_\mathcal{V}
	\label{eq:tkebal}
\end{gather}
where
\begin{gather}
	k\equiv\frac{1}{2}\langle\mbf v'\cdot\mbf v'\rangle
\end{gather}
(apostrophe ``$\prime$'' indicates fluctuating components), $\langle\cdot\rangle$ represents average over $x$ and $z$, and $\langle\cdot\rangle_\mathcal{V}$ represents volume average.
The three terms on the RHS are the volume averages of
\begin{gather}
	\mathcal{P}^k\equiv
		-\langle v_x'v_y'\rangle\frac{\partial\langle v_x\rangle}{\partial y}\\
	\epsilon_\text{v}^k\equiv
		\frac{2\beta}{\mathrm{Re}}\left\langle\mbf{\Gamma}':\mbf{\Gamma}'\right\rangle\\
	\epsilon_\text{p}^k\equiv
		\frac{2(1-\beta)}{\mathrm{ReWi}}
		\left\langle\mbf{\tau}_\text{p}':\mbf{\Gamma}'\right\rangle
	\label{eq:epsilonp}
\end{gather}
which represent contributions from turbulence production (through the inertial mechanism), viscous dissipation, and the conversion rate between TKE and polymer elastic energy, respectively~\citep{Zhu_Xi_JNNFM2019,Xi_POF2019,Zhu_Xi_arXiv2020}.
Time series of $-\langle\epsilon_\text{p}^k\rangle_\mathcal{V}$ are plotted in \cref{fig:epsilonp}.
For IDT, $-\langle\epsilon_\text{p}^k\rangle_\mathcal{V}$ stays negative, as polymer stress suppresses turbulence and converts TKE to polymer elastic energy. 
For EIT, $-\langle\epsilon_\text{p}^k\rangle_\mathcal{V}$ is positive, indicating that polymer stress is fueling instability.

As to the effects of AD, for IDT, $-\langle\epsilon_\text{p}^k\rangle_\mathcal{V}$ from SM+GAD has the same magnitude, at least statistically, as the HM (no AD) case.
Within the limited time window shown, two strong (negative) overshoots are seen in the SM+GAD case, whereas temporal fluctuations in the HM case appears milder.
These overshoots typically follow the so-called bursting events in the flow field, which are intermittent in nature~\citep{Zhu_Xi_JNNFM2019}.
Therefore, comparison based on this short time window is not conclusive.
Such discrepancy in $-\langle\epsilon_\text{p}^k\rangle_\mathcal{V}$ fluctuations, if real, is likely associated with polymer stress fluctuations, which are slightly affected by AD (discussed below).

At EIT, effects of GAD are more consequential. At $\mathrm{Sc}=0.5$, GAD quickly suppresses EIT and causes the laminarization of the flow.
Using a separate FD code and for comparable parameter settings, \citet{Sid_Terrapon_PRFluids2018} reported that GAD with $\mathrm{Sc}<9$ cannot sustain 2D EIT.
Although the threshold $\mathrm{Sc}$ likely depends on the specific numerical scheme and spatial resolution, it is almost certain that at $\mathrm{Sc}<\RevisedText{\bigO}(1)$, which is required for numerical stability in SM, self-sustaining EIT cannot be captured.
This shows the importance of shock-capturing capability of the numerical scheme, particularly the numerical treatment of the $\mbf v\cdot\mbf\nabla\mbf\alpha$ term, for DNS in the EIT regime.
As such, for the rest of the paper, AD effects will only be discussed for the IDT regime and EIT solutions can only be obtained from HM.
Shock-capturing capabilities of SM+GAD, TVD, and other common FD schemes are further compared in \ref{Sec_schmComp}.

\begin{figure}
	\centering
	\includegraphics[width=0.5\linewidth, trim=0mm 0mm 0mm 0mm, clip]{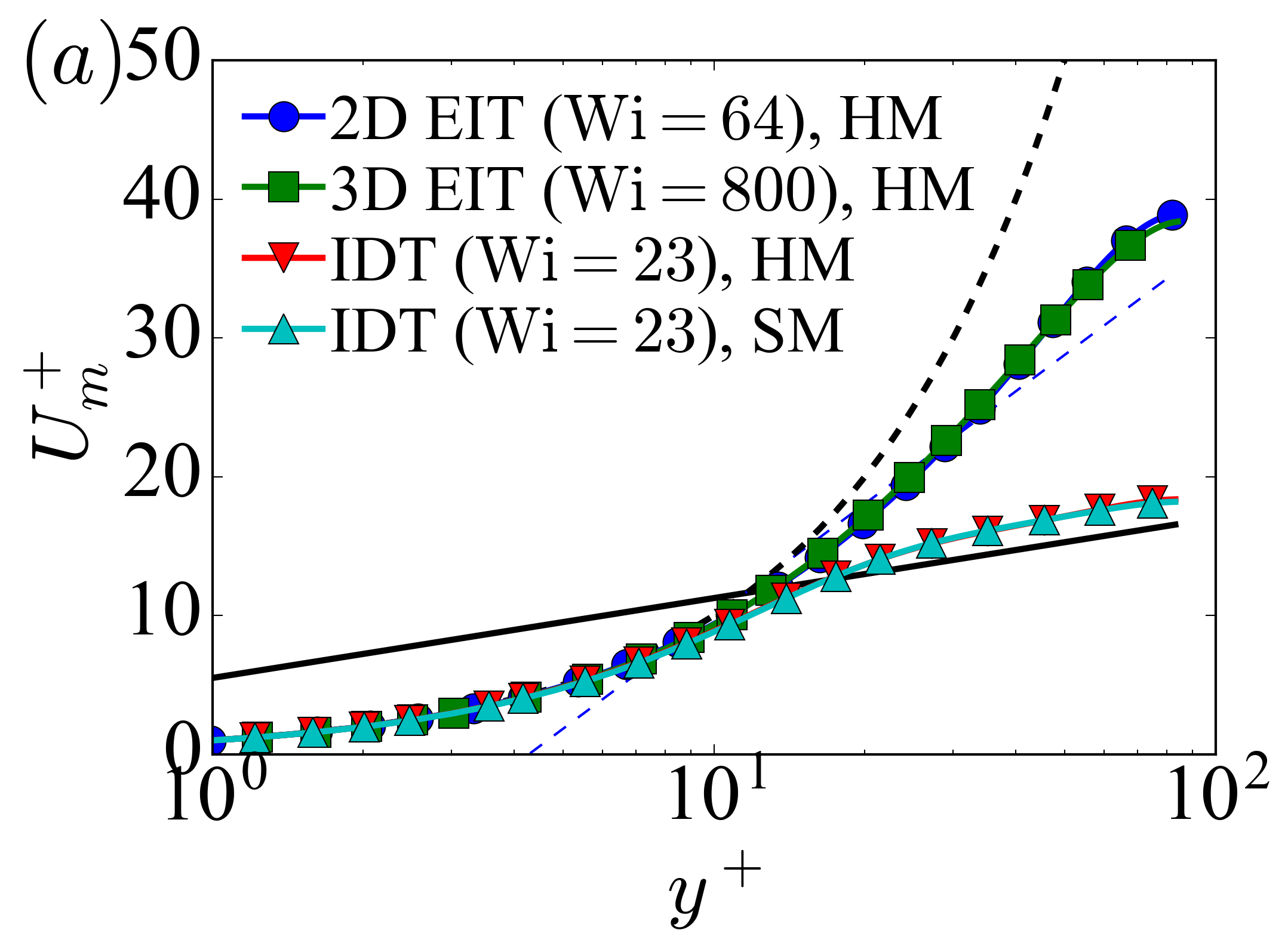}%
	\includegraphics[width=0.5\linewidth, trim=0mm 0mm 0mm 0mm, clip]{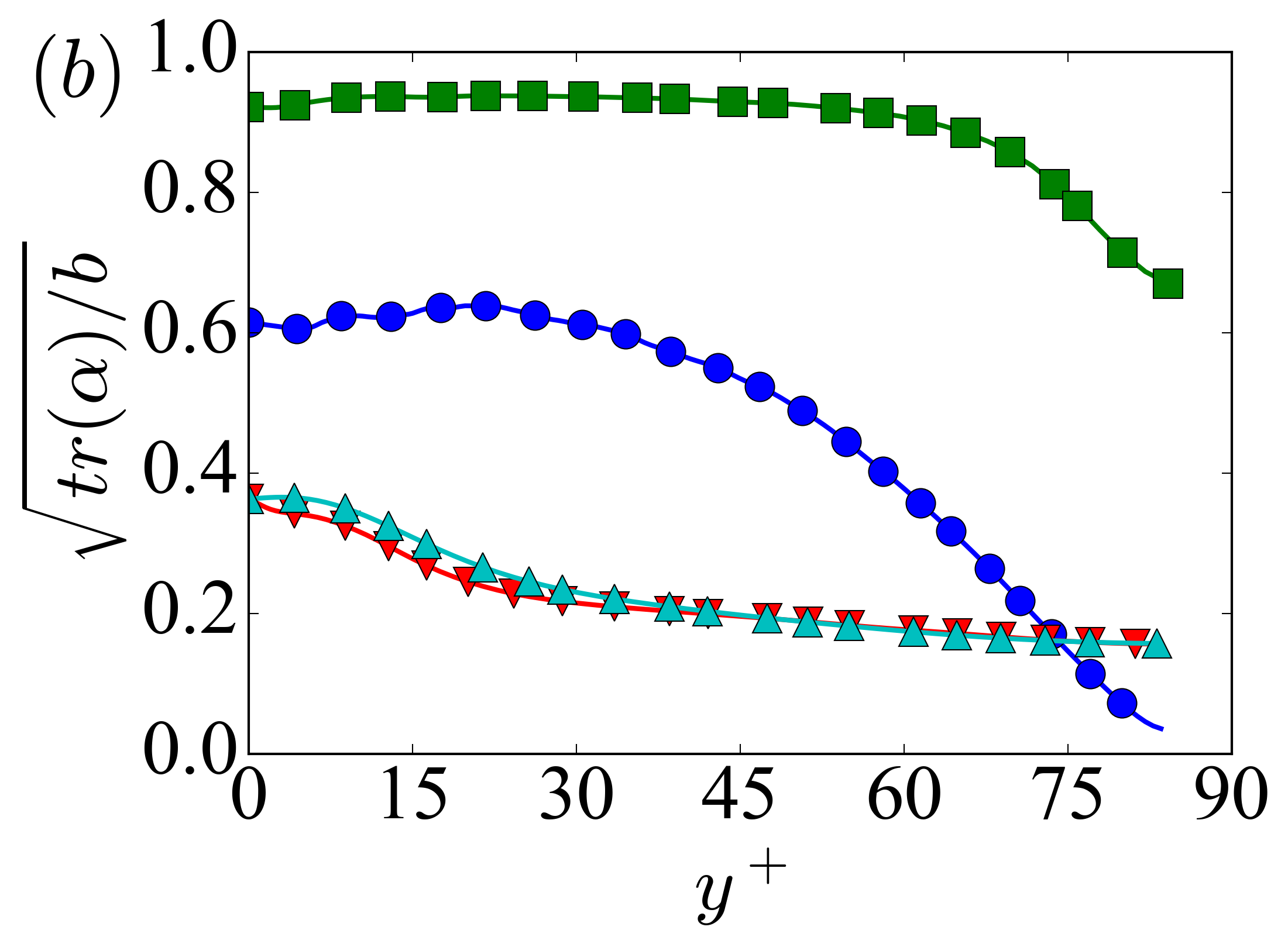}
	\caption{\RevisedText{Mean profiles of (a) streamwise velocity and (b) extent of polymer stretching of IDT (SM with $\mathrm{Sc}=0.5$ vs. HM with no AD), 2D EIT ($\mathrm{Wi}=64$; HM with no AD), and 3D EIT ($\mathrm{Wi}=800$; HM with no AD).
	All 3D solutions use MFU.}
	References lines are viscous sublayer $U_\text{m}^+=y^+$, von~K\'arm\'an (Newtonian) log law $U_\text{m}^+=2.5\ln y^++5.5$~\citep{Kim_Moin_JFM1987}, and Virk MDR $U_\text{m}^+=11.7\ln y^+-17.0$~\citep{Virk_AIChEJ1975}.} 
	\label{fig:mean_vel}
\end{figure}

Mean velocity profiles of these solutions are plotted in \cref{fig:mean_vel}(a).
At IDT, mean velocity profiles from SM+GAD and HM are in excellent agreement and both are slightly elevated compared with the Newtonian von~K\'arm\'an law, which reflects a moderate level of DR at $\mathrm{Wi}=23$.
DR in MFUs in this regime using the SM+GAD scheme was thoroughly studied in \citet{Xi_Graham_JFM2010}.
Comparison with HM results here shows that adding AD at the tested level ($\mathrm{Sc}=0.5$) apparently has little effect.
The \RevisedText{profiles of 2D (at $\mathrm{Wi}=64$) and 3D EIT (at $\mathrm{Wi}=800$) are} much higher and also of a different shape, which clearly \RevisedText{belong} to a different flow regime.
\RevisedText{These two profiles apparently overlap, which is, however, only a coincidence.
As shown in \citet{Zhu_Xi_arXiv2020}, with increasing $\mathrm{Wi}$, the 2D EIT profile continues to drop as instability intensifies, while the 3D EIT mean velocity converges to a constant level -- i.e., MDR behavior.
The underlying dynamics behind this behavior will not be further discussed here. }


The non-dimensional end-to-end distance of polymer chains is measured by $\sqrt{\mathrm{tr}(\mbf\alpha)}$.
After normalization by its upper limit $\sqrt{b}$, $\sqrt{\mathrm{tr}(\mbf\alpha)/b}$ measures the extent of polymer stretching, with $\sqrt{\mathrm{tr}(\mbf\alpha)/b}=1$ being the fully stretched limit.
As shown in \cref{fig:mean_vel}\RevisedText{(b)},
at IDT \RevisedText{of $\mathrm{Wi}=23$}, profiles of $\sqrt{\mathrm{tr}(\mbf\alpha)/b}$ from the two numerical schemes are largely consistent, with some small discrepancies in the near-wall region.
\RevisedText{For 2D EIT at $\mathrm{Wi}=64$}, polymers are highly stretched near the wall but the stretching is much less near the channel center, suggesting that the instability is driven by near-wall polymer stress.
\RevisedText{The 3D EIT solution shown in \cref{fig:mean_vel}(b) is from a much higher $\mathrm{Wi}$ of $800$. Thus polymers are close to the full-extension limit except near the channel center where the mean shear vanishes owing to the flow symmetry.}

\begin{figure}
	\centering
	\includegraphics[width=0.8\linewidth, trim=10 5 5 0, clip]{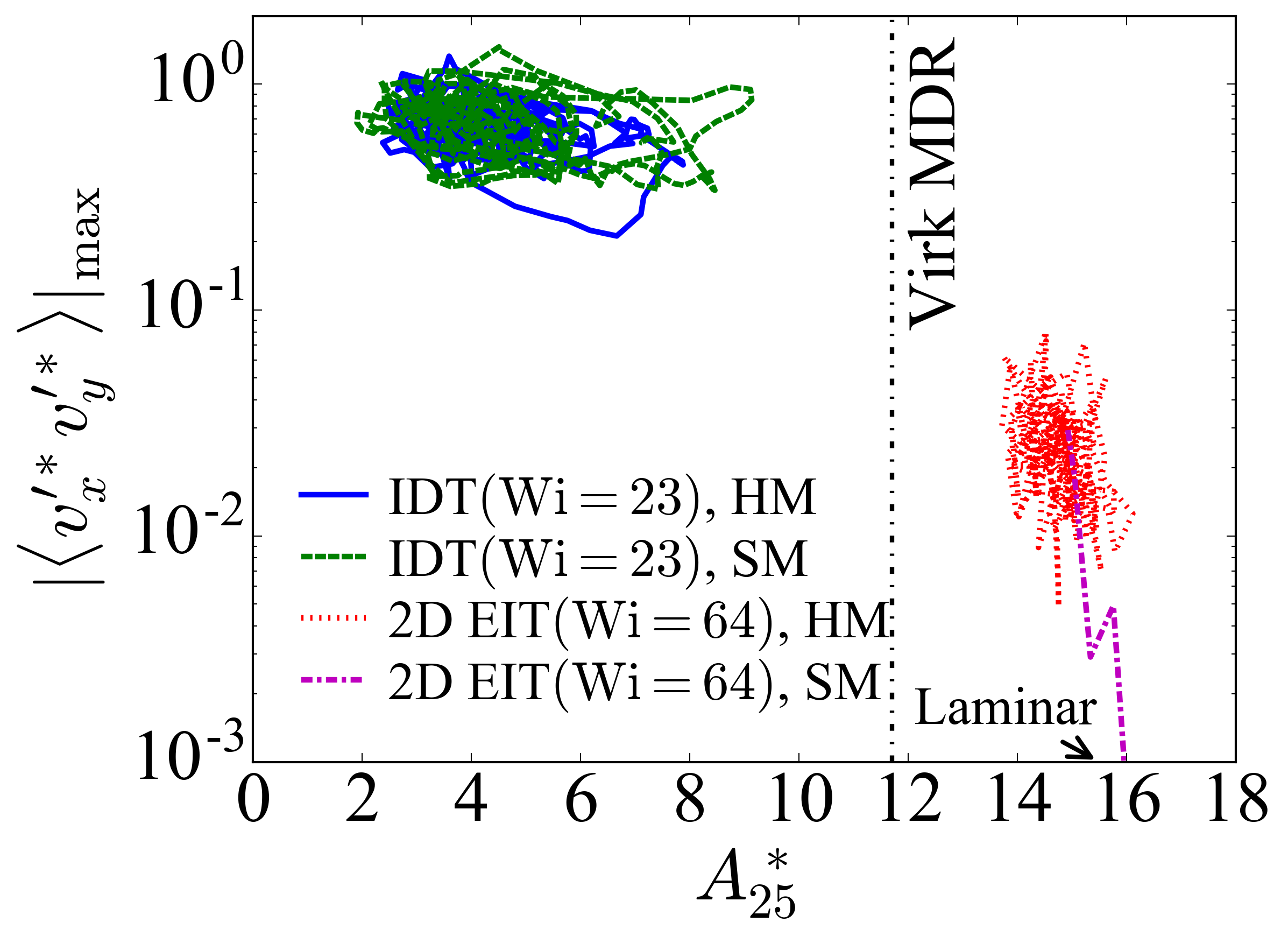}
	\caption{Projection of solution trajectories onto the $|\langle v_{x}'^{*}v_{y}'^{*}\rangle|_{\max} - A^\ast_{25}$ plane.}
	\label{fig:statSpace_ss}
\end{figure}

Dynamical trajectories of the same solutions are projected to a 2D plane in \cref{fig:statSpace_ss}.
One of the axes uses $A^\ast_{25}$, which is the value of the logarithmic slope function of the instantaneous mean velocity profile 
\begin{gather}
	A^\ast \equiv y^\ast\frac{\partial U^\ast_m}{\partial y^\ast}
	\label{eq:Aast}
\end{gather}
measured at $y^\ast=25$.
(Side notes: (1) ``$\ast$'' marks quantities in instantaneous turbulent inner scales -- i.e., same as the ``$+$'' units but using the instantaneous wall shear stress $\tau_\text{w}^*$ for the scaling~\citep{Xi_Graham_PRL2010,Xi_Graham_JFM2012,Xi_POF2019};
(2) \cref{eq:Aast} is obtained by taking the derivative of $U_m^\ast=A^\ast\ln y^\ast + B^\ast$ w.r.t. $\ln y^\ast$.)
The other uses the peak magnitude of the instantaneous Reynolds shear stress (RSS) profile $|\langle v_{x}'^{*}v_{y}'^{*}\rangle|_{\max}$.
The projection shows that IDT and EIT occupy distinctly different regions in the state space.
IDT sits in the upper-left region in the view whereas EIT is found near the lower-right (high velocity and low RSS) corner.
Note that the logarithmic scale is used for the $|\langle v_{x}'^{*}v_{y}'^{*}\rangle|_{\max}$ axis -- the RSS magnitude of EIT is nearly two orders of magnitude lower than that of IDT.
Since RSS is essential for turbulence production through the inertia-driven mechanism, vanishingly low RSS reflects the different self-sustaining mechanism of EIT~\citep{Xi_POF2019}.

Comparing IDT trajectories from the two numerical schemes, there is no discernible difference in their distribution patterns in the state space.
Both trajectories densely sample the same ``core'' region but sporadic excursions to its right are found in both cases.
The excursions are known as ``hibernating turbulence'' (so-named to contrast the regular active turbulence that is stronger in intensity) events, which exist in Newtonian turbulence but become unmasked at sufficiently high $\mathrm{Wi}$~\citep{Xi_Graham_PRL2010,Xi_Graham_JFM2012}.
The current results show that this phenomenon, previously studied mostly using SM+GAD, is insensitive to the use of AD.
Detailed comparison of time series and flow images of active and hibernating turbulence between SM+GAD and HM schemes was provided in \citet{Zhu_PhD2019}.

\begin{figure}
	\centering
	\includegraphics[width=.5\linewidth, trim=0mm 0mm 0mm 0mm, clip]{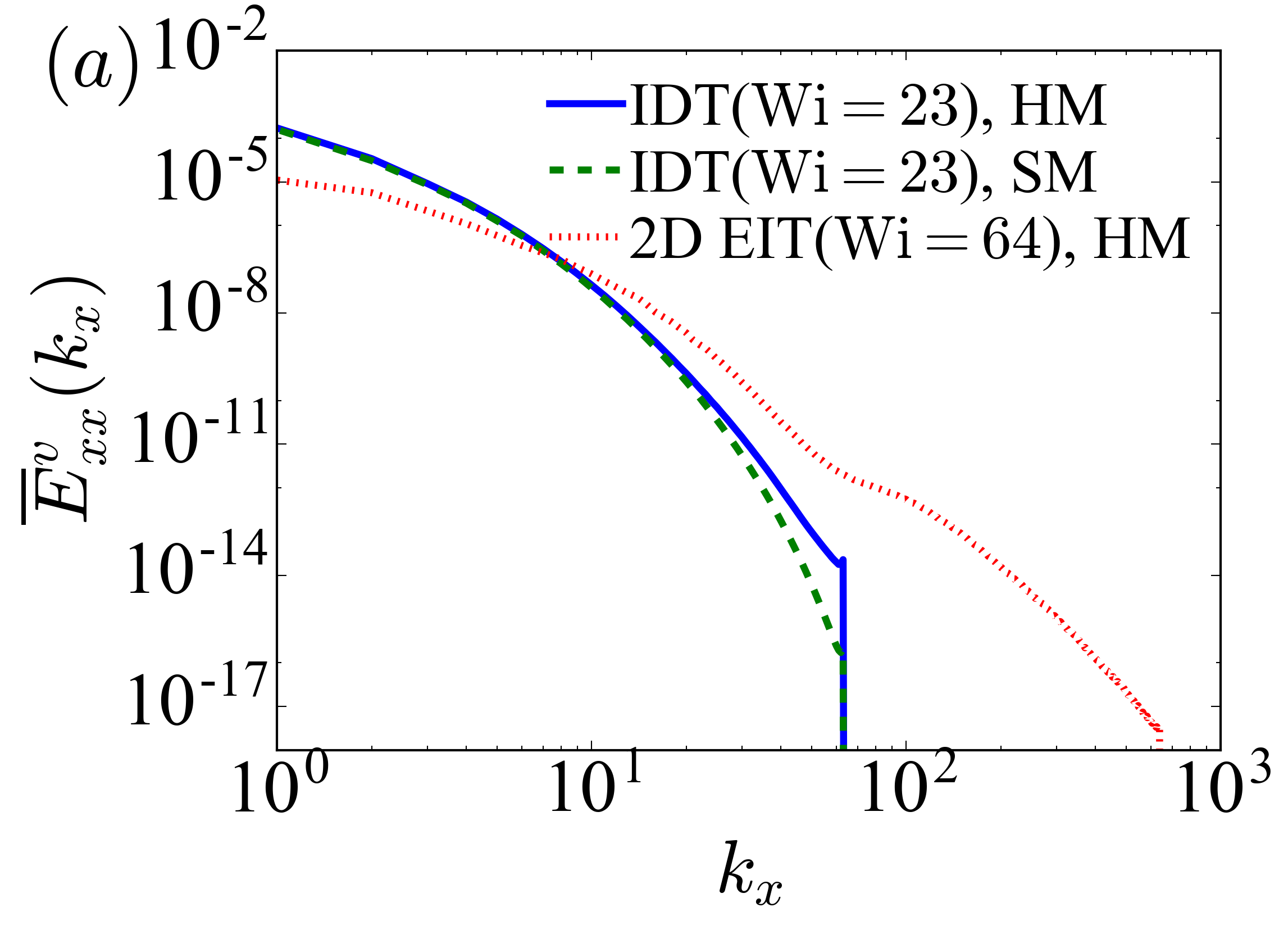}%
	\includegraphics[width=.5\linewidth, trim=0mm 0mm 0mm 0mm, clip]{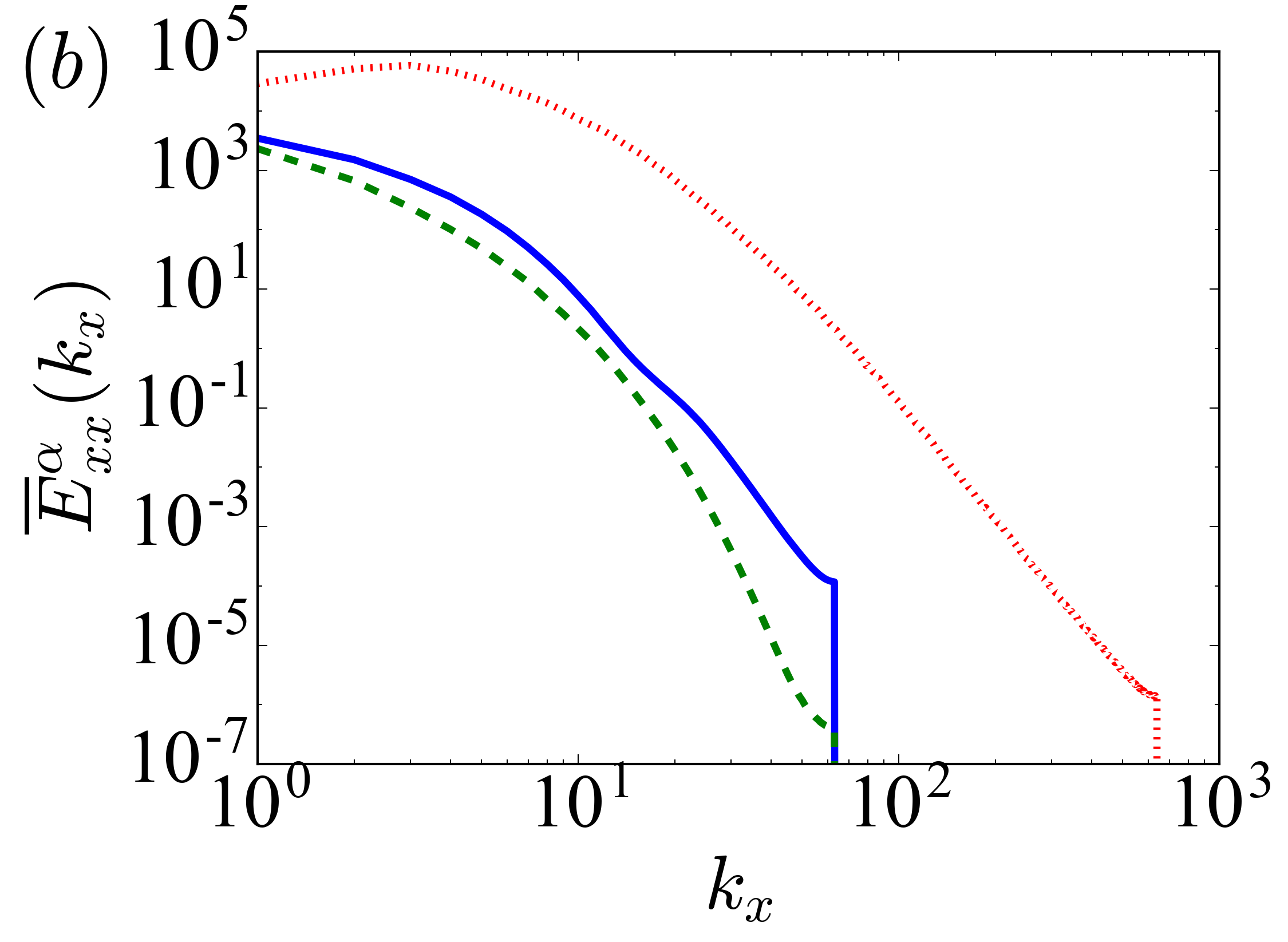}
	\caption{One-dimensional spectra of (a) streamwise velocity and (b) the $xx$-component of the polymer conformation tensor of IDT (MFU; SM with $\mathrm{Sc}=0.5$ vs. HM with no AD) and EIT (2D; \RevisedText{HM with no AD}) solutions.}
	\label{fig:energy_spec}
\end{figure}

\Cref{fig:energy_spec} shows the one-dimensional (along the $x$ direction) spectra of streamwise velocity and the $xx$-component of the polymer conformation tensor.
The spectrum of the $w$-component of velocity is defined as
\begin{gather}
	\overline{E}^v_{ww}(k_x)=\frac{1}{2l \Delta T}\int_{t_0}^{t_0+\Delta T}\int_{-l}^l
		\sum_{k_z}\tilde{v}_w^{\prime \circledast} \tilde{v}_w^{\prime}dydt
	\label{Eq:energyspec_v}
\end{gather}
while that of $\alpha_{xx}$ is
\begin{gather}
	\overline{E}^\alpha_{xx}(k_x)=\frac{1}{2l \Delta T}\int_{t_0}^{t_0+\Delta T}\int_{-l}^l
		\sum_{k_z}\tilde\alpha_{xx}^{\prime \circledast} \tilde\alpha_{xx}^{\prime}dydt
	\label{Eq:energyspec_alpha}
\end{gather}
(``$\circledast$'' denotes complex conjugate; $\Delta T$ is the time window for averaging). 
The velocity spectra of IDT from the two numerical schemes overlap one another for a broad range of $k_x$ and differ only near the end of the spectrum at $k_x\gtrsim 30$ (\cref{fig:energy_spec}(a)): i.e., GAD (used in SM) suppresses structures only at the smallest scales. 
For $L_x^+=360$ used in the MFU, the corresponding length scale is $\lesssim 12$ wall units.
To put it in perspective, typical DNS in the pre-EIT era used the numerical mesh size of $\delta_x^+\sim 10$.
Therefore, GAD at the tested level ($\mathrm{Sc}=0.5$ or $1/\mathrm{Pe}=\num{5.56e-4}$) does not affect the most important structures as far as IDT is concerned.
This conclusion is consistent with the observations above that both the mean velocity and temporal intermittency in the velocity field of IDT are not influenced by GAD in any appreciable way.
For the polymer conformation tensor (\cref{fig:energy_spec}(b)), the SM+GAD profile is slightly lower that of HM -- AD seems to slightly suppress the spatial fluctuation in polymer stress.
However, for all flow statistics and dynamical patterns investigated in this study, none appears to be affected by this small discrepancy in polymer stress fluctuation.
Discrepancy between the two profiles widens at $k_x\gtrsim 20\sim 30$, coinciding with that in the velocity spectra.

Suppression of small-scale fluctuations by AD is consistent with its smearing effects on sharp gradients, which is also the reason why EIT cannot be captured with GAD of this level.
Compared with IDT, the velocity spectrum of 2D EIT (\cref{fig:energy_spec}(a); obtained from HM) contains much lower energy in large scales (low $k_x$) but higher energy at the small scale (large $k_x$) end.
The corresponding profile for $\alpha_{xx}$ (\cref{fig:energy_spec}(b)) is significantly higher than that of IDT (of a lower $\mathrm{Wi}$) across all length scales.
Large fluctuations in polymer stress over very small length scales are the reason why EIT poses different requirements on the numerical method.

\begin{figure}
	\centering
	\includegraphics[width=.8\linewidth, trim=0mm 0mm 0mm 0mm, clip]{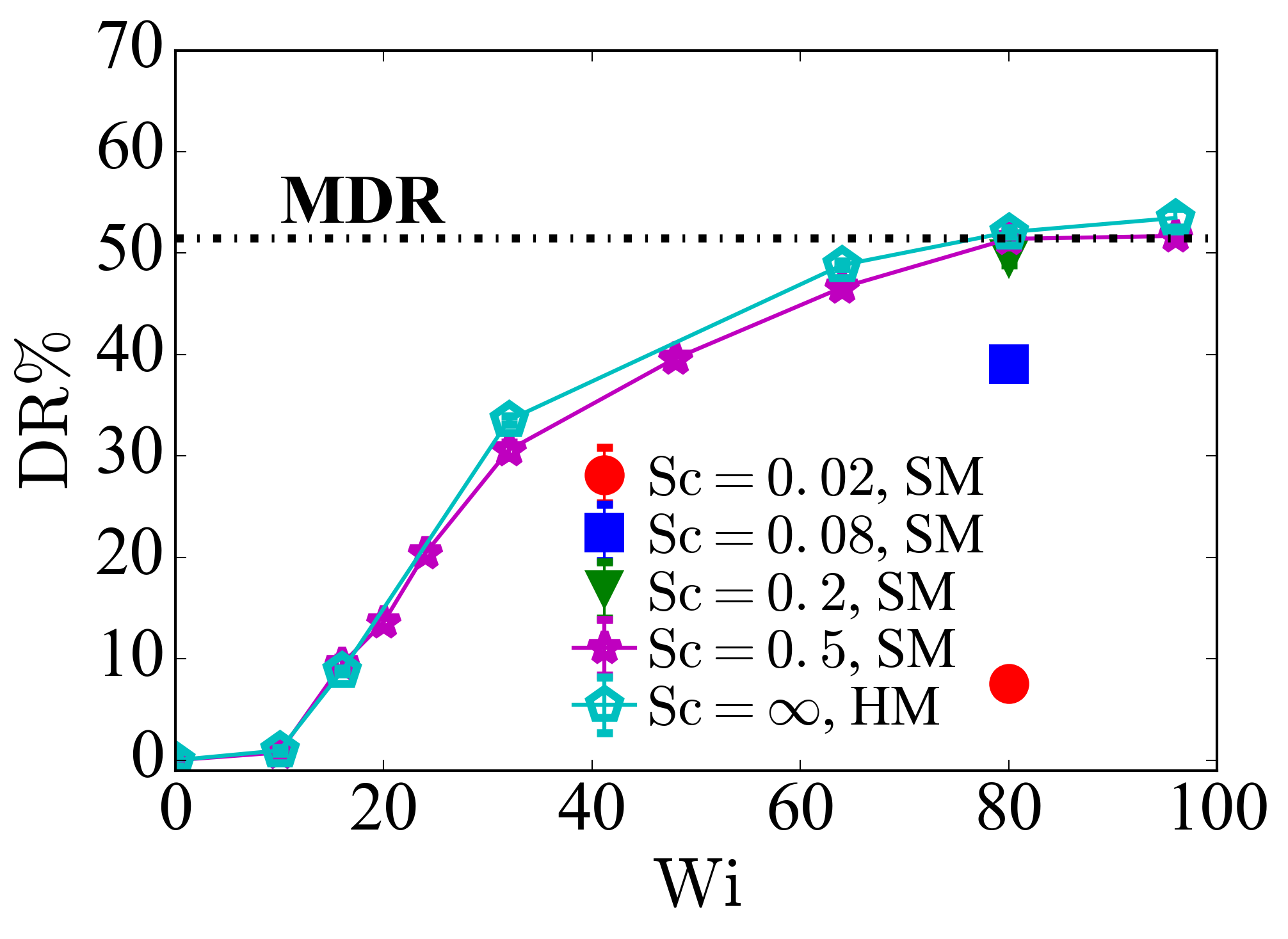}
	\caption{Dependence of $\mathrm{DR\%}$ of IDT on $\mathrm{Wi}$ in an extended simulation domain: comparison between HM (no AD) and SM (various $\mathrm{Sc}$).}
	\label{fig:DR_wi}
\end{figure}


So far, the conclusion that IDT flow structures, dynamics, and mean statistics are not strongly affected by GAD is based on the DNS of a single moderate $\mathrm{Wi}=23$ in a MFU.
We now move to an extended flow domain of $L_x^+\times L_z^+=4000\times 800$ where IDT solutions can be found at much higher $\mathrm{Wi}$~\citep{Zhu_Xi_JNNFM2018}.
In \cref{fig:DR_wi}, we plot the percentage of drag reduction
\begin{gather}
	\mathrm{DR\%}\equiv \frac{C_{f}-C_{f,\text{Newt}}}{C_{f,\text{Newt}}}
\end{gather}
as a function of $\mathrm{Wi}$. Here, $C_{f}$ is the friction factor defined as 
\begin{gather}
	C_f\equiv \frac{2\tau_w}{\rho U^2_{\mathrm{avg}}}
\end{gather}
($U_\text{avg}$ denotes the $(x,y,z,t)$-averaged streamwise velocity)
and $C_{f,\text{Newt}}$ is its value in Newtonian turbulence at the same $\mathrm{Re}$.
Contrary to the common claim in the literature, that GAD results in the under-prediction of $\mathrm{DR}\%$ and delayed onset of DR~\citep{Yu_Kawaguchi_JNNFM2004,Vaithianathan_Collins_JNNFM2006}, we find that with $\mathrm{Sc}=0.5$, results from SM are in reasonable agreement with the HM prediction for the whole $\mathrm{Wi}$ range tested, including accurately predicting $\mathrm{Wi}_\text{onset}$.

After testing several $\mathrm{Sc}$ at $\mathrm{Wi}=80$, we find that $\mathrm{DR}\%$ remains the same as the $\mathrm{Sc}=\infty$ (no AD) limit down to $\mathrm{Sc}=0.2$ ($1/\mathrm{Pe}=\num{1.39e-3}$), after which $\mathrm{DR}\%$ decays quickly with decreasing $\mathrm{Sc}$.
At $\mathrm{Sc}=0.08$ ($1/\mathrm{Pe}=\num{3.47e-3}$), $\mathrm{DR}\%$ is underestimated by $\approx 10$ percentage points (or $\approx 20\%$ relative error),
while at $\mathrm{Sc}=0.02$ ($1/\mathrm{Pe}=\num{1.38e-2}$), the obtained $\mathrm{DR}\%$ is unrealistically low.
As such, our result does not contradict the earlier observation of \citet{Yu_Kawaguchi_JNNFM2004}, which showed that adding GAD with $1/\mathrm{Pe}=\num{e-2}$ underestimates the mean velocity by $\approx 10\%$ at a lower $\mathrm{Wi}=30$ (also different $\mathrm{Re}$, $\beta$, and $b$ than ours).
Evidently, the level of GAD that they tested is already beyond the acceptable range for prediction accuracy.
\citet{Vaithianathan_Collins_JNNFM2006} reported a relative error of $\approx 10\%$ in $\mathrm{DR}\%$ at $\mathrm{Sc}=1/3$, which, however, was in a non-bounded homogeneous shear flow.
Smaller AD effects at similar $\mathrm{Sc}$ observed in our case may be attributed to the dominance of wall-generated coherent structures of larger scales in channel flow.

\begin{figure}
	\centering
	\includegraphics[width=.7\linewidth, trim=0mm 0mm 0mm 0mm, clip]{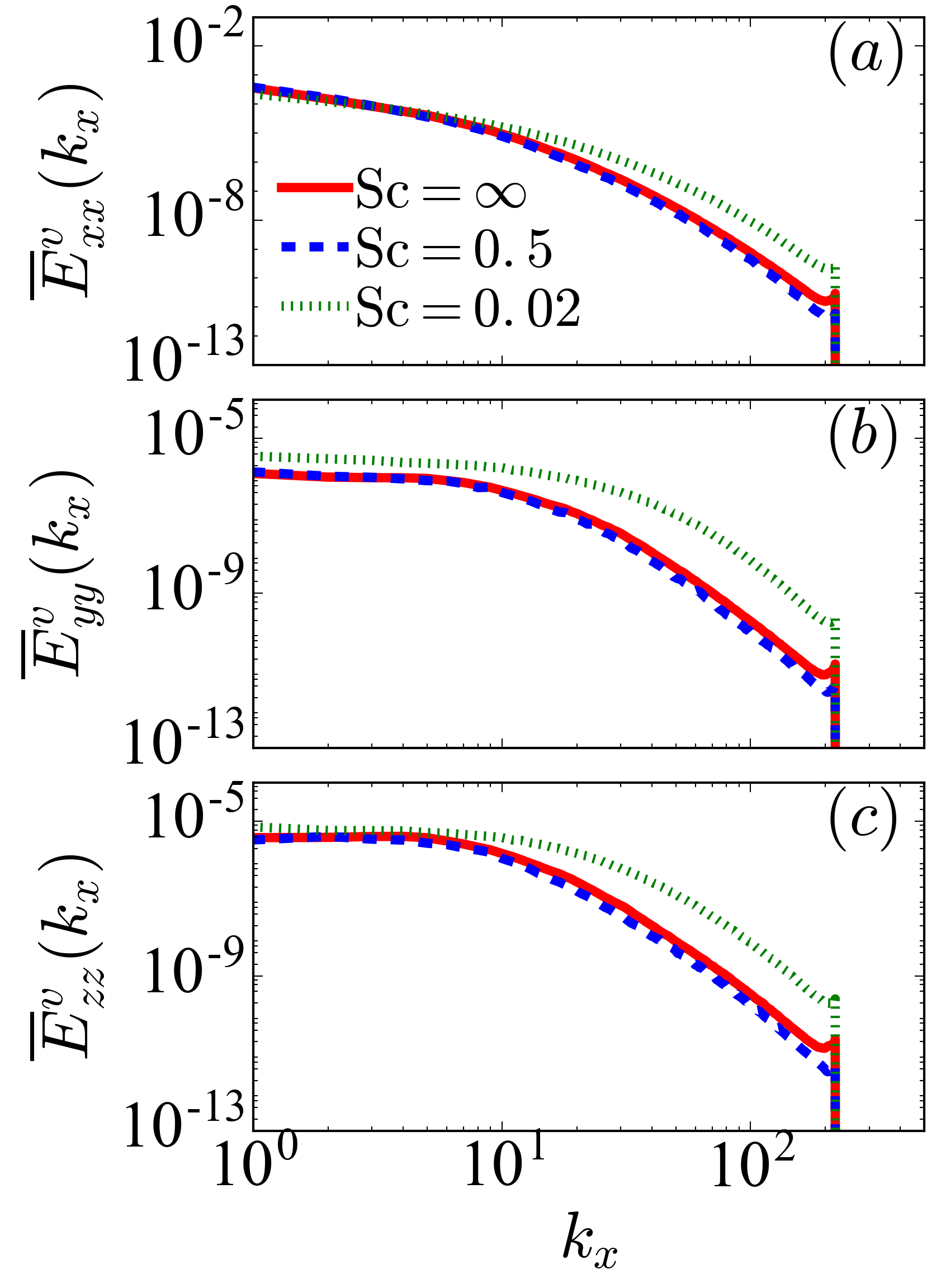}
	\caption{One-dimensional velocity spectra of IDT at $\mathrm{Wi}=80$ (extended domain): comparison between HM ($\mathrm{Sc}=\infty$) and SM with different $\mathrm{Sc}$.}
	\label{fig:eneSpe_Sc}
\end{figure}

In \cref{fig:eneSpe_Sc}, we examine the effects of GAD on the spectra of all three velocity components (\cref{Eq:energyspec_v}) at $\mathrm{Wi}=80$.
Compared with the $\mathrm{Sc}=\infty$ (HM with no AD) limit, $\mathrm{Sc}=0.5$ has little effect on all three spectra.
Note that the mesh size $\delta_x^+=9.09$ used in this extended domain (\cref{tab:resolution_3d}; which is typical compared with previous literature) is on a par with the length scale where the suppression of small-scale velocity fluctuation by AD is expected (as observed in our MFU simulation using a finer grid; see \cref{fig:energy_spec}).
Such effect is thus not observed here.
With $\mathrm{Sc}=0.02$, however, instead of suppressing fluctuation, AD enhances velocity spectra in both transverse directions ($\overline{E}^v_{yy}$ and $\overline{E}^v_{zz}$) across all scales.
For streamwise velocity, enhanced fluctuation occurs at most length scales except the very largest ones. The spectrum is higher than the $\mathrm{Sc}=\infty$ case for $k_x\gtrsim 4$ -- for $L_x^+=4000$, this covers all length scales below 1000 wall units.
The seeming opposite effects between small- and large-magnitude GAD can be rationalized considering our recent finding in \citet{Zhu_Xi_JNNFM2019} that in an IDT-dominated flow, small-scale EIT-like structures can still occur in near-wall regions.
Small-magnitude GAD ($\mathrm{Sc}=0.5$) suppresses such structures, just like their impact on 2D EIT (\cref{fig:epsilonp}), but leaves the dominant IDT structures largely intact,
whereas large-magnitude GAD under-predicts polymer stress and DR effects overall, resulting in higher velocity fluctuations and lower mean flow (thus lower $\overline{E}^v_{xx}$ at $k_x<4$).

As our overall takeaway, the initial assumption by \citet{Sureshkumar_Beris_JNNFM1995}, that with reasonably low GAD DNS results are not strongly affected, remains valid at least in the IDT regime and when EIT-related structures and dynamics are not of interest.
However, the range of acceptable $\mathrm{Sc}$ must be carefully validated with a $\mathrm{Sc}$-sensitivity analysis. The acceptable range likely also varies with flow parameters.
For high $\mathrm{Wi}$, $1/\mathrm{Pe}=\RevisedText{\bigO}(\num{e-2})$, which was sometimes seen in earlier SM studies, may lead to misbehaving results.

\subsection{Effects of spatial resolution on HM simulation}\label{Sec_Mesh}

This section focuses on the determination of the numerical mesh resolution for the new HM scheme.
Similar to the case of AD, effects of numerical resolution on IDT and EIT are also expected to differ because of their different natures and structural characteristics.
They will thus be discussed separately.

\begin{figure}
	\centering
	\includegraphics[width=.5\linewidth, trim=0mm 0mm 0mm 0mm, clip]{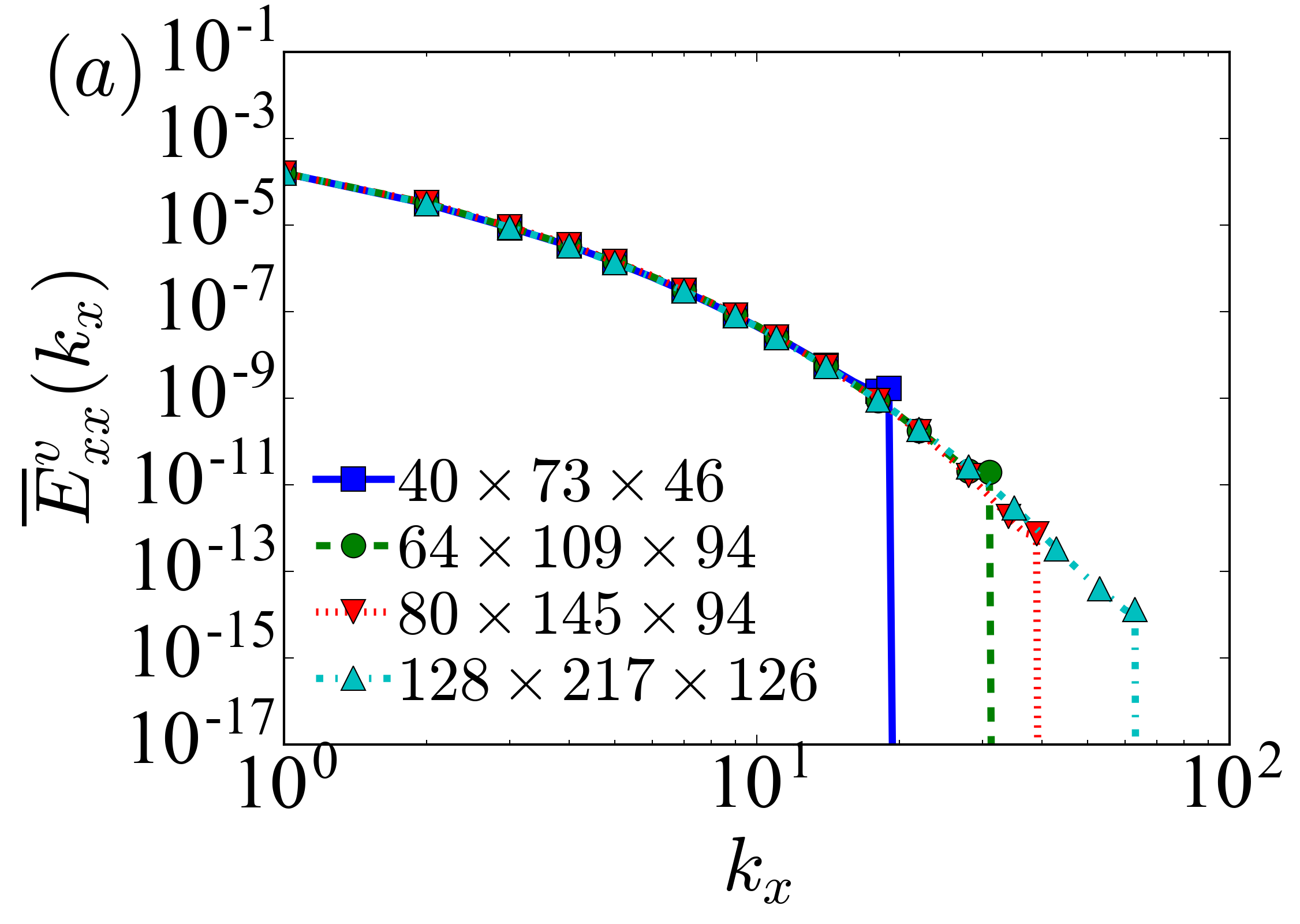}%
	\includegraphics[width=.5\linewidth, trim=0mm 0mm 0mm 0mm, clip]{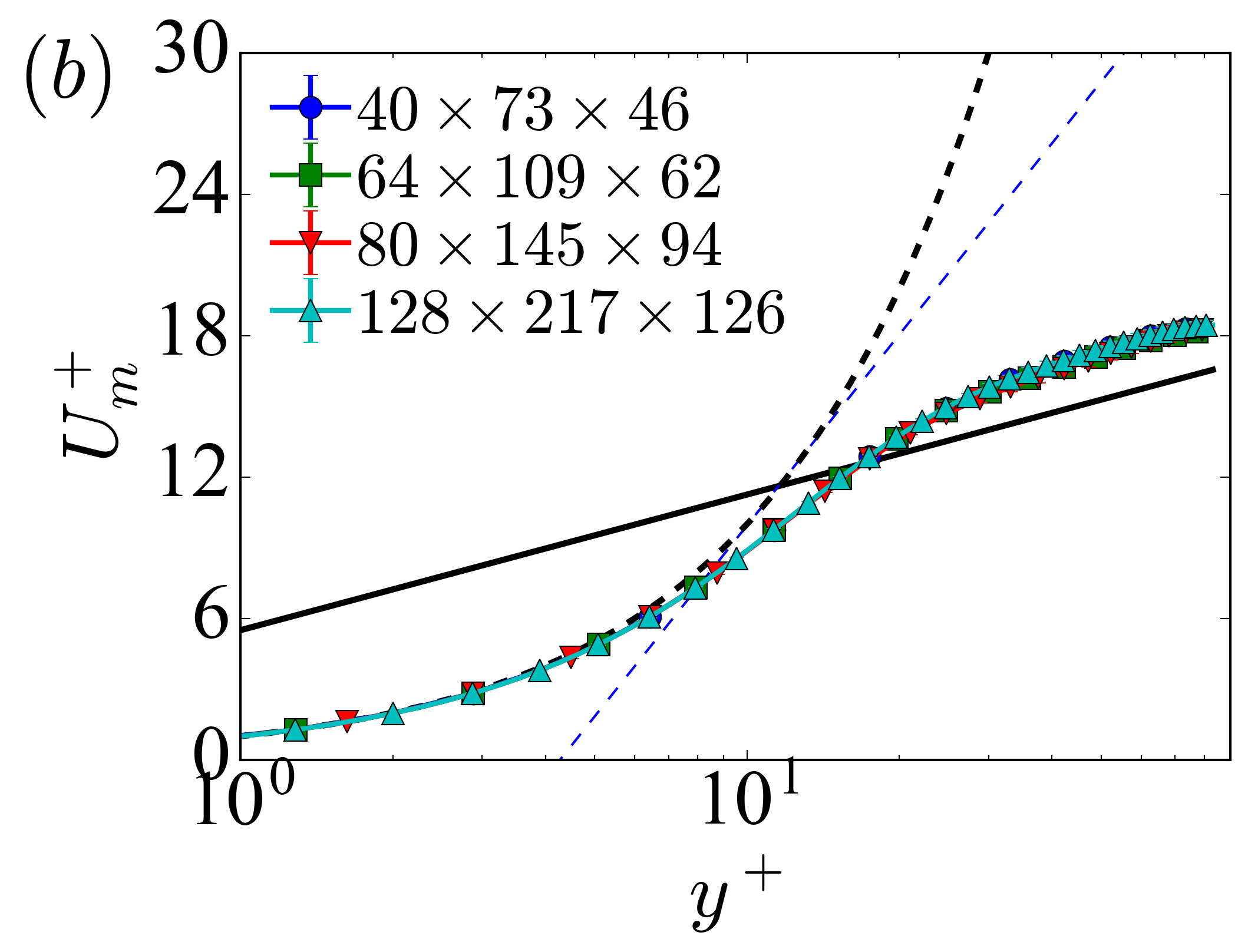}
	\\
	\includegraphics[width=.5\linewidth, trim=0mm 0mm 0mm 0mm, clip]{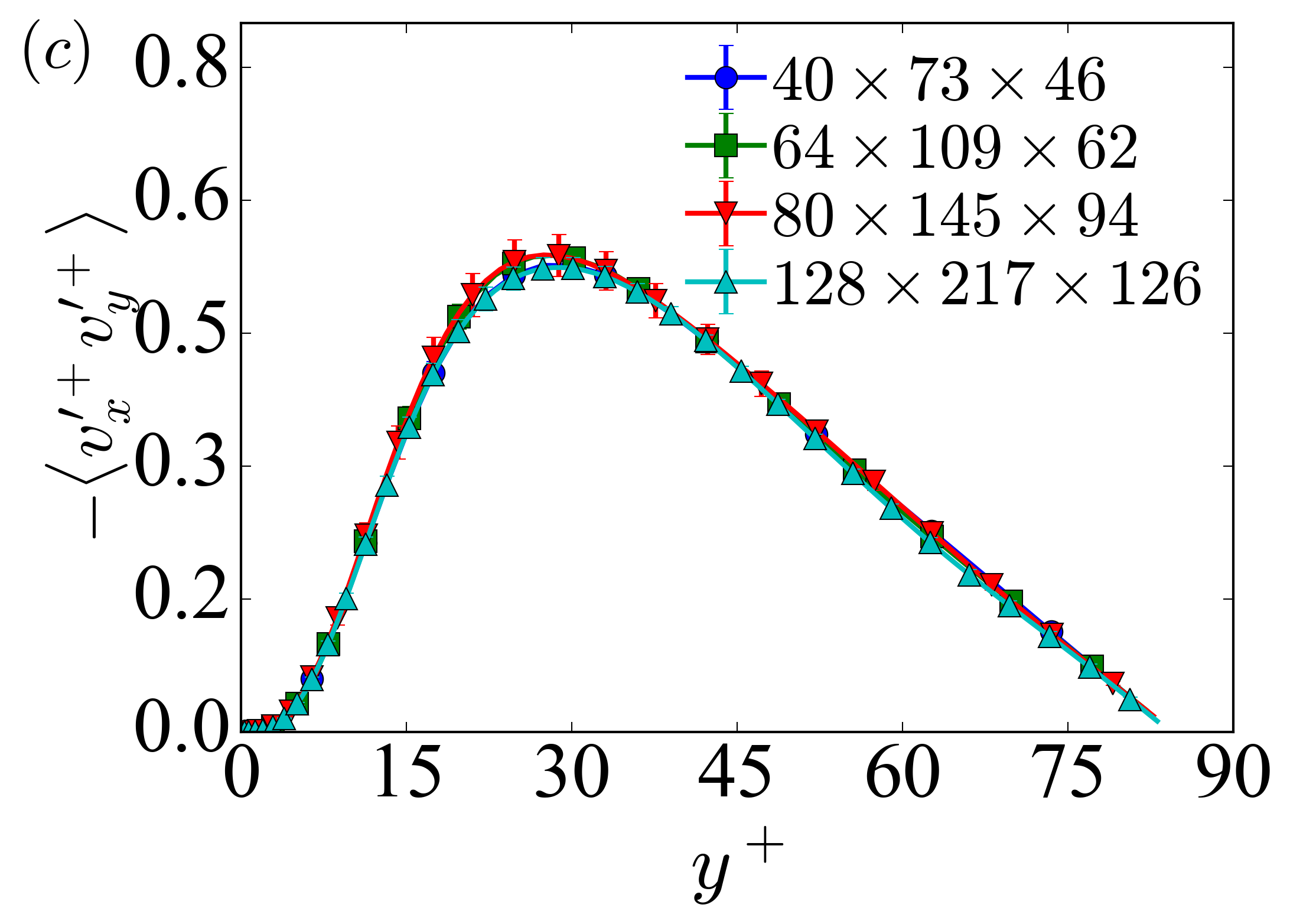}%
	\includegraphics[width=.5\linewidth, trim=0mm 0mm 0mm 0mm, clip]{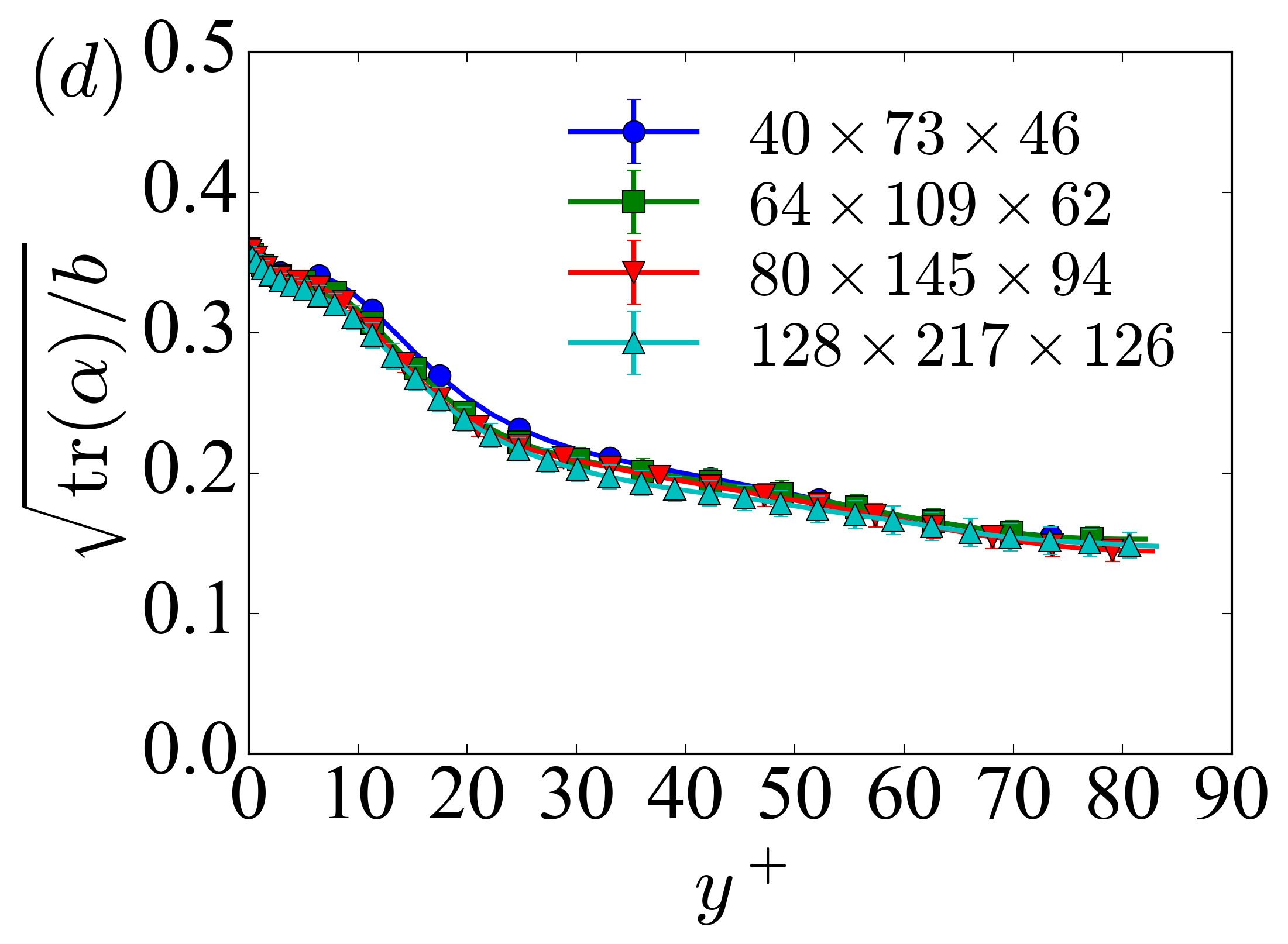}
	\caption{Statistical quantities of IDT ($\mathrm{Wi}=23$; HM with no AD) in the MFU using different resolutions $N_x\times N_y\times N_z$:
	(a) one-dimensional spectrum of the streamwise velocity,
	(b) mean velocity profile (same reference lines as \cref{fig:mean_vel}(a)),
	(c) Reynolds shear stress profile,
	(d) extent of polymer stretching.}
	\label{fig:stat_resol_idt}
\end{figure}

For IDT, we have tested four resolution levels in the same MFU reported above and resolution effects on different quantities are summarized in \cref{fig:stat_resol_idt}.
The velocity spectra from all 4 tested cases well overlap one another, except that with increasing resolution, fluctuations at increasingly small scales are captured (\cref{fig:stat_resol_idt}(a)).
Excellent agreement between these resolutions is also found in mean velocity, RSS, and the extent of polymer stretching measured again by $\sqrt{\mathrm{tr}(\mbf\alpha)/b}$ (\cref{fig:stat_resol_idt}(b)-(d)).
Therefore, even the lowest resolution $N_x\times N_y\times N_z = 40\times 73\times 46$ tested here can be deemed adequate for typical DNS application in the IDT regime.
This resolution level, corresponding to $\delta_x^+=9.0$ and $\delta_z^+=5.43$ (\cref{tab:resolution_3d}), is commensurate with our earlier DNS studies using SM~\citep{Xi_Graham_JFM2010,Zhu_Xi_JNNFM2019}.
Thus, adopting the new HM does not bring in any additional computational burden associated with its resolution requirement.

\begin{figure}
	\centering
	\includegraphics[width=.5\linewidth, trim=0mm 0mm 0mm 0mm, clip]{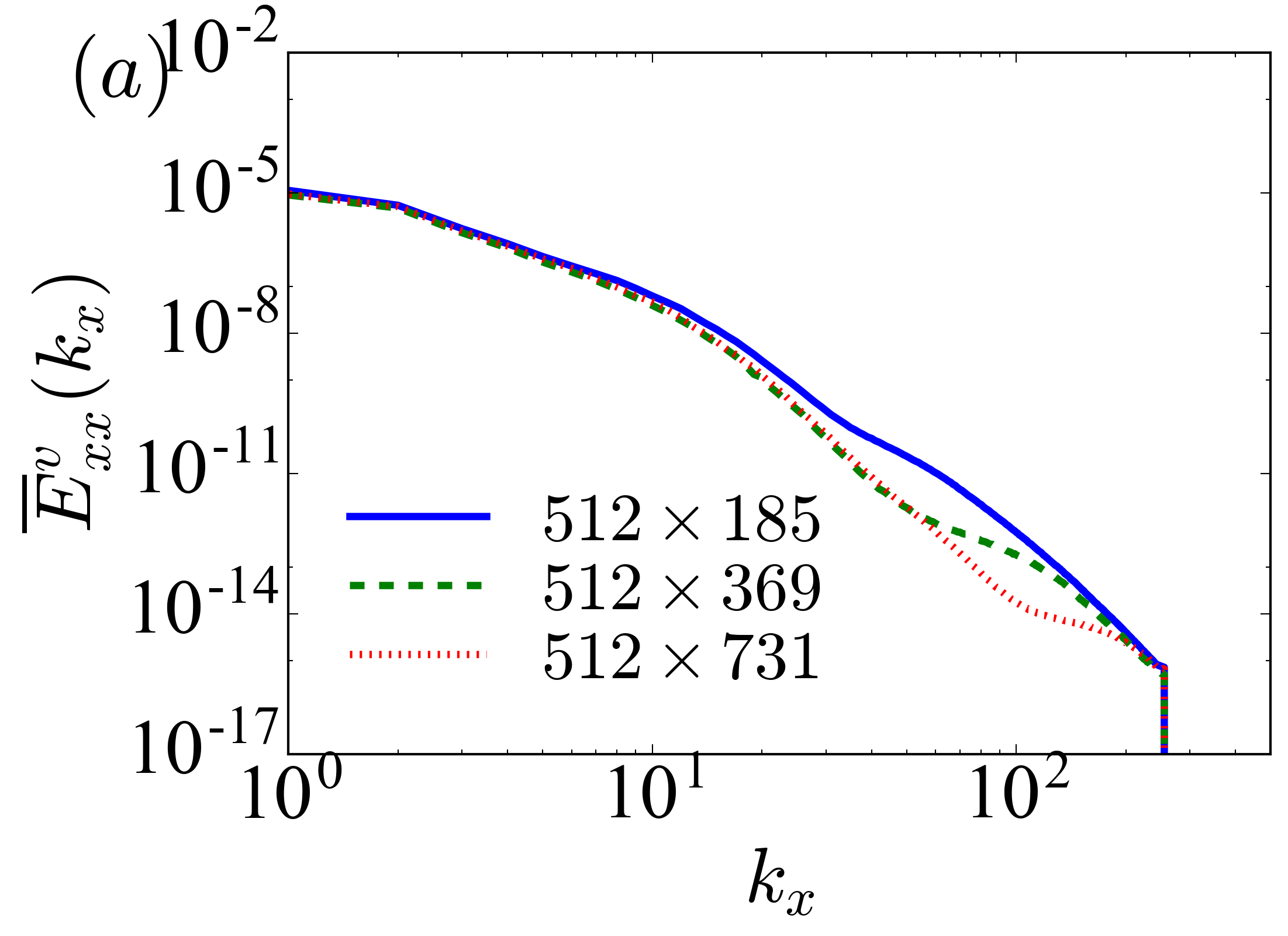}%
	\includegraphics[width=.5\linewidth, trim=0mm 0mm 0mm 0mm, clip]{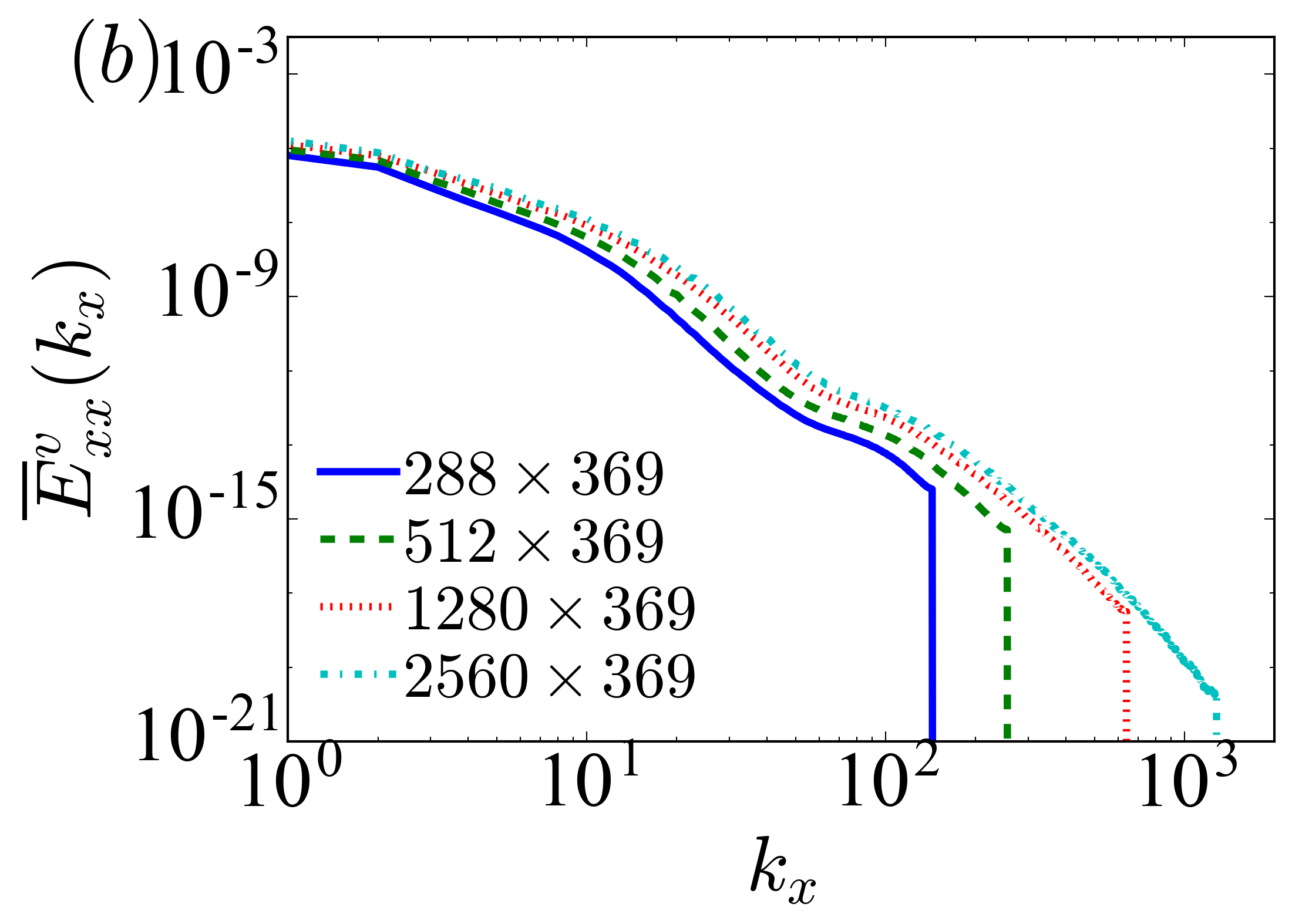}\\
	\includegraphics[width=.5\linewidth, trim=0mm 0mm 0mm 0mm, clip]{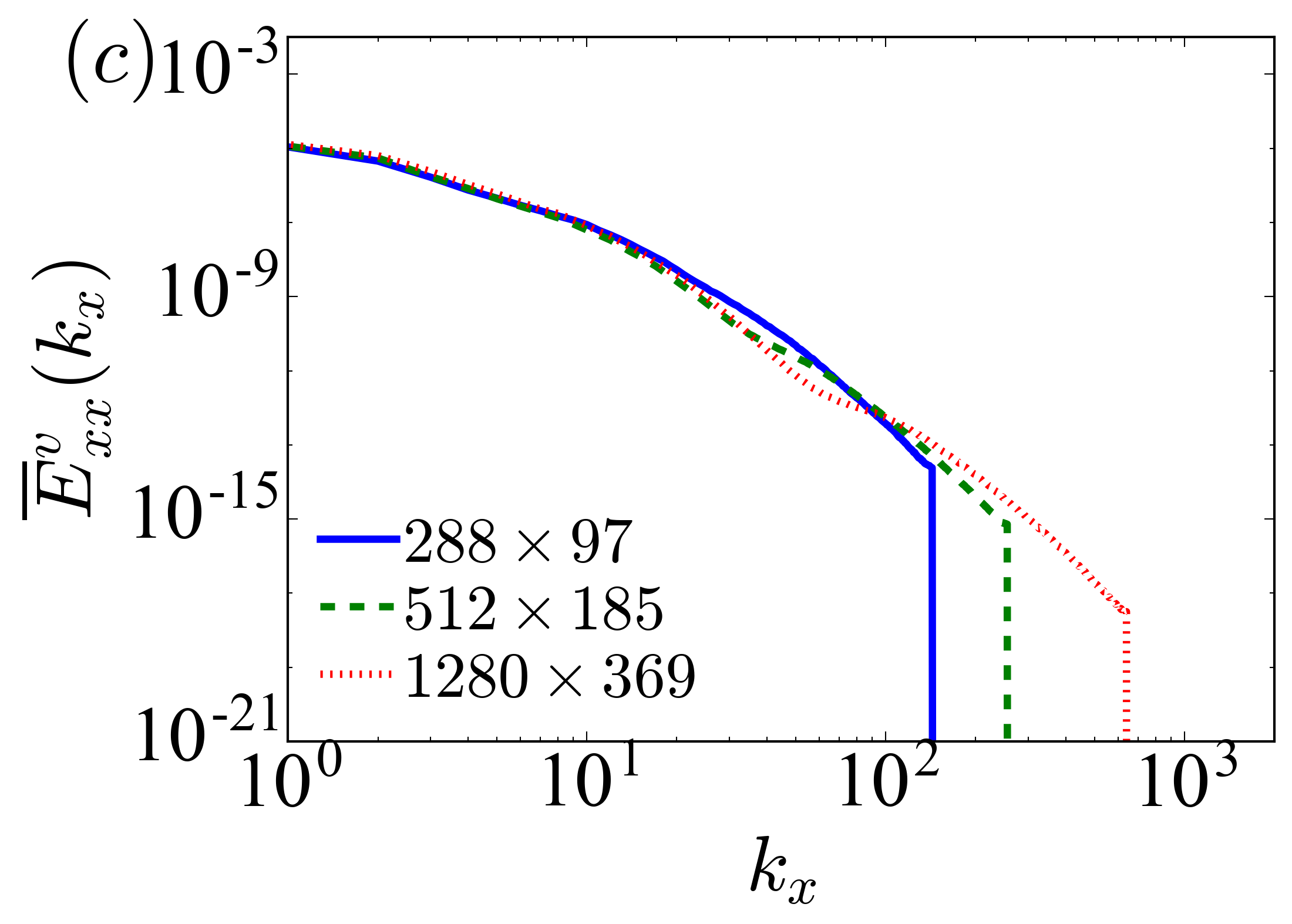}%
	\includegraphics[width=.5\linewidth, trim=0mm 0mm 0mm 0mm, clip]{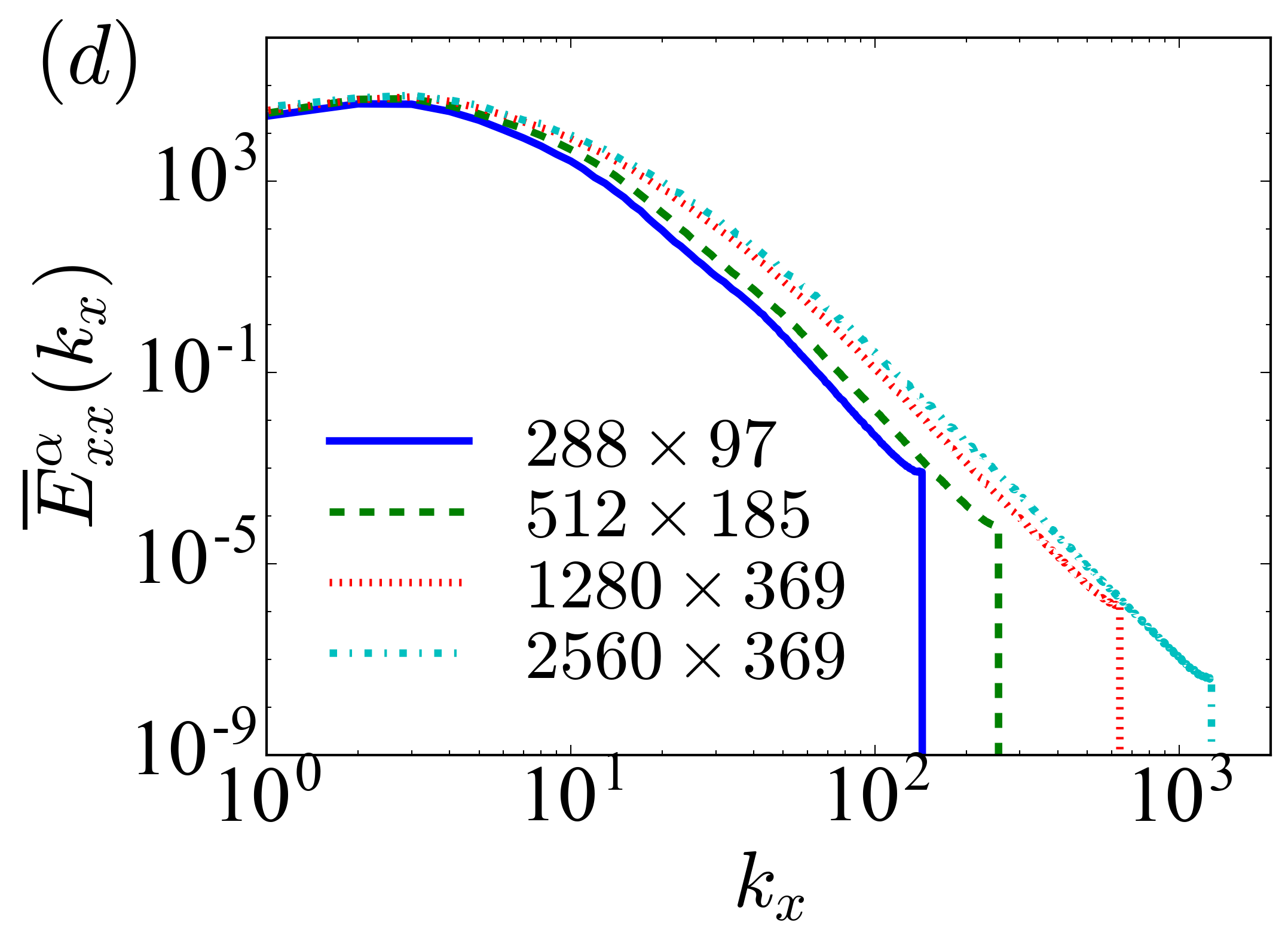}
	\caption{Effects of spatial resolution $N_x\times N_y$ on the one-dimensional spectra of (a)--(c) streamwise velocity and (d) $\alpha_{xx}$ of 2D EIT at $\mathrm{Wi}=64$ (all using HM with no AD).}
	\label{fig:stat_resol_eit}
\end{figure}


Achieving numerical convergence at EIT, if at all possible, is expected to require much higher resolution. This is a natural consequence of its sharp stress gradients and small-scale variations.
FENE-P is known to generate sheets of high polymer stress, known in experiments as birefringent strands, with sharp edges along the diverging direction of extensional flow~\citep{Xi_Graham_JFM2009}.
It was recently proposed, by \citet{Shekar_Graham_PRL2019}, that the same mechanism is responsible for the titled polymer sheets observed in EIT.
It would take infinitely refined numerical meshes to fully resolve a perfectly sharp stress shock front.
However, through carefully validated resolution settings, reasonably accurate solution can be obtained

Velocity spectra of EIT using different resolutions are compared in \cref{fig:stat_resol_eit}(a)--(c). 
Starting with a fixed $N_x=512$ (for $L_x^+=720$ used here, it gives $\delta_x^+=1.41$ -- less than $1/6$ of the mesh size recommended above for IDT), \cref{fig:stat_resol_eit}(a) compares the streamwise velocity spectra $\overline{E}^v_{xx}$ of three different levels of $y$-resolution: $N_y=\;$\numlist{185;369;731}.
The profiles overlap for $k_x\lesssim 30$.
At $k_x\approx 30$ (length scale $\approx 24$ wall units), the $N_y=185$ profile separates from the other two. Separation between $N_y=369$ and $731$ profiles occurs at $k_x\approx 60$ ($\approx 12$ wall units).
Finite resolution in $y$ seems to cause the over-prediction of fluctuations at a certain small-scale range, which is manifested as a mild bump in the velocity spectrum (compared with higher-resolution profiles) near the numerical truncation.
The critical streamwise length scale for the start of the bump apparently decreases in proportion to the decreasing $y$ mesh size.
In \cref{fig:stat_resol_eit}(b), $N_y=369$ is fixed. With increasing $N_x$, $\overline{E}^v_{xx}$ noticeably increases -- i.e., a coarse grid in $x$ underestimates velocity fluctuations for a substantial $k_x$ range.
The trend clearly slows down at $N_x=1280$ -- the profile of $1280\times 369$ is very close to that of $2560\times 369$.

Based on this analysis, we propose the $1280\times 369$ grid for our standard EIT simulation.
Numerical convergence is achieved with increasing $N_x$, whereas for $N_y$ mesh-independence is found at most length scales except the smallest ones.
For comparison, \citet{Sid_Terrapon_PRFluids2018} simulated EIT in the same 2D domain using a third-order WENO scheme for the convection term (and second-order FD for N-S) and found that a $1024\times 288$ grid captures most TKE and polymer stress fluctuations. They also reported the spurious accumulation of energy at the small-scale end near the numerical truncation as a finite-mesh effect, which based on our analysis above (\cref{fig:stat_resol_eit}(a)) depends primarily on wall-normal resolution.
Their mesh-size dependence analysis was done at $\mathrm{Re}_\tau=40$ and $\mathrm{Wi}=310$ (same $\beta$ and $b$ as ours).  
For their lower $\mathrm{Re}$ and higher $\mathrm{Wi}$, elasticity is expected to play a relatively more important role in EIT and the problem is expected to be more difficult to resolve numerically.
Using a $1280\times 369$ grid in our case is thus on the more conservative side.

Considering the oblique arrangement of polymer sheets at EIT (\cref{fig:eit:images}), we also explore the resolution dependence while keeping the mesh aspect ratios approximately fixed.
In \cref{fig:stat_resol_eit}(c), $N_x/N_y$ is kept close to the ratio of $3:1$ (following our final chosen grid). 
Strikingly, improved convergence (compared with increasing resolution in either direction independently) is found even at resolution as low as $288\times97$.
We also tested a $3:2$ ratio (not shown here), which does not converge as easily as the $3:1$ ratio.
\RevisedText{This finding can probably be attributed to the well-defined spatial orientation of the stress shocks, which is also relatively steady over time in 2D EIT.
The optimal mesh aspect ratio can potentially be determined from the shock-front orientation and the resulting stress derivative ratio between $x$ and $y$ directions, which most likely would not happen to be exactly $3:1$.}

Fluctuation in the polymer conformation field is more sensitive to resolution.
As shown in \cref{fig:stat_resol_eit}(d), for the same $288\times97$, $512\times185$, and $1280\times369$ cases (whose velocity spectra agree well except near the numerical truncation according to \cref{fig:stat_resol_eit}(c)), discrepancy in the $\overline{E}^\alpha_{xx}$ profiles is noticeable at most scales.
Nevertheless, $1280\times369$ agrees well with $2560\times369$ -- the highest resolution we have.


\begin{figure}
	\centering
	\includegraphics[width=.5\linewidth, trim=0mm 0mm 0mm 0mm, clip]{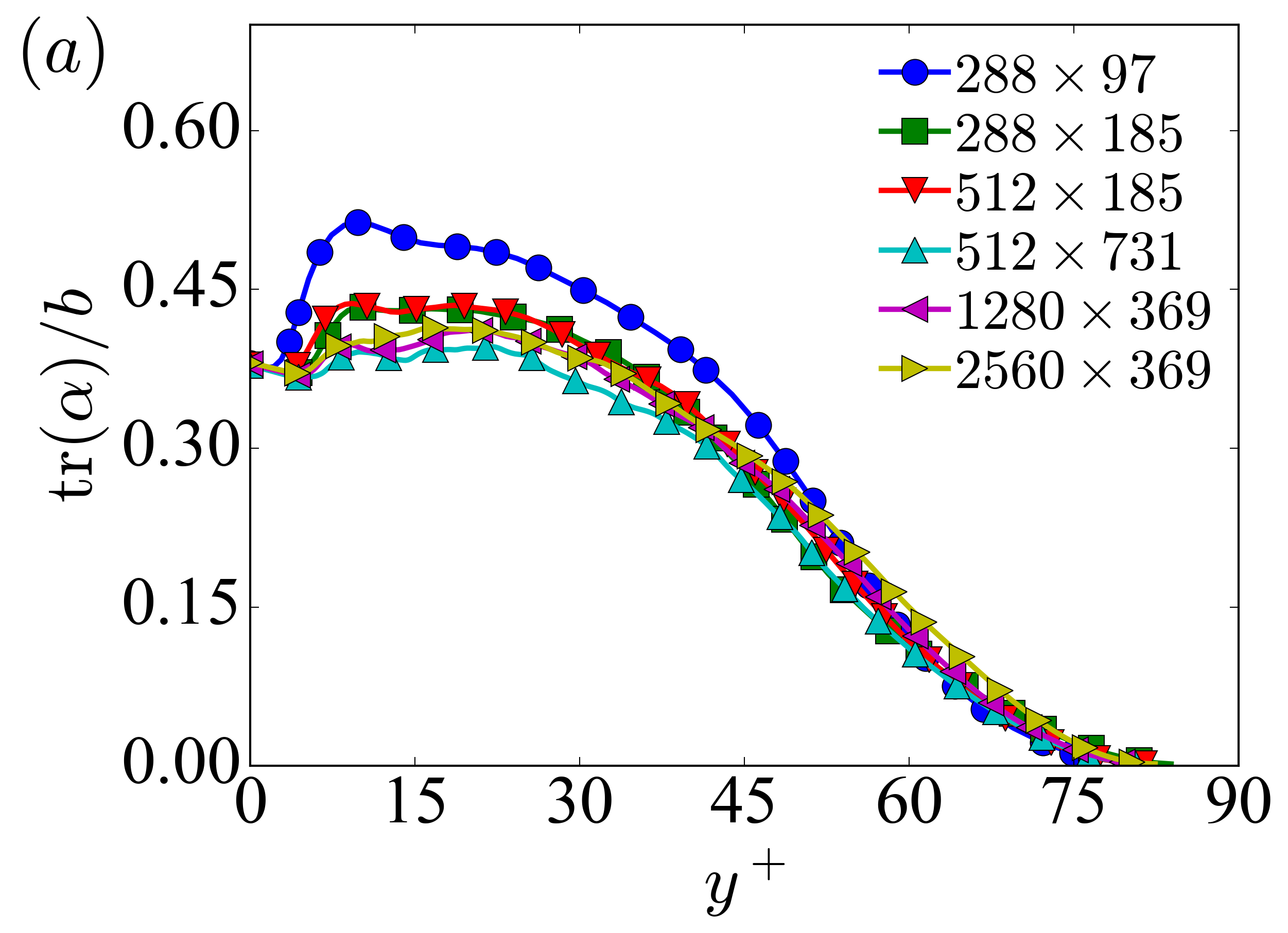}%
	\includegraphics[width=.5\linewidth, trim=0mm 0mm 0mm 0mm, clip]{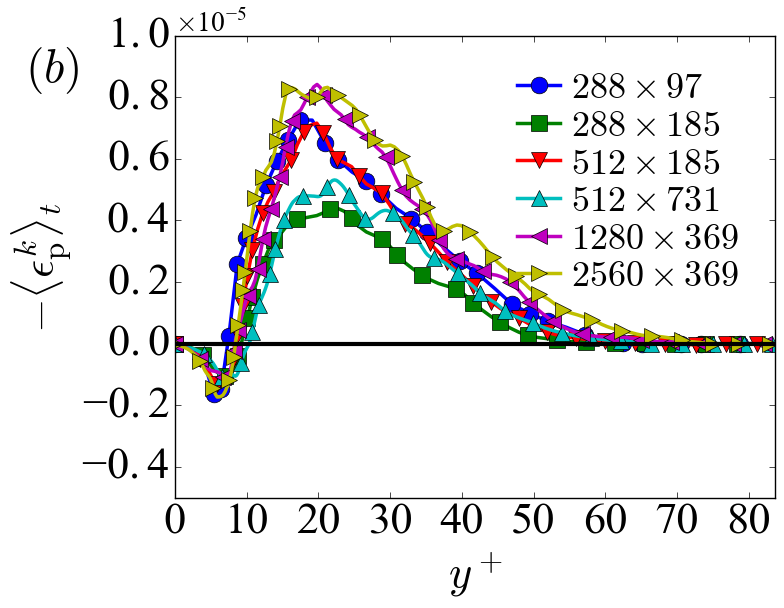}\\
	\caption{Effects of spatial resolution $N_x\times N_y\times N_z$ on the profiles of (a) $\mathrm{tr}(\mbf\alpha)$ (normalized by $b$) and (b) polymer elastic energy conversion rate $-\langle\epsilon_\text{p}^k\rangle_t$ of 2D EIT at $\mathrm{Wi}=64$ (all using HM with no AD).}
	\label{fig:profile_resol_eit}
\end{figure}

Resolution sensitivity of the $\mbf\alpha$ field is better reflected in the wall-normal profiles of $\mathrm{tr}(\mbf\alpha)/b$ (\cref{fig:profile_resol_eit}(a)).
Interestingly, the profile appears to be only dependent on $N_y$ and insensitive to changing $x$-resolution.
Numerical convergence is reached at $N_y=369$ -- two cases at this $N_y$ (with different $N_x$) shown in the figure are very close to the only $N_y=731$ case.
At lower $N_y$, a peak appears in the near-wall region. The $N_y=97$ profile in \cref{fig:profile_resol_eit}(a) closely resembles the $\mathrm{tr}(\mbf\alpha)/b$ profile reported in \citep{Sid_Terrapon_PRFluids2018} using $N_y=288$ (for $\mathrm{Wi}=40$, $\mathrm{Re}_\tau=85$, and the same box size as ours).
Since our HM uses non-uniform CGL grids in $y$, when compared at the same $N_y$, our $\delta_y^+$ values near the walls are much finer than that of a uniform grid.
The exact grid point distribution used in \citep{Sid_Terrapon_PRFluids2018} was not specified.
However, for their pure FD scheme, it is unlikely that any kind of Chebyshev grids was used.
We thus believe that the near-wall peak is an effect of inadequate $y$-resolution and conclude that refined resolution in the near-wall region is particularly important for capturing the polymer conformation (stress) field at EIT.

Finally, in \cref{fig:profile_resol_eit}(b), we plot the wall-normal profiles of polymer elastic energy conversion rate $-\langle\epsilon_\text{p}^k\rangle_t$ (see \cref{eq:epsilonp}; $\langle\cdot\rangle_t$ denotes time average) from different resolutions.
We expect this quantity to be most sensitive to spatial discretization since it involves both polymer stress fluctuation $\mbf\tau_\text{p}'$ and fluctuations in velocity gradients.
Profiles from the two highest resolutions tested are almost inseparable.
Their $-\langle\epsilon_\text{p}^k\rangle_t$ magnitudes are also the highest among all cases, which is expected -- since EIT relies on sharp stress shocks, numerically resolved stress variation is critical for fully capturing the driving force for this instability (i.e., $-\epsilon_\text{p}^k$).
Interestingly, the $288\times97$ and $512\times185$ cases, both of which have the desired $3:1$ mesh number ratio, predict $-\langle\epsilon_\text{p}^k\rangle_t$ profiles very close to the ultimate magnitudes from higher resolutions. This explains their superior performance of capturing the velocity spectra at lower resolution (\cref{fig:stat_resol_eit}(c)).
By contrast, the $512\times731$ case under-predicts the peak $-\langle\epsilon_\text{p}^k\rangle_t$ magnitude by $\sim 35\%$, even though its total number of meshes is comparable to the $1280\times 369$ case.
Also, the $288\times 185$ case is far less accurate than the $288\times 97$ case despite its higher $y$-resolution.
Both examples suggest that increasing the resolution in one direction is less effective than holding a proper mesh aspect ratio.

\section{Conclusions}\label{Sec_conclude}

In this study, a hybrid pseudo-spectral/finite-difference numerical algorithm for the DNS of viscoelastic turbulent channel flow is presented. 
It uses a TVD finite difference scheme to discretize the convection term of the FENE-P equation in space, whereas other spatial derivatives are discretized using the standard Fourier-Chebyshev-Fourier spectral scheme, by which spectral accuracy is retained to the maximal extent.
Numerically stable solutions are obtained for $\mathrm{Wi}$ up to $800$ (which is the highest level tested) without the need of artificial diffusion (either local or global AD).
It is also computationally efficient in comparison with not only other high-order FD schemes, but also the standard pseudo-spectral scheme.
Correctness of the code is validated against an earlier SM code in transient STG simulation.


Comparing simulation results between the new HM algorithm and our earlier SM algorithm, we conclude that discussion regarding the effects of GAD must differentiate between the IDT and EIT flow regimes.
For EIT, any level of GAD necessary for stabilizing the SM algorithm will be too large to capture the sharp stress shocks required for its sustenance. As such, a proper FD scheme for the polymer convection term is required.
For IDT, there is indeed a range of acceptable AD levels with which flow statistics and dynamics are reliably preserved.
Within this range, turbulent fluctuations are accurately captured across most of the spectrum, except the smallest scales which are mostly below the mesh size used in typical IDT simulation.
Fluctuations at the affected scales are reduced by AD. They likely reflect the incidental EIT-like structures occurring alongside the dominant IDT structures, which are not captured by SM with GAD.
As the level of AD exceeds the acceptable range, prediction can quickly deteriorate. In that case, AD results in larger (not reduced) velocity fluctuations across most length scales and leads to the underestimation of $\mathrm{DR}\%$ compared with the HM (no AD) case.

Effects of numerical resolution on the HM algorithm also differs between IDT and EIT regimes.
The required resolution for IDT is comparable to that of SM.
EIT, on the other hand, is very challenging to resolve completely.
With increasing streamwise resolution, numerical convergence is achieved at $\delta_x^+\sim0.5$, whereas in the wall-normal direction, finite mesh size results in the spurious accumulation of energy at the small-scale end for all resolution levels tested.
The mean $\mathrm{tr}(\mbf\alpha)$ profile is sensitive to wall-normal resolution and requires highly-refined meshes in near-wall regions, which is made easier by the non-uniform CGL grid used in HM. 
Interestingly, there appear to be a certain range of optimal mesh aspect ratios (adjusted by the $N_x/N_y$ mesh number ratio) that allow a much lower resolution to closely approximate the results from a highly-refined grid.


\section*{Acknowledgments}
The authors gratefully acknowledge the financial support from the Natural Sciences and Engineering Research Council of Canada (NSERC) through its Discovery Grants Program (No.~RGPIN-2014-04903) as well as the computing resource allocated by Compute/Calcul Canada.
This work is made possible by the facilities of the Shared Hierarchical Academic Research Computing Network (SHARCNET: \texttt{www.sharcnet.ca}).
We are grateful to John~F.~Gibson (U. New Hampshire), Tobias~M.~Schneider (EPFL), and others for making the Newtonian \texttt{Channelflow} codes available~\citep{Gibson_ChFlowCode,Gibson_ChFlowCode2p0}.
Assistance from Tobias~M.~Schneider, Hecke~Schrobsdorff (Max Planck Inst.), and Tobias Kreilos (EPFL) for implementing MPI in our SM code is also acknowledged.

\appendix
\section{Influence-matrix method}\label{appendix:tau}
As shown in \cref{Sec_TimeIntNS}, the Navier-Stokes equation for velocity and pressure fields, after discretization in time (AB/BD3) and $x$ and $z$ spatial dimensions (FFT), becomes a problem of solving \cref{eq:tau:vp,eq:tau:divv,eq:tau:bc} for each $(k_x,k_z)$ wavenumber pair.
Those equations are rewritten here in a more general form (after dropping superscripts $\dagger$, $n$, and $n+1$)
\begin{gather}
	\upsilon\frac{d^2\tilde{\mbf v}}{d y^2}-\lambda\tilde{\mbf v}
		- \widetilde{\mbf\nabla}\tilde{p} = - \widetilde{\mbf R}
	\label{eq:taugen:vp}\\
	\widetilde{\mbf\nabla}\cdot\tilde{\mbf v}=0
	\label{eq:taugen:divv}\\
	\left.\tilde{\mbf v}\right\vert_{y=\pm 1}=0
	\label{eq:taugen:bc}
\end{gather}
where
\begin{gather}
	\upsilon\equiv\frac{\beta}{\mathrm{Re}},\\
	\lambda\equiv\upsilon\varkappa^2+\frac{\zeta}{\delta_t},\\
	\varkappa^2\equiv4\pi^2\left(\frac{k_x^2}{L_x^2}+\frac{k_z^2}{L_z^2}\right).
\end{gather}
We follow the \citet{Kleiser_Schumann_Proc3GAMM1980} influence-matrix method to find numerical solutions of $\tilde{\mbf v}(y)$ and $\tilde p(y)$ to \cref{eq:taugen:vp} that consistently satisfies the constraint of \cref{eq:taugen:divv} and the boundary conditions of \cref{eq:taugen:bc}.
The method is discussed in detail in section~7.3.1 of \citet{Canuto_Hussaini_1988} and implemented in Newtonian \texttt{Channelflow} codes~\citep{Gibson_ChFlowCode,Gibson_ChFlowCode2p0}.

The problem narrows down to solving the following so-called
\begin{gather}
A\text{-problem:}\quad
\begin{dcases}
	\frac{d^2\tilde p}{dy^2}-\varkappa^2\tilde p
		= \widetilde{\mbf\nabla}\cdot\widetilde{\mbf R}
	\\
	\left.\frac{d\tilde v_y}{d y}\right\vert_{y=\pm 1} = 0
	\\
	\upsilon\frac{d^2\tilde v_y}{d y^2}-\lambda\tilde v_y
		- \frac{d\tilde p}{d y} = - \widetilde R_y
	\\
	\left.\tilde v_y\right\vert_{y=\pm 1}=0
\end{dcases}.
\label{eq:probA}
\end{gather}
It is clear that the last two equations are just the $y$-components of \cref{eq:taugen:vp,eq:taugen:bc}, while the first two are obtained by taking the divergence of \cref{eq:taugen:vp,eq:taugen:bc} and applying \cref{eq:taugen:divv}, noting that the Laplacian (according to \cref{eq:diffop:lap}) is simply
\begin{gather}
	\widetilde\nabla^2=\frac{d^2}{d y^2}-\varkappa^2.
\end{gather}
Once the $A$-problem is solved, $\tilde p(y)$ will be a known function and the $x$- and $z$-components of \cref{eq:taugen:vp}, with boundary conditions \cref{eq:taugen:bc}, are simply inhomogeneous Helmholtz equations to be solved with the Chebyshev-tau method~\citep{Canuto_Hussaini_1988} (the $y$-component will have already been solved as part of the $A$-problem).

The $A$-problem is still not readily solvable because it has two differential equations for $\tilde p$ and $\tilde v_y$, respectively, but both boundary conditions are for $\tilde v_y$.
If the Neumann boundary conditions for $\tilde v_y$ can be replaced by a pair of Dirichlet boundary conditions for $\tilde p$, the resulting hypothetical
\begin{gather}
B\text{-problem:}\quad
\begin{dcases}
	\frac{d^2\tilde p}{dy^2}-\varkappa^2\tilde p
		= \widetilde{\mbf\nabla}\cdot\widetilde{\mbf R}
	\\
	\left.\tilde p\right\vert_{y=\pm 1}=\widetilde P_\pm
	\\
	\upsilon\frac{d^2\tilde v_y}{d y^2}-\lambda\tilde v_y
		- \frac{d\tilde p}{d y} = - \widetilde R_y
	\\
	\left.\tilde v_y\right\vert_{y=\pm 1}=0
\end{dcases}
	\label{eq:probB}
\end{gather}
would be much easier to solve. Of course, the boundary values $\widetilde P_+$ and $\widetilde P_-$ are not known \textit{a priori}, but we know that the general solution to the $B$-problem can be formulated as a linear combination between one particular solution $(\tilde p_0,\tilde v_{y,0})$ from the
\begin{gather}
B_0\text{-problem:}\quad
\begin{dcases}
	\frac{d^2\tilde p_0}{dy^2}-\varkappa^2\tilde p_0
		= \widetilde{\mbf\nabla}\cdot\widetilde{\mbf R}
	\\
	\left.\tilde p_0\right\vert_{y=\pm 1}=0
	\\
	\upsilon\frac{d^2\tilde v_{y,0}}{dy^2}-\lambda\tilde v_{y,0}
		- \frac{d\tilde p_0}{dy} = - \widetilde R_y
	\\
	\left.\tilde v_{y,0}\right\vert_{y=\pm 1}=0
\end{dcases}
	\label{eq:probB0}
\end{gather}
and the basis of solutions $[(\tilde p_+,\tilde v_{y,+}), (\tilde p_-,\tilde v_{y,-})]$ from the
\begin{gather}
B_+\text{-problem:}\quad
\begin{dcases}
	\frac{d^2\tilde p_+}{dy^2}-\varkappa^2\tilde p_+ = 0
	\\
	\left.\tilde p_+\right\vert_{y=-1}=0
	\\
	\left.\tilde p_+\right\vert_{y=+1}=1
	\\
	\upsilon\frac{d^2\tilde v_{y,+}}{dy^2}-\lambda\tilde v_{y,+}
		- \frac{d\tilde p_+}{dy} = 0
	\\
	\left.\tilde v_{y,+}\right\vert_{y=\pm 1}=0
\end{dcases}
	\label{eq:probBplus}
\end{gather}
and
\begin{gather}
B_-\text{-problem:}\quad
\begin{dcases}
	\frac{d^2\tilde p_-}{dy^2}-\varkappa^2\tilde p_- = 0
	\\
	\left.\tilde p_-\right\vert_{y=-1}=1
	\\
	\left.\tilde p_-\right\vert_{y=+1}=0
	\\
	\upsilon\frac{d^2\tilde v_{y,-}}{dy^2}-\lambda\tilde v_{y,-}
		- \frac{d\tilde p_-}{dy} = 0
	\\
	\left.\tilde v_{y,-}\right\vert_{y=\pm 1}=0
\end{dcases}
	\label{eq:probBminus}
\end{gather}
which can be written as
\begin{gather}
	\begin{pmatrix}
		\tilde p
	\\
		\tilde v_{y}
	\end{pmatrix}
	=
	\begin{pmatrix}
		\tilde p_0
	\\
		\tilde v_{y, 0}
	\end{pmatrix}
	+
	\delta_+
	\begin{pmatrix}
		\tilde p_+
	\\
		\tilde v_{y, +}
	\end{pmatrix}
	+
	\delta_-
	\begin{pmatrix}
		\tilde p_-
	\\
		\tilde v_{y, -}
	\end{pmatrix}.
	\label{eq:gensol}
\end{gather}
Each of \cref{eq:probB0,eq:probBplus,eq:probBminus} consists of two Helmholtz equations solvable again by the Chebyshev-tau method.
In practice, the tau correction is applied in solving Helmholtz equations to counter discretization errors and improve numerical stability~\citep{Canuto_Hussaini_1988}.
Note that the $B_+$-  and $B_-$-problems are invariant over time (assuming the same constant $\delta_t$ is used) and only need to be solved once at the beginning for each $(k_x,k_z)$ with the results $[(\tilde p_+,\tilde v_{y,+}), (\tilde p_-,\tilde v_{y,-})]$ stored for the entire simulation duration.
Coefficients $\delta_+$ and $\delta_-$ are determined by subjecting the general solution \cref{eq:gensol} to the Neumann boundary conditions of the $A$-problem (which were replaced in the $B$-problem): i.e.
\begin{gather}
	\begin{pmatrix}
		\left.\frac{d\tilde v_{y, 0}}{dy}\right\vert_{y=+1} \\
		\left.\frac{d\tilde v_{y, 0}}{dy}\right\vert_{y=-1}
	\end{pmatrix}
	+
	\begin{pmatrix}
		\left.\frac{d\tilde v_{y, +}}{dy}\right\vert_{y=+1}
			& \left.\frac{d\tilde v_{y, -}}{dy}\right\vert_{y=+1} \\
		\left.\frac{d\tilde v_{y, +}}{dy}\right\vert_{y=-1}
			& \left.\frac{d\tilde v_{y, -}}{dy}\right\vert_{y=-1}
	\end{pmatrix}
	\begin{pmatrix}
		\delta_\mathrm{+} \\
		\delta_\mathrm{-}
	\end{pmatrix}
	=
	\begin{pmatrix}
		0\\
		0
	\end{pmatrix}.
\end{gather}
The $2\times2$ matrix in the second term is called the influence matrix.

\section{Comparison of numerical differentiation schemes in a benchmark convection problem}\label{Sec_schmComp}

\begin{figure}
	\centering
	\includegraphics[width=0.75\linewidth, trim=0mm 0mm 0mm 0mm, clip]{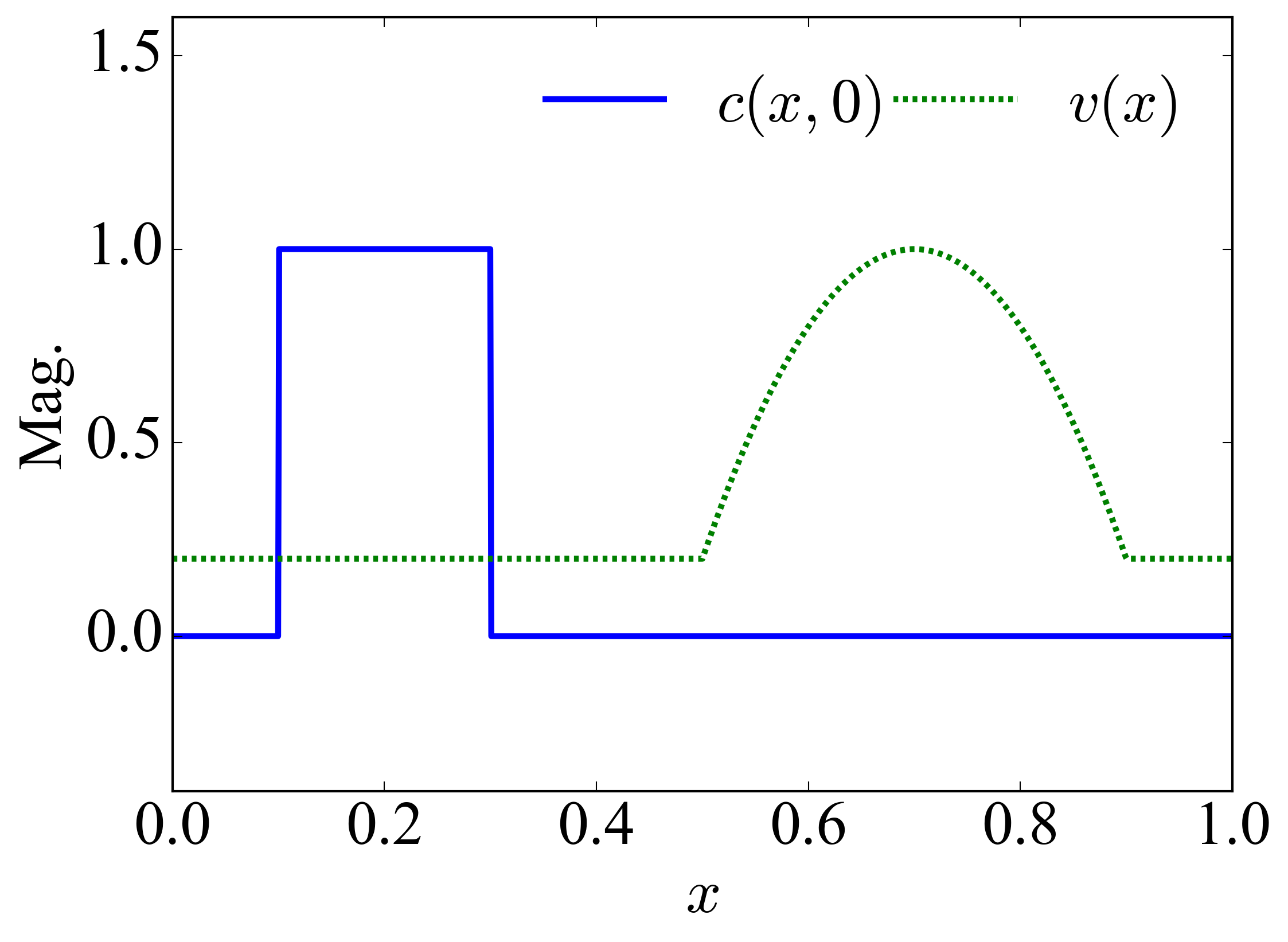}
	\caption{The initial concentration profile $c(x,0)$ and convective velocity profile $v(x)$ of the benchmark problem.}
	\label{fig:bench_ini}
\end{figure}

To compare the performance of several commonly-used numerical schemes for hyperbolic problems, we use a simple pure convection problem
\begin{equation}
	\frac{\partial c}{\partial t}+\frac{\partial (vc)}{\partial x}=0
\label{equ_bench}
\end{equation}
in a one-dimensional periodic domain of length $L$ as a benchmark.
\Cref{equ_bench} describes the time evolution of a concentration profile $c(x,t)$ with an, in our case, temporally-invariant convective velocity $v(x)$.
We may have $x$ nondimensionalized by $L$, $v$ nondimensionalized by the maximum velocity magnitude $U$, and $c$ nondimensionalized by the maximum initial concentration. The natural time unit is then $L/U$.

The initial condition of nondimensional $c$,
\begin{equation} 
	c(x,0)=
	\begin{dcases}
		1	&	0.1<x\leq 0.3\\
		0	&	\text{elsewhere}
	\end{dcases},
\label{equ_ben_iniCond}
\end{equation}
is a square wave. The marching speed of the concentration profile is determined by the nondimensional convective velocity,
\begin{equation} 
	v(x)=
	\begin{dcases}
		0.2+20(x-0.5)(0.9-x)	&	0.5<x\leq 0.9\\
		0.2						&	\text{elsewhere}
	\end{dcases}.
\label{equ_ben_convV}
\end{equation}
Although technically-speaking, for 1D flow of an incompressible fluid, the velocity must be invariant over $x$ (because of the continuity constraint $\partial v/\partial x=0$), we purposefully impose a spatially-varying velocity $v(x)$ to test the shock-capturing capability of different schemes in a non-uniform flow field.
\Cref{fig:bench_ini} shows the initial concentration profile $c(x,0)$ and the velocity profile $v(x)$ of the benchmark problem.

Time advancement of \cref{equ_bench} uses a third-order semi-implicit Adams-Bashforth/backward-differentiation scheme (AB/BD3)~\citep{peyret2002spectral}, which is consistent with that used in our DNS.
After temporal discretization, \cref{equ_bench} is written as
\begin{equation}
	\frac{\zeta}{\delta_t}c^{n+1}=-\sum^{2}_{j=0}\left(
			\frac{a_j}{\delta_t}c^{n-j}
			+b_j\frac{\partial\left(vc^{n-j}\right)}{\partial x}
		\right)
\label{equ_ben_timeDic}
\end{equation}
where numerical coefficients $\zeta$, $a_j$ and $b_j$ are provided in \cref{tab:BDAB3_coeff}.
Six different numerical differentiation schemes commonly used for the convection term in viscoelastic constitutive equations are tested here for the $\partial (vc)/\partial x$ term: (I) a second-order MINMOD TVD scheme (TVD2)~\citep{sweby1984high,Zhang_Jiang_JCompPhys2015}, (II) pseudo-spectral schemes (SM) without and (III) with GAD~\citep{Sureshkumar_Beris_POF1997,Xi_PhD2009}, (IV) a fifth-order WENO scheme (WENO5)~\citep{shu1998essentially,shu2009high}, (V) a third-order compact upwind scheme (CUD3)~\citep{Min_Choi_JNNFM2001,Dubief_Lele_FTC2005}, and (VI) a second-order upwind scheme (UD2)~\citep{Zhang_Jiang_JCompPhys2015}.
A brief description of each scheme is summarized below.

The TVD2 scheme implemented in the benchmark problem is the same as that described in \cref{Sec_PolyConv}. Let $F\equiv vc$ be the numerical flux. The convection term at the grid point $q$ can be discretized by following \cref{equ_ben_conv2,equ_ben_conv3,equ_ben_conv4,equ_ben_conv5,equ_ben_tvd,equ_ben_tvd2,equ_ben_tvd3}.

Same as TVD2, WENO5 and UD2 schemes also adopt LLFS  (\cref{equ_ben_conv3}) to guarantee the upwindness in numerical differentiation.
They differ in the specific FD expressions used to approximate the fluxes at cell edges.
UD2 uses the same formula as \cref{equ_ben_tvd} except that the flux limiter function $\phi=1$, which, for the $q+1/2$ edge, becomes
\begin{equation}
	\begin{dcases}
		F^+_{q+1/2}=F^+_q+\frac{1}{2}\left(F^+_q-F^+_{q-1}\right)\\
		F^-_{q+1/2}=F^-_{q+1}+\frac{1}{2}\left(F^-_{q+1}-F^-_{q+2}\right)
	\end{dcases}.
\label{equ_ben_ud2}
\end{equation}
In WENO5, the edge flux (again taking $q+1/2$ for illustration) is expressed as a weighted sum in the form of
\begin{equation}
	F^+_{q+1/2} = \omega_1 F^{+,(1)}_{q+1/2}+
			\omega_2 F^{+,(2)}_{q+1/2}+\omega_3 F^{+,(3)}_{q+1/2}
\label{equ_ben_weno1}
\end{equation}
where
\begin{equation} 
	\begin{dcases}
		F^{+,(1)}_{q+1/2} = \frac{1}{3}F^{+}_{q-2}
			-\frac{7}{6}F^{+}_{q-1}+\frac{11}{6}F^{+}_{q}\\
		F^{+,(2)}_{q+1/2} = -\frac{1}{6}F^{+}_{q-1}
			+\frac{5}{6}F^{+}_{q}+\frac{1}{3}F^{+}_{q+1}\\
		F^{+,(3)}_{q+1/2} = \frac{1}{3}F^{+}_{q}
			+\frac{5}{6}F^{+}_{q+1}-\frac{1}{6}F^{+}_{q+2}
	\end{dcases}
\label{equ_ben_weno2}
\end{equation}
are three different third-order FD expressions for the edge flux.
Weight coefficients $\omega_1$, $\omega_2$, and $\omega_3$ are determined by a so-called smoothness analysis which measures the relative smoothness of the solution from each flux expression in the local stencil and assign higher weights to expressions giving smoother solutions~\citep{shu1998essentially,shu2009high}.
The resulting combination \cref{equ_ben_weno1} is fifth-order accurate.
Note that for the positive flux $F^+_{q+1/2}$, the combined local stencil $[q-2,q-1,q,q+1,q+2]$ is left-biased in relation to $q+1/2$, which reflects the upwindness of the scheme.
For the negative flux $F^-_{q+1/2}$, a right-biased stencil ($[q-1,q,q+1,q+2,q+3]$) is used and the corresponding expressions are mirror images of \cref{equ_ben_weno1,equ_ben_weno2} w.r.t. to $q+1/2$.

The CUD3 and SM schemes do not use the conservative form of \cref{equ_bench}, which is rewritten as
\begin{equation}
	\frac{\partial c}{\partial t}+v\frac{\partial c}{\partial x}
		+c\frac{\partial v}{\partial x}=0.
\label{equ_bench_noncons}
\end{equation}
Since the convective velocity (\cref{equ_ben_convV}) is independent of time, $\partial v/\partial x$ in the last term is analytically evaluated.
In CUD3, $\partial c/\partial x$ in the second term is obtained by solving
\begin{equation}
\begin{split}
	\left(2-3s_q\right)\frac{\partial c_{q+1}}{\partial x}
		&+8\left(\frac{\partial c_q}{\partial x}\right)
		+\left(2+3s_q\right)\frac{\partial c_{q-1}}{\partial x}
	\\&
	=\frac{6}{\delta_x}\left(
		\left(1-s_q\right)c_{q+1}+2s_q c_q+\left(1+s_q\right)c_{q-1}
	\right).
\end{split}
\label{equ_bench_cud3}
\end{equation}
for all $q$, where
\begin{gather}
	s_q\equiv\mathrm{sign}\left({v_q}\right)
\end{gather}
is the sign of the velocity at point $q$.

As to SM with GAD, the nondimensional governing equation becomes
\begin{gather}
	\frac{\partial c}{\partial t}=-N_c+\frac{1}{\mathrm{Pe}}\frac{\partial^2 c}{\partial x^2}
	\label{eq:benchmark:gad}
\end{gather}
where
\begin{gather}
	N_c\equiv v\frac{\partial c}{\partial x}+cv'(x)
	\label{eq:benchmark:Nc}
\end{gather}
groups the nonlinear terms and
\begin{gather}
	\mathrm{Pe}\equiv\frac{UL}{D}
\end{gather}
is the Peclet number ($D$ is the numerical diffusivity) and $1/\mathrm{Pe}$ can be viewed as the nondimensional numerical diffusivity.
Note that the notation
\begin{gather}
	v'(x)\equiv\frac{\partial v}{\partial x}
\end{gather}
is used to indicate that this derivative is analytically evaluated.
Time integration and numerical differentiation in $x$ are performed in the Fourier space
\begin{gather}
	\frac{\zeta}{\delta_t}\tilde c^{n+1}_{k_x}
	+\frac{4\pi^2}{\mathrm{Pe}}\frac{k_x^2}{L_x^2}\tilde c^{n+1}_{k_x}
	=-\sum^{2}_{j=0}\left(
			\frac{a_j}{\delta_t}\tilde c^{n-j}_{k_x}
			+b_j\widetilde N_{c,k_x}^{n-j}
		\right)
\end{gather}
while $\widetilde N_{c,k_x}$ is calculated by first evaluating the spatial derivative in the Fourier space
\begin{gather}
	\widetilde{\frac{\partial}{\partial x}}\tilde c_{k_x} = 2\pi i\frac{k_x}{L_x}\tilde c_{k_x},
\end{gather}
applying inverse FFT on the result, calculating $N_c$ according to \cref{eq:benchmark:Nc}, and applying FFT on $N_c$.
The domain length $L_x=1$ in the nondimensionalized system.
Same as DNS (\cref{sec:SM}), the GAD term is treated implicitly in time integration. The term is simply discarded for the no-GAD SM scheme.

\begin{figure}
	\centering
	\includegraphics[width=\linewidth, trim=0mm 0mm 0mm 0mm, clip]{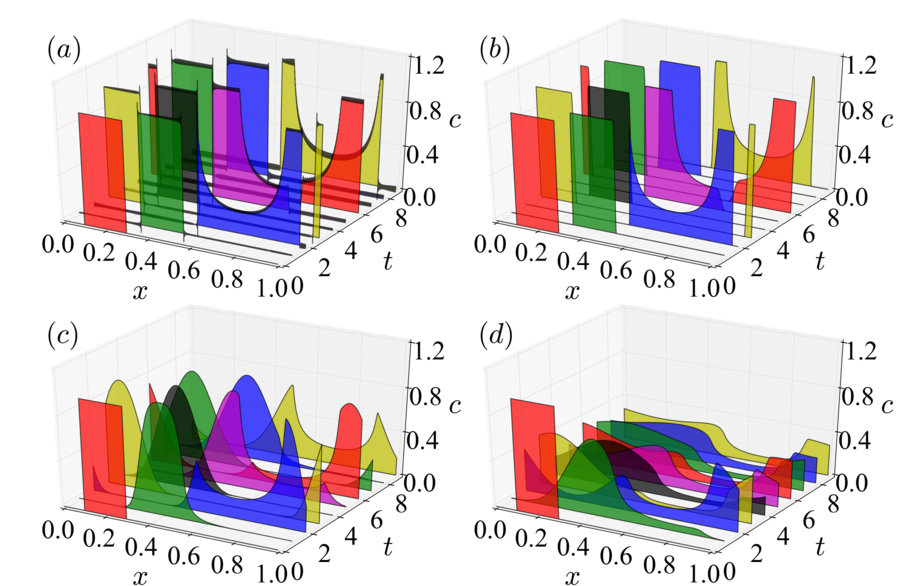}
	\caption{Temporal evolution of the concentration profile $c(x,t)$ using the SM:
	(a) $1/\mathrm{Pe}=0$ (no GAD);
	(b) $1/\mathrm{Pe}=\num{5e-7}$;
	(c) $1/\mathrm{Pe}=\num{5e-4}$;
	(d) $1/\mathrm{Pe}=\num{5e-3}$.}
	\label{fig:SM:evol}
\end{figure}

\begin{figure}
	\centering
	\includegraphics[width=0.85\linewidth, trim=0mm 0mm 0mm 0mm, clip]{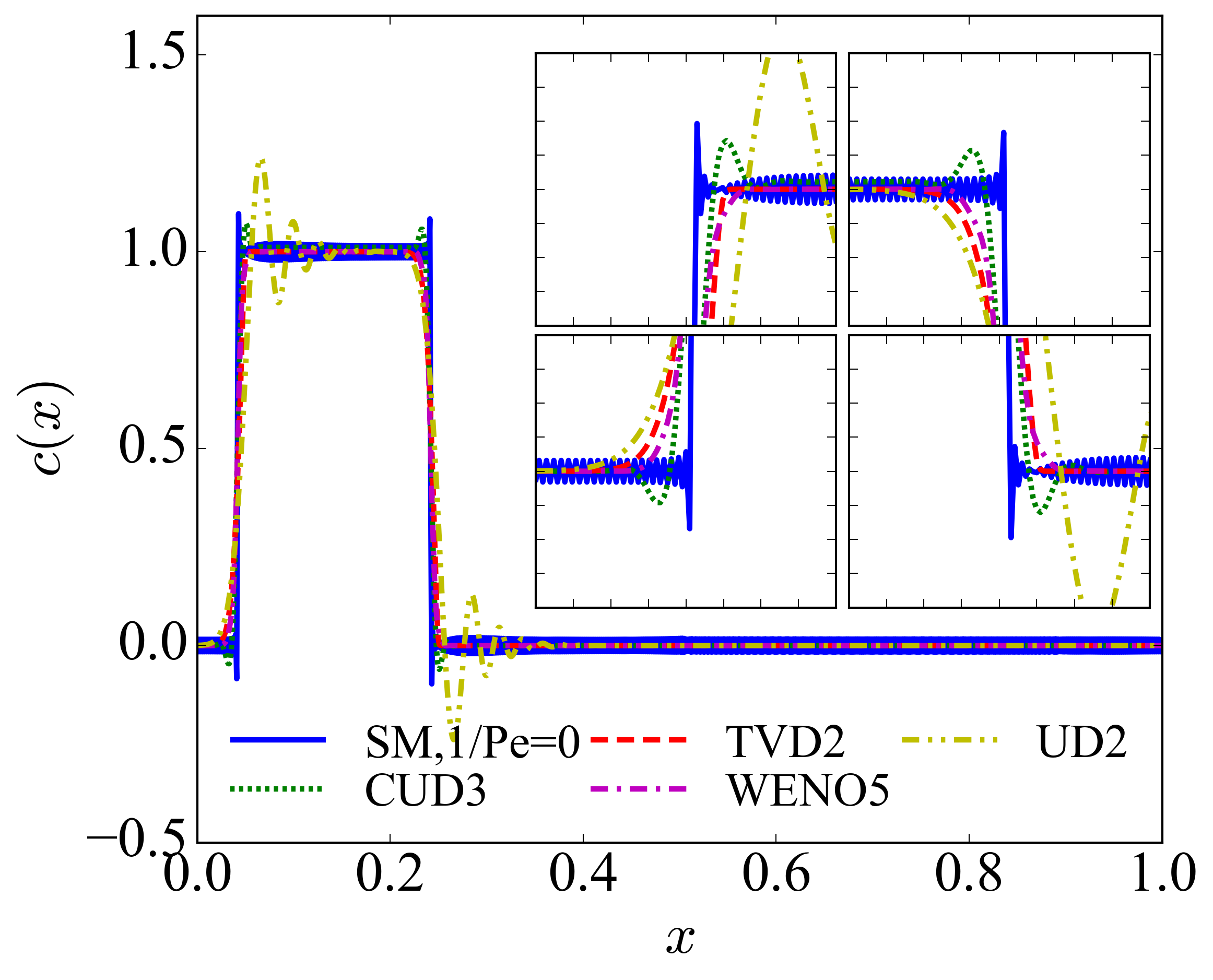}
	\caption{Comparison of the concentration profile at $t=7$ using different spatial discretization schemes. Insets show enlarged views of all four corners of the square wave (with grid spacings of $0.01$ and $0.05$ in the $x$ and $c$ axes, respectively -- i.e., magnification of the $x$ axis is $\approx 2.5$ times that of the $c$ axis).}
	\label{fig:bench_comp}
\end{figure}

\begin{table}
\begin{center}
	\begin{tabular}{@{}c@{}|@{}ccccc@{}}
		\hline	
		Scheme	&TVD2	&\parbox{8ex}{\centering SM/\par no-GAD}	&UD2		&WENO5	&CUD3\\
		\hline
		CPU time (\si{s})	&	$2.18$	&	$1.92$	&	$2.04$		&	$16.80$		&	$15.64$\\
		\hline
	\end{tabular}
	\caption{CPU time required by each scheme to run a $10^4$ time-step benchmark simulation on a single core of an \texttt{Intel}\textsuperscript\textregistered \SI{1.80}{GHz} i3-3217U CPU.}
	\label{tab:cpu_scheme}
\end{center}	
\end{table}

A uniform mesh with $\delta_x=\num{9.77e-4}$ and time step of $\delta_t=\num{9.77e-5}$ are used for all results reported here.
Evolution of the resulting concentration profile from SM without GAD ($1/\mathrm{Pe}=0$) and with different levels of GAD are compared in \cref{fig:SM:evol}.
The wave travels from left to right repeatedly across the periodic domain. As it enters the $x\in(0.5,0.9]$ region, its sharp boundaries are stretched into curves by the nonuniform velocity field.
Without numerical errors, a sharp square wave will be restored after it fully passes the region, which is clearly observed in the no-GAD case.
Indeed, as shown more clearly in \cref{fig:bench_comp}, even compared with FD schemes, the no-GAD SM scheme most accurately preserves the sharpness of the shock fronts.
(Note that $t=7$ shown in \cref{fig:bench_comp} is after the wave has passed the non-uniform velocity region twice.)
Approximating solutions containing discontinuities with smooth continuous Fourier series, however, leads to the emergence of spurious oscillation across the domain which is strongest around the shocks. This is the well-known Gibbs phenomenon~\citep{Canuto_Hussaini_1988}.
In DNS, no-GAD SM tends to lose numerical stability for sufficiently high $\mathrm{Wi}$, which covers most cases of interest in DR study.

Applying GAD can effectively suppress numerical oscillations at the expense of shock-capturing accuracy, as the concentration profile becomes smeared over time.
The smearing process accelerates with increasing $1/\mathrm{Pe}$, but at the lowest $1/\mathrm{Pe}=\num{5e-7}$ presented in \cref{fig:SM:evol}, sharp shock fronts are preserved up to $t=9$ shown in the figure.
In 3D turbulence, stress shocks must be transient in nature. Therefore, effects of GAD would be minimal if large stress gradients are not substantially smeared over their lifetime.
The standard $\mathrm{Sc} = 0.5$, which corresponds to $1/\mathrm{Pe}=1/(\mathrm{ReSc})=\num{5.56e-4}$, used in our IDT DNS does not strongly affect the results (see, e.g., \cref{fig:DR_wi}).
Meanwhile, at a similar $1/\mathrm{Pe}=\num{5e-4}$ (\cref{fig:SM:evol}(c)), the square wave in the benchmark problem is significantly smeared within the flow-through time.
This suggests that stress gradients at IDT are likely not as steep and sharp stress changes, if important in IDT dynamics, appear only for short time periods.

All FD schemes result in slightly reduced slopes at the shock front compared with SM with no-GAD (\cref{fig:bench_comp}).
The effect is strongest in UD2, which also comes with strong oscillations near the shocks.
CUD3 provides good approximation to the shock fronts themselves but it also shows unphysical over- and under-shoots nearby, which may explain why LAD was necessary in DNS using this scheme~\citep{Min_Choi_JNNFM2001,Dubief_Lele_FTC2005}.
Interestingly, the result from TVD2 is comparable to that of WENO5 despite the latter's higher order of accuracy.
Computational cost is also an important consideration, since DNS of viscoelastic fluids is already a computationally expensive problem.
As shown in \cref{tab:cpu_scheme}, the efficiency of TVD2 and UD2 is comparable to that of SM, while WENO5 and CUD3 take about $8$ times as much CPU time to complete the same number of time steps.
As such, after balancing efficiency and accuracy considerations, TVD2 is adopted in this study.






%
%
%
\bibliographystyle{elsarticle-num-names}
\bibliography{Zhu_bibtex,General,FluidDyn,Polymer}
\end{document}